\documentclass[acmsmall]{acmart}
\usepackage{booktabs}
\usepackage{graphicx}  
\usepackage{subfigure}
\usepackage{epstopdf}
\usepackage{amsmath}
\usepackage{cases}
\usepackage{makecell}
\usepackage{multirow}
\usepackage{array}
\usepackage{amsfonts}
\usepackage{algorithmic}
\usepackage[ruled]{algorithm2e} 

\AtBeginDocument{%
  \providecommand\BibTeX{{%
    \normalfont B\kern-0.5em{\scshape i\kern-0.25em b}\kern-0.8em\TeX}}}

\setcopyright{acmcopyright}
\copyrightyear{2021}
\acmYear{2021}
\acmDOI{10.1145/1122445.1122456}

\acmJournal{TKDD}
\acmVolume{0}
\acmNumber{0}
\acmArticle{0}
\acmMonth{0}

\begin{document}

\title{Recurrent Coupled Topic Modeling over Sequential Documents} 


\author{Jinjin Guo}
\email{yb57414@um.edu.mo}
\orcid{} 
\affiliation{%
	\institution{State Key Lab of IoTSC and Department of Computer and Information Science, University of Macau}
	\streetaddress{}
	\city{Macau S.A.R}
	\postcode{999078}
	\country{China}
}

\author{Longbing Cao}
\email{LongBing.Cao@uts.edu.au}
\affiliation{%
	\institution{Data Science Lab, University of Technology Sydney}
	\streetaddress{Ultimo, New South Wales 2007}
	\city{Sydney}
	\country{Australia}
}

\author{Zhiguo Gong}
\email{fstzgg@um.edu.mo}
\affiliation{%
	\institution{State Key Lab of IoTSC and Department of Computer and Information Science, University of Macau}
	\streetaddress{}
	\city{Macau S.A.R}
	\postcode{999078}
	\country{China}
}

\begin{abstract}

The abundant sequential documents such as online archival, social media and news feeds are streamingly updated, where each chunk of documents is incorporated with smoothly evolving yet dependent topics. Such digital texts have attracted extensive research on dynamic topic modeling to infer hidden evolving topics and their temporal dependencies. However, most of the existing approaches focus on single-topic-thread evolution and ignore the fact that a current topic may be coupled with multiple relevant prior topics. In addition, these approaches also incur the intractable inference problem when inferring latent parameters, resulting in a high computational cost and performance degradation. In this work, we assume that a current topic evolves from all prior topics with corresponding coupling weights, forming the \textit{multi-topic-thread evolution}. Our method models the dependencies between evolving topics and thoroughly encodes their complex multi-couplings across time steps. To conquer the intractable inference challenge, a new solution with a set of novel data augmentation techniques is proposed, which successfully discomposes the multi-couplings between evolving topics. A fully conjugate model is thus obtained to guarantee the effectiveness and efficiency of the inference technique. A novel Gibbs sampler with a backward-forward filter algorithm efficiently learns latent time-evolving parameters in a closed-form. In addition, the latent Indian Buffet Process (IBP) compound distribution is exploited to automatically infer the overall topic number and customize the sparse topic proportions for each sequential document without bias. The proposed method is evaluated on both synthetic and real-world datasets against the competitive baselines, demonstrating its superiority over the baselines in terms of the low per-word perplexity, high coherent topics, and better document time prediction.

\end{abstract}


%


\maketitle
\newcommand{\stitle}[1]{\vspace*{0.2em}\noindent{\bf #1\/}}

\section{Introduction}
The all-the-time update of abundant digital documents, such as Google News, Twitter and Flickr, have generated large amounts of sequential temporal-tagged documents, exhibiting complex temporal dependencies across the time steps. Such temporal-tagged digital documents have attracted extensive studies on the \textit{time-evolving} nature of topics. A successful way for such a task is to divide the collection into a sequence of document chunks and each chunk corresponds to a time-slice incorporated with topics in the temporal period \cite{blei2006dynamic,ahmed2008dynamic,iwata2009topic,schein2016poisson,acharya2018dual,guo2020deep}. Then, the problem of topic evolution could be addressed by studying relationships between topics crossing two adjacent time slices.   

Though fruitful results have been obtained in this area, most of the existing approaches are contracted with the single-topic-thread assumption, i.e., a topic in the current time-slice can only develop into a single topic in the subsequent slice \cite{blei2006dynamic,wang2008continuous,iwata2009topic,liang2016dynamic, acharya2018dual}. Obviously, this assumption cannot well align with the reality. Taking the news about COVID-19 as an example, the topic of coronavirus outbreak not only develops itself with intensive reports along the time but also triggers other topics such as the shortage of medical masks, shutdown of entertainment venues, and flight suspension. On the other hand, a new topic (e.g., work resumption) could be coupled with multiple prior topics (e.g., the effective control of coronavirus pandemic and market pressure). Such multi-topic coupling relationships over time are quite common and complex in the real world \cite{WangSC13,ChengMWC13,ipm_Cao15,HaoSNC18}, which pose significant challenges to the existing dynamic topic modeling techniques \cite{blei2006dynamic,wang2008continuous,iwata2009topic,liang2016dynamic, acharya2018dual}. This paper investigates this multi-topic coupling nature by assuming the \textit{multi-topic-thread evolution}, and proposes the \textit{recurrent Coupled Topic Model} (rCTM) to learn the multiple probabilistic dependencies between topics.



\subsection{Limitations of the Existing Work}
Two limitations of the existing work on topic evolution relevant to this paper are discussed in this section.

\subsubsection{Single-Topic-Thread Evolution}
A well-known mechanism for analyzing the temporal evolution of topics is the state space model for the dynamic topic modeling \cite{blei2006dynamic,wang2008continuous}, where the temporal dependency between evolving topics is captured by Gaussian distributions. Another widely used mechanism exploits a Dirichlet distribution \cite{iwata2009topic,liang2016dynamic} to encode the temporal dependency. Despite their difference in encoding the temporal development of topics, one common limitation lies in their single-topic-thread assumption as mentioned above. This violates the nature of many real cases.

\begin{figure}[!htb]
	\centering
	\includegraphics[width=.99\textwidth]{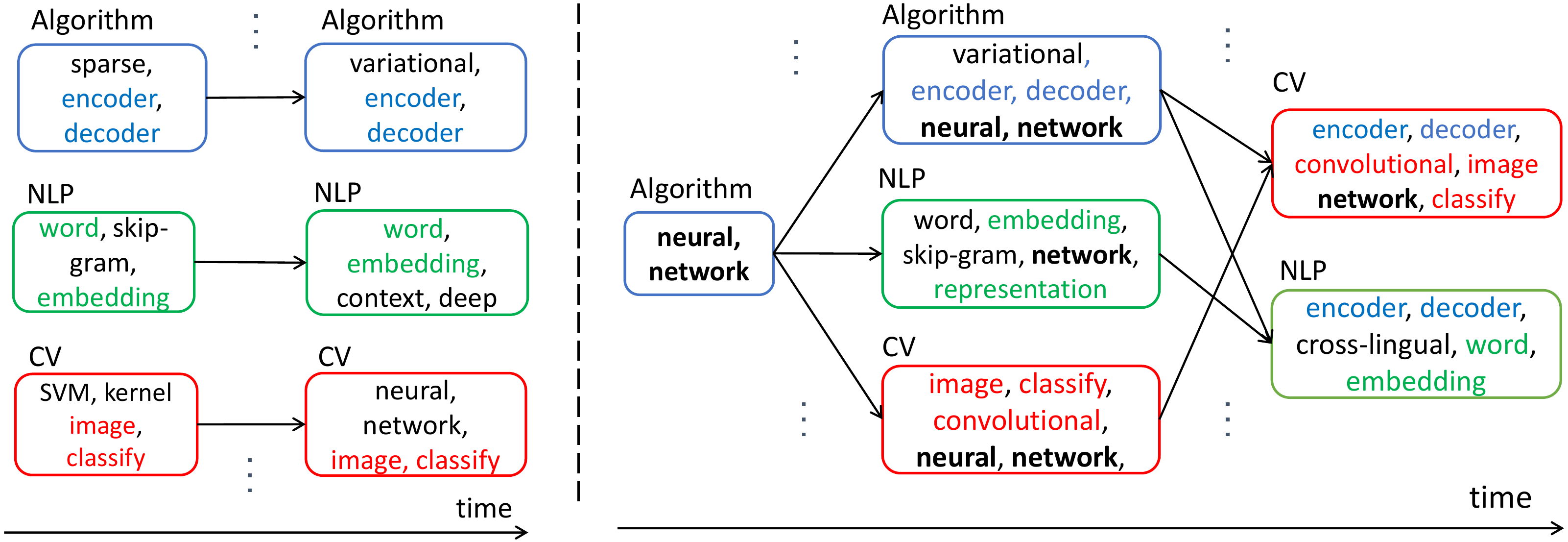}	
	\caption{An example of topic evolving process in terms of the conventional dynamic modeling (left) and the proposed recurrent Coupled Topic Modeling (right), where the left side denotes topics evolving under the single-topic-thread assumption, while the right side corresponds to the multi-topic-thread evolution.}
	\label{fig:intro}
\end{figure}

As noted in Fig. \ref{fig:intro}, the left side presents a topic evolutionary process following the single-topic-thread assumption, where each topic develops itself in a single thread, and the description words evolve in different slices.
For example, the evolution of content about Algorithm depends only on its own past state and ignores the influence of other prior topics such as Natural Language Process (NLP) and Computer Vision (CV).
This oversimplified evolution model does not reflect the reality in the real world. In contrast, the right side in the figure corresponds to an example of multi-thread-dependent evolutionary process, where the content on Algorithm not only develops itself but also significantly influences NLP and CV. 
Further, the content on CV in the last slice evolves not only from its past content but also being influenced by Algorithm. Such multi-thread influence on the posterior topics is reinforced by the highlighted common words, for example, the content on CV in the last slice shares common words from the prior topics of Algorithm and CV.

This example reinforces the fact that the development of topics is not constrained in one thread, rather, multiple topics are interactively coupled with each other \cite{ChengMWC13,HaoSNC18}. Without thoroughly encoding the complex temporal dependencies between evolving topics, the detected topic sequence from those conventional dynamic models might be defective. Therefore, our work aims to address this problem and learn multiple probabilistic dependencies between topics.

\subsubsection{Fixed Topic Number in Documents}
With some old topics phasing out and new ones coming in, the number of topics in each time-slice can be significantly different. Though existing studies \cite{Teh04hierarchicaldirichlet,acharya2018dual} are enabled to automatically learn the overall topic number for a collection of documents, they ignore the fact that each specific document from the collection may only involve a very small subset of those topics. Associating all topics with each individual document may cause the topic sparsity problem. Taking the collection of the published conference papers in a year as an example, the overall involved topics are diverse and numerous, where each individual paper is only related to very few of those topics, and the topics vary from paper to paper. It is clear that the traditional practice of assigning all topics to each document is very inappropriate, resulting in noisy topics assigned to a document and thus degrading the performance. Typically, the problem gets worse in the task of sequential short texts \cite{li2017enhancing}. The prior work \cite{yin2018model,liang2016dynamic} mitigates the sparsity problem to some extent by restricting one-topic assignment for a document, however, such a setting ignores the case of long documents which may contain more topics.
Therefore, in terms of topic settings for both document chunks and individual documents, a unified and powerful mechanism is required to simultaneously infer both the overall topic number for each slice and the sparse topic number for individual documents.

\subsection{Our Contributions}
Motivated by the above discussion, this paper introduces a new Bayesian sequential model \textit{recurrent Coupled Topic Modeling} (rCTM) over sequential documents through the following proposals.

First, we assume that a topic $\mathbf{\phi}_{k_t}$ ($k_t \in \{1, \cdots, K_t\}$) at slice $t$ evolves from all prior topics $ \mathbf{\Phi}_{t-1}$ of slice $t-1$ with the corresponding coupling weight $\beta_{k_{t-1} k_t}$ w.r.t. a Dirichlet distribution, and the distinguishable weights associated with the prior topics are learned from hierarchical Gamma distributions. The proposal induces a new and more flexible framework that topic $\mathbf{\phi}_{k_t}$ jointly depends on multiple prior topics, and a prior topic $\mathbf{\phi}_{k_{t-1}}$ could also contribute to multiple topics at step $t$, which breaks the single-topic-thread limitation of  the existing dynamic topic modelings \cite{blei2006dynamic,wang2008continuous,iwata2009topic,liang2016dynamic, acharya2018dual}. Hence, the complex multi-topic-thread dependencies between time-evolving topics are thoroughly encoded by this proposal.

Second, the above new proposal of the multi-coupling relationships between evolving topics induces an unexplored and intractable inference problem, which significantly challenges the existing inference techniques. To fully solve this problem, we propose a novel solution with a set of novel data augmentation and marginalization techniques, which is the main novel contributions of this paper. Our solution also discloses that the coupling weight between consecutive topics is indeed indicated by their shared latent word occurrences, and accordingly  a novel negative binomial distribution is incorporated into the inference framework to obtain the latent word occurrences. Finally, with novel data augmentation, the joint multi-dependency between topics is discomposed into separated relationships and each coupling weight turns to be measurably independent, leading to a fully conjugate and interpretable Bayesian model.

Third, with the update of sequential document chunk, no one knows its optimal topic setting at each time-slice. In addition, each document only talks about a sparse number of topics, which remains unknown and varies from document to document. To fully tackle these problems, we leverage a nonparametric prior, a latent Indian Buffet Process (IBP) compound distribution \cite{williamson2010ibp,doshi2015graph}, to solve the sparsity problem over the document-topic matrix. In addition to the unbound topic number at each slice, the mechanism of IBP allows each document to contain its customized latent topics without bias.

With the aid of novel data augmentation and marginalization techniques, a new Gibbs sampler with a backward-forward filter algorithm is proposed to approximate latent time-evolving parameters. In this algorithm, at each iteration latent word counts are propagated backward from slice $T$ to the initial slice, and the latent parameters are drawn forward from the initial slice to slice $T$ with updated word counts. To validate the significance of multi-topic coupled dependencies from the prior topics, we design a variant model injected by a dropout technique from neural networks to prune the couplings with the prior topics.
We explore both synthetic and real-world datasets with varying document lengths to evaluate the performance of rCTM against the competitive baselines. The extensive experimental results confirm the superiority of rCTM in terms of the low per-word perplexity, high topic coherence and better document time prediction.

To our best knowledge, this is the first paper to address the \textit{coupled topic modeling} problem, to which we make the following novel contributions:
\begin{itemize}
	\item A new and general framework of encoding multi-topic-thread evolution is proposed for sequential document analysis, where a topic in the current slice may be flexibly influenced by multiple prior topics, and also develop into multiple threads with corresponding weights in the subsequent slice.	
	\item A novel solution with data augmentations is presented to solve the unexploredly intractable problem and thoroughly decode the complex multi-dependencies between topics. rCTM thus enjoys a full conjugacy, where not only the evolution of topics across the slices but also their coupling relationships are efficiently captured in a closed-form.
	\item Without the manual setting of the topic number, a nonparametric mechanism, a latent IBP compound distribution, is leveraged to automatically learn the whole topic number for a document chunk as well as the sparse topic numbers for individual documents. Such a mechanism solves the topic sparsity problem and 
	flexibly accommodates both long and short documents. 
\end{itemize}


\section{The Proposed Model}
We discretize a collection of temporally sequential documents into $T$ time slices $\{\mathbf{d}_t|1\leq t\leq T\}$, where $\mathbf{d}_t$ is the document chunk of the $t$-th slice with $|\mathbf{d}_t|$ documents,  and each document in the chunk is represented by a bag-of-words with $t$-th timestamp. Given the sequential documents, the word dictionary with $V$ unique words is predefined. Before introducing our multi-topic-thread model, we define some notations and functions.

In what we present below, vectors and matrices are denoted by bold-faced lowercase and capital letters respectively and scalar variables are written in italic. $Dir_V()$, $Gam()$, $Mult()$, $Pois()$ and $Bern()$ stand for the $V$-dimensional Dirichlet, Gamma, multinomial, Poisson and Bernoulli distribution respectively. For a tensor $X \in \mathbb{Z}^{K_1 \times K_2 \times K_3}$ the $(k_1, k_2, k_3)$ entry is denoted by $x_{{k_1}{k_2}{k_3}}$. Also $x_{{k_1}{k_2}{\cdot}} = \sum_{k_3}^{K_3}x_{{k_1}{k_2}{k_3}}$ and $x_{k_1\cdot\cdot} = \sum_{k_2}^{K_2}\sum_{k_3}^{K_3}x_{{k_1}{k_2}{k_3}}$.

\subsection{The Multi-Topic-Thread Generative Process}
\label{sub}

The proposed \textit{recurrent Coupled Topic Modeling} with multiple threads consists of two important integrated components: (1) the topic proportion learning, which automatically determines the total number of topics over the slice and sparsifies the affinity between topics and documents, and (2) the multi-topic-thread evolution, which incorporates the joint multiple dependencies between consecutive topics. 

\stitle{Topic proportion learning.} 
Given a document chunk $\mathbf{d}_t$ of slice $t$, the hidden topics not only evolve from prior slice $t-1$, but may also come as new. Hence, the topic number $K_t$ may change from slice to slice. In the existing work, the Hierarchical Dirichlet Process (HDP) \cite{Teh04hierarchicaldirichlet} is widely used to determine the topic number. However, the HDP induces a rich-gets-richer problem, such that the infrequent topics are always overwhelmed by the popular ones \cite{williamson2010ibp}. For example, an article from the conference paper collections on Bayesian Network could be dominated by the popular topic of Neural Network in its topic assignment. Furthermore, the HDP ignores the topic sparsity problem for individual documents, which may bring noise topic intruding in the topic assignment. 

To resolve the above mentioned problems, the latent Indian Buffet Process (IBP) Compound Distribution is exploited in the proposed model to get rid of the rich-gets-richer harm and boost the rare topics in the topic assignment of documents. In addition, it also enables a sparsity mechanism for each document to select its customized topics via the Bernoulli technique.

\begin{figure}[!htb]
	\centering
	\subfigure[The sparse document-topic affinity matrix $\overline{\mathbf{\theta}}$ for document chunk $\mathbf{d}_t$.]{
		\includegraphics[width=.53\textwidth]{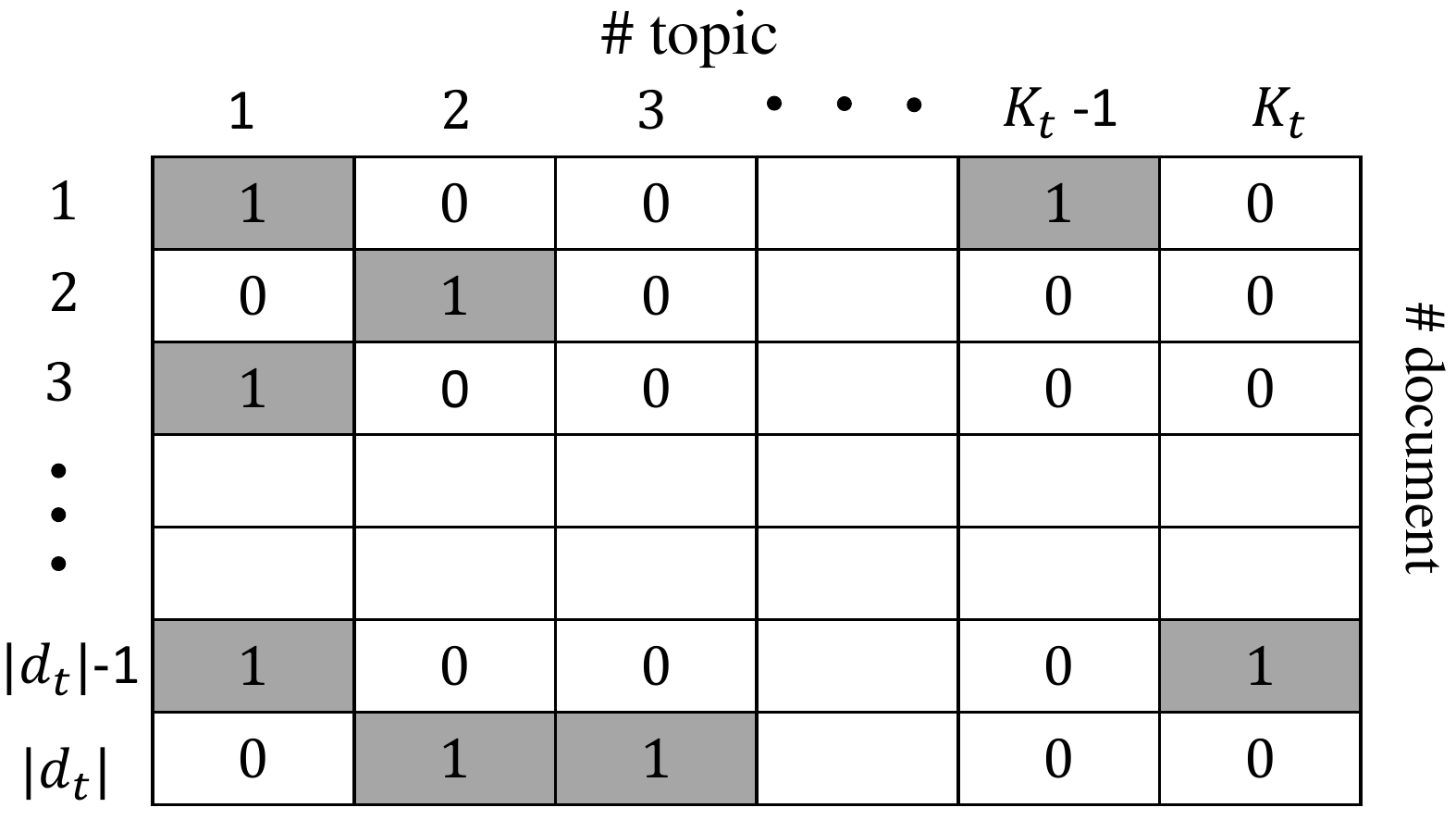}}
	\hspace{0.2in}	
	\subfigure[Graphical representation of topic proportions at slice $t$.]{
		\includegraphics[width=.37\textwidth]{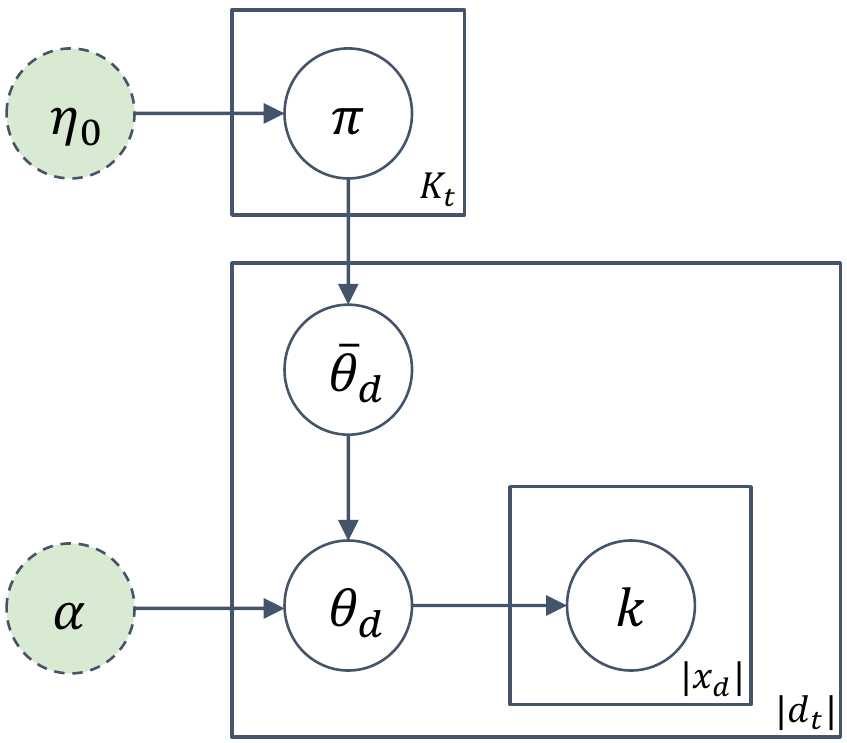}}
	\caption{The construction of topic proportions at slice $t$. In Fig.(a), each document in $\mathbf{d}_t$ contains its customized topics marked with $1$ in the shaded color, and those topics excluded are left as $0$ in blank, which are determined via the IBP mechanism. Fig.(b) presents the graphical representation of topic proportion construction, where the circles with dash lines indicate the specified hyper-parameters, and the rest denote latent variables.}
	\label{fig:table}
\end{figure}


In detail, as shown in Fig. \ref{fig:table} (a) the sparsification of document-topic affinity is specified by a $|\mathbf{d}_t|\times K_t$ matrix $\overline{\theta}$, entries of which are stochastic variables that entry=$1$ indicates the affinity is true, otherwise false.  
Hence, not only the overall topic number $K_t$ but also the affinity matrix $\overline{\theta}$ are stochastic variables which need to be inferred simultaneously in the learning process. 

The generative process of topic proportion follows the procedure  below:
\begin{alignat}{3}
& \mathbf{\pi} \sim IBP(\eta_0), \qquad \qquad \qquad 
&& \overline{\theta}_{dk_t} \sim Bern(\mathbf{\pi}), \nonumber \\
&\mathbf{\theta}_d  \sim Dir_K(\overline{\mathbf{\theta}}_d \odot \mathbf{\alpha}), \qquad \qquad \qquad
&& k_t \sim Mult(\mathbf{\theta}_d),
\end{alignat}
where $\odot$ is the element-wise Hadamard product and IBP is the Indian Buffet Process \cite{griffiths2011indian, doshi2015graph}, and other notations are presented in Table \ref{tab:notation}. The process first generates a probability matrix $\mathbf{\pi}$ via the IBP mechanism (the principles will be introduced below); next, taking $\mathbf{\pi}$ as the prior, a sparse document-topic affinity matrix $\overline{\mathbf{\theta}}$ is produced via the Bernoulli distribution, indicating document $d$ selects topic $k_t$ if $\overline{\theta}_{dk_t} = 1$, otherwise they have no affinity; then, drawing the topic distribution $\theta_{d}$ for document $d$ via the Dirichlet distribution by taking $\overline{\theta}_{d}$ as the prior; after that, drawing topic $k_t$ for document $d$ via the multinomial distribution; finally drawing words via the multinomial distribution with the word distribution $\mathbf{\phi}_{k_t}$, which is introduced in the following component.

Now, we introduce in detail the first step of how to obtain the probability $\mathbf{\pi}$ via the IBP. 
Assume there are $N$ customers in the restaurant, and each customer encounters a buffet consisting of infinitely many dishes arranged in a line. The first customer starts at the left of the buffet and takes a serving from each dish, stopping after $Pois(\eta_0)$ number of dishes as his plate is full. The $i$-th customer moves along the buffet and samples dishes with proportion to their popularity $\frac{m_k}{i}$, where $m_k$ is the number of previous customers who have taken the $k$-th dish. At the end of all previously sampled dishes, the $i$-th customer tries $Pois(\frac{\eta_0}{i})$ number of new dishes.

By analogy to the IBP, the sparse document-topic affinity matrix $\overline{\mathbf{\theta}}$ with $|\mathbf{d}_t|$ documents corresponds to the $N$ customers' specific choices over infinite dishes by taking the limit $K_t \to \infty$. The probability matrix $\mathbf{\pi}$ generating $\overline{\theta}$ corresponds to the probabilities of all customers' selection of dishes. Based on $\mathbf{\pi}$, each document is thus allowed to sequentially select its customized topics via the Bernoulli distribution. Topic proportions $\mathbf{\theta}_d$ are generated by the Hadamard product between $\overline{\theta}_{d}$ and hyper-parameter $\mathbf{\alpha}$ via a Dirichlet distribution. That means only those selected topics ($\overline{\theta}_{dk_t} = 1$) are endowed with weight $\alpha$ to constitute the topic proportion for document $d$ (i.e. sparsified the document-topic affinity matrix). The graphical representation of topic proportion construction is presented in Fig.~\ref{fig:table} (b).

\begin{figure}[!htb]
	\centering
	\includegraphics[width=.58\textwidth]{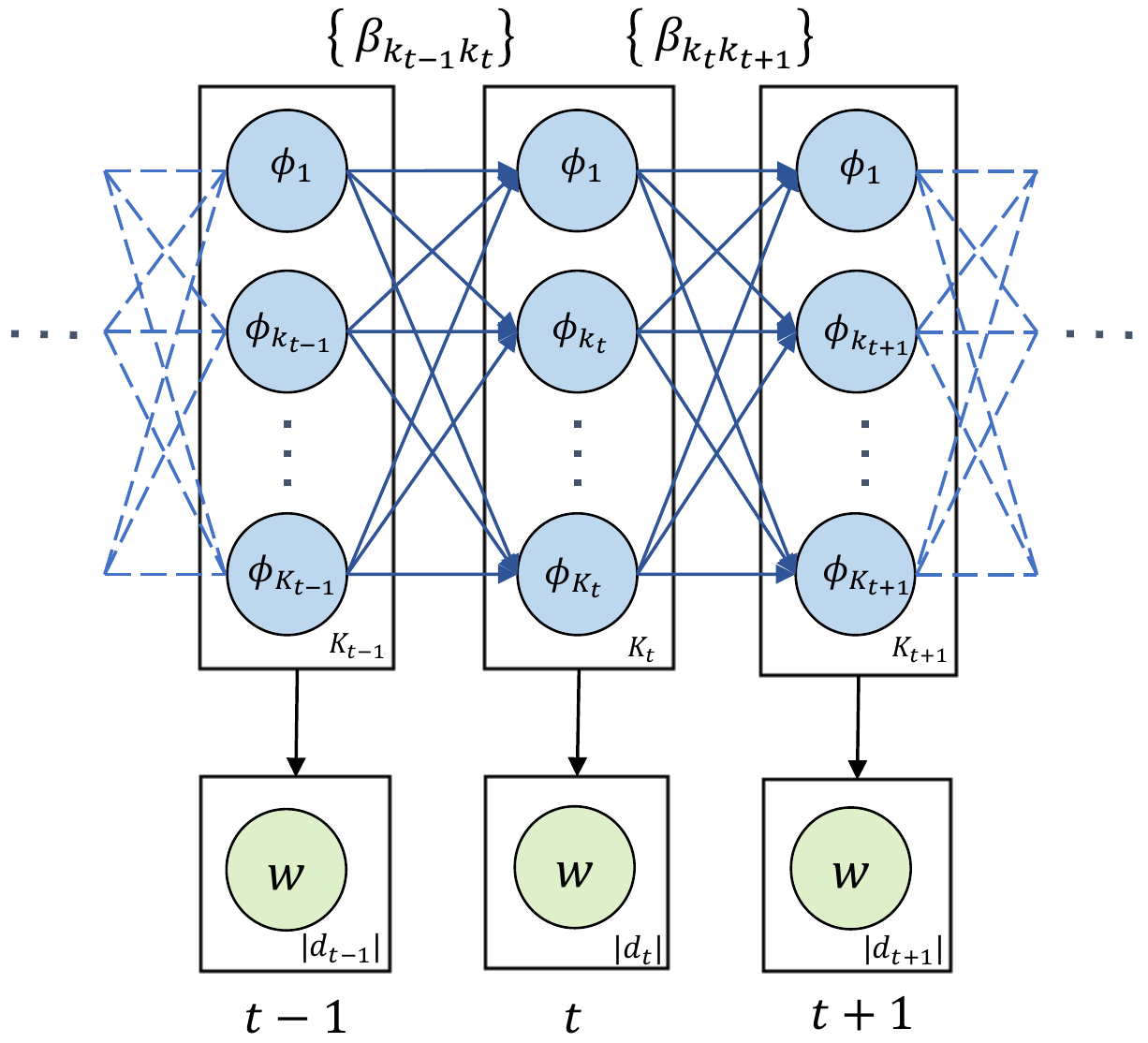}
	\caption{Graphical representation of recurrent coupled topic evolution crossing three consecutive slices from $t-1$ to $t+1$, where $\mathbf{\phi}_{k_t} (k_t \in \{1, 2,\cdots, K_t\})$ represents the hidden topic $k$ at time $t$ denoted by blue circles, the coupling relationships $ \{\beta_{k_{t-1} k_t}\}$ between consecutive topics marked by the blue arrows denote their temporal dependencies, and $\mathbf{w}$ in green color represents the observed words from document chunk $\mathbf{d}_t$. }
	\label{fig:example}
\end{figure}

\begin{table}[tb]
	\centering
	\caption{Summary of notations.}
	\resizebox{0.92\textwidth}{!}{%
		\begin{tabular}{l | l}
			\hline 
			\textbf{Symbol} & \textbf{Description} \\ \hline
			$\eta_0$ & the hyper-parameter of the Indian Buffet Process (IBP)  \\ \hline
			$\alpha$ & the hyper-parameter of Dirichlet distribution \\ \hline
			$\pi$ & the parameter of the Bernoulli distribution \\ \hline
			$\overline{\mathbf{\theta}}_d$ & the vector of sparse topic affinity with document $d$ \\ \hline
			$\mathbf{\theta}_d$	& topic proportions for document $d$  \\ \hline
			$\mathbf{\phi}_{k_t}$ & word distribution of topic $k_t$ at slice $t$ \\ \hline
			$\beta_{k_{t-1}k_t}$ & evolutionary coupling weight between topic $k_{t-1}$ and $k_t$  \\ \hline
			$\mathbf{\Phi}_{t}$ & topic set at slice $t$ \\ \hline
			$\mathbf{B}_{t-1,t}$ & coupling matrix for topics between slice $t-1$ and $t$ \\ \hline
			$\eta$ & hyper-parameter of Dirichlet distribution at slice $1$ \\ \hline
			$K_t$ & the inferred total topic number at slice $t$  \\ \hline 
			
			$x_{dw}$ & the observed word $w$'s occurrence in the document $d$  \\ \hline
			$x_{dwk_t}$ & the word $w$'s occurrence in the document $d$ assigned to the topic $k_t$   \\ \hline
			$\mathbf{x}_{k_t}$ & the vector representation of word occurrence assigned to topic $k_t$  \\ \hline
			$y_{wk_t}$ & the auxiliary latent word $w$'s occurrence assigned topic $k_t$ via $CRT$ \\ \hline
			$y_{w k_t k_{t-1}}$ & the propagated word $w$'s occurrence from topic $k_t$ to $k_{t-1}$ \\ \hline
			$\mathbf{z}_{k_t}$ & the vector representation of propagated word occurrence from $t+1$ slice \\ \hline
			$y_{\cdot k_t k_{t-1}}$ & the sum of propagate word counts from topic $k_t$ to $k_{t-1}$  \\ \hline 
			
			$\xi_{k_t}$ & auxiliary variable from the beta distribution \\ \hline 
			$r_{k_t}$ & shape parameter of the Gamma distribution \\ \hline
			$c_t, c_0$ & rate parameter of the Gamma distributions \\ \hline		
			$a_0, b_0, e_0, d_0, r_0$ & hyper-parameters of the Gamma distributions \\ \hline 
		\end{tabular}%
	}
	\label{tab:notation}
\end{table}

\stitle{Multi-Topic-Thread evolution.} 
The other important component of the generative process is how to encode multiple topic dependencies crossing slices. Fig.~\ref{fig:example} presents a simple scenario of coupled topic evolution crossing three consecutive slices.

At the initial slice $t=1$, without any prior dependency, the topic $\mathbf{\phi}_{k_1}$ $(k_1 \in \{1,\cdots, K_1\})$ is sampled from the Dirichlet distribution parameterized by $\mathbf{\eta}$. At slice $t$, the topic $\mathbf{\phi}_{k_t}$ $(k_t \in \{1,\cdots, K_t\})$ is assumed to evolve from the prior topics with the corresponding coupling weights $\beta_{k_{t-1} k_t}$ via the Dirichlet distribution, where $\beta_{k_{t-1} k_t}$ is drawn from a Gamma distribution to measure the evolutionary closeness to the topic in the prior slice. At the final slice $t=T$, topic $\mathbf{\phi}_{k_T}$ $(k_T \in \{1,\cdots, K_T\})$ evolves depending on the prior topics at $T-1$. Given the defined recurrent topics, words $\mathbf{w}$ from the document chunk $\mathbf{d}_t$ are accordingly generated via the multinomial distributions at each slice.

The recurrent coupled topic sequences smoothly evolve across the $T$ slices according to the following generative process,
\begin{equation}
\label{eq:topic_sequence}
\begin{split}
& (x_{\cdot wk_1})_{w=1}^V \sim Mult(\mathbf{\phi}_{k_1}), \quad  \mathbf{\phi}_{k_1} \sim Dir_V(\eta),  \\ 
& \qquad \quad \cdots   \\
& (x_{\cdot wk_t})_{w=1}^V \sim Mult(\mathbf{\phi}_{k_t}), \quad \mathbf{\phi}_{k_t} \sim Dir_V( \sum_{k_{t-1}=1}^{K_{t-1}}\beta_{k_{t-1} k_t} \mathbf{\phi}_{k_{t-1}}),  \quad \beta_{k_{t-1} k_t} \sim Gam(r_{k_{t-1}}, 1/c_t),   \\
& \qquad \quad  \cdots  \\  
& (x_{\cdot wk_T})_{w=1}^V \sim Mult(\mathbf{\phi}_{k_T}), \quad \mathbf{\phi}_{k_T} \sim Dir_V(\sum_{k_{T-1}=1}^{K_{T-1}}\beta_{k_{T-1} k_T} \mathbf{\phi}_{k_{T-1}}),  \quad \beta_{k_{T-1} k_T} \sim Gam(r_{k_{T-1}}, 1/c_T),  \\
\end{split}
\end{equation}
where Gam(-,-) is the Gamma distribution with shape and scale parameters. We further impose Gamma priors on the following variables: $r_{k_{t-1}} \sim Gam(r_0/K_{t-1}, 1/c_0)$, $c_t \sim Gam(e_0, 1/d_0)$ and $c_0 \sim Gam(a_0, 1/b_0)$, where $a_0, b_0, d_0, e_0$ and $r_0$ are specified hyper-parameters.


The idea of our recurrent modeling of multiple coupled topic sequences is summarized as follows:
\begin{itemize}
	\item[--] From a forward-backward view, the proposed model resembles the stochastic feedforward network \cite{tang2013learning}, where the input is a topic set $\{\mathbf{\phi}_{k_1}\}_{k_1 =1}^{K_1}$ at slice $1$, the output is topics $\{\mathbf{\phi}_{k_T}\}_{k_T =1}^{K_T}$, the weight matrices are $\{\beta_{k_{t-1} k_t}\}_{{k_{t-1}},{k_t}}^{{K_{t-1}},{K_t}}$ and the activation functions are Dirichlet distributions. 
	\item[--] When learning the word distribution of topic $k_t$ ($t>1$), the mixture of prior topics $\{\mathbf{\phi}_{k_{t-1}}\}_{k_{t-1} =1}^{K_{t-1}}$ serves as the prior knowledge to initialize topic $k_t$ via a Dirichlet distribution, and the coupling weights $\{\beta_{ k_{t-1} k_t}\}_{k_{t-1}}^{K_{t-1}}$ identify the different contributions of prior topics to topic $k_t$. 
	
	\item[--] According to the expectation of a Dirichlet distribution, topic $\mathbf{\phi}_{k_t}$ is expected to be the weighted arithmetic mean of prior topics at slice $t-1$, $E(\mathbf{\phi}_{k_t}) = \frac{\sum_{k_{t-1}=1}^{K_t}\beta_{k_{t-1} k_t}\mathbf{\phi}_{k_{t-1}}} {\sum_{k_{t-1}=1}^{K_t}\beta_{k_{t-1} k_t}} $, implying that the evolution of topic $\mathbf{\phi}_{k_t}$ jointly depends on multiple prior topics with the corresponding weights rather than on a single past topic, and the prior topic $\mathbf{\phi}_{k_{t-1}}$ ($k_{t-1} \in \{1, \cdots, K_{t-1}\}$) also contributes to multiple topics at slice $t$. The coupling weight $\beta_{k_{t-1}k_t}$ is noted to play an important role in measuring the evolutionary distance between two word distributions of topic $k_{t-1}$ and $k_t$. In addition, this expectation indicates coupling weights $\{\beta_{k_{t-1}k_t}\}_{k_{t-1}=1}^{K_{t-1}}$ associated with topic $k_{t}$ are not shared with other parallel topics, which allows topics at slice $t$ to evolve differently with flexible dependency on the common priors. 	
	\item[--] The coupling weight $\beta_{k_{t-1} k_t}$ is drawn from a hierarchical Gamma prior (its shape parameter $r_{k_{t}}$ is also drawn from a Gamma). Such a hierarchical design leads to more distinguishable and sparse coupling weights associated with topic $\mathbf{\phi}_{k_t}$ \cite{zhao2018dirichlet}. 
\end{itemize}

\subsection{A Dropout Technique}
In the context of topic evolution with multiple threads, one question is naturally raised about how to validate the significance of multi-dependencies between evolving topic sequences, since each topic evolves from all prior topics. Further, one may argue that the salient coupling connections of one topic during evolving process are a small set and sparsely distributed in practice, e.g., in light of diverse and enormous topics inferred from the computer science articles last year, the topic about Bayesian network only connects with a small number of relevant topics by the salient coupling weight, while the weights with most unrelated topics are small. Thus, could the proposed model distinguish the salient coupled topics from the less related ones by the weights? To answer this question, we develop a variant of the proposed model named rCTM-D as a comparison to rCTM.

In this approach, we borrow the dropout mechanism from the neural network and inject it into our Bayesian framework. Dropout \cite{srivastava2014dropout} is one of the most popular and successful regularizers for deep neural network. It randomly drops out each neuron with a predefined probability at each iteration of stochastic gradient descent, to avoid the overfitting problem and reinforce the performance. In our solution, at each iteration of inference process, the topic node $\mathbf{\phi}_{k_t}$ is attached with a probability $\rho$ to drop out coupling connection with prior topics $\mathbf{\Phi}_{t-1}$, which is denoted as:
\begin{equation}
\label{eq:sparse_dropout}
	\mathbf{\phi}_{k_{t}} \sim Dir_V(\mathbf{\psi}_{k_t}), \qquad \mathbf{\psi}_{k_t} = \sum_{k_{t-1}=1}^{K_{t-1}} (\beta_{ k_{t-1} k_t}(1- m_{k_{t-1}})) \mathbf{\phi}_{k_{t-1}}, \qquad m_{k_{t-1}} \sim Bern(\rho),
\end{equation}
where $m_{k_{t-1}}$ is the dropout indicator drawn from a Bernoulli distribution with parameter $\rho$. If $m_{k_{t-1}}=0$, the coupling connection from prior topic $k_{t-1}$ is preserved with its original weight; otherwise, the connection is dropped out and this prior topic would not participate in the inference to posterior topics. 

Let's consider the dropout probability in two extreme cases. (1) If we set the dropout probability $\rho=0$, then $m_{k_{t-1}} =0$ $(k_{t-1} \in \{1,\cdots, K_{t-1}\})$, it means all coupling connections are preserved and rCTM-D is recovered to rCTM. (2) If $\rho=1$, then $m_{k_{t-1}}=1$ $(k_{t-1} \in \{1,\cdots, K_{t-1}\})$, rCTM-D is thus degraded to $T$ separated topic modelings at each time-slice without any connections. Hence, we would give the dropout probability $\rho$ within the range $(0,1)$ in the rCTM-D, to see its performance with different ratios of coupling connection dropped out.


\section{The Posterior Inference}
Since we induce \textit{multi-topic-thread evolution}, the main challenge for the proposed rCTM is to solve the intractable problem and obtain a closed-form inference to recurrent topics $\mathbf{\Phi}_t $ as well as their coupling matrix $\mathbf{B}_{t-1,t}$ at each time slice. Such a task has never been explored before. To tackle this problem, a set of auxiliary variables and data augmentation techniques are introduced. In this section, we propose a novel Gibbs sampler with a backward-forward filter algorithm to implement its inference process.

\textbf{\textit{Sampling $\overline{\mathbf{\theta}}$}}: the sparse document-topic affinity matrix $\overline{\mathbf{\theta}}$ could be sampled by marginalizing out $\mathbf{\theta}$ and $\pi_k$. First, we note that if the word count $x_{dwk_t} > 0$, then $\overline{\theta}_{dk_t}$ must be 1 because it implies there exists at least one word assigned to topic $k_t$. Let vector $\overline{\mathbf{\theta}}_{d(0)} $ represent the $d$-th row vector of $\overline{\mathbf{\theta}}$ with entries of 0, and vector $\overline{\mathbf{\theta}}_{k(0)}$ denote the $k$-th column vector of $\overline{\mathbf{\theta}}$ with entries of 0.
If $x_{d\cdot k_t} = 0$, the probability $\overline{\theta}_{dk_t} = 1$ is marginalized as,
\begin{equation}
\label{eq:theta_}
P(\overline{\theta}_{dk_t} = 1|\alpha, \eta_0) =
\frac{B(\alpha |\overline{\mathbf{\theta}}_{d(0)}| + |\overline{\mathbf{\theta}}_{d(0)}|, \alpha) (|\overline{\mathbf{\theta}}_{k(0)}| + \eta_0)} {B(\alpha  |\overline{\mathbf{\theta}}_{d(0)}|, \alpha) (|\mathbf{d}_t| - |\overline{\mathbf{\theta}}_{k(0)}| + \eta_0)},
\end{equation}
where $B(-,-)$ denotes a beta distribution,  $|\mathbf{d}_t|$ records the document number at slice $t$,  $|\overline{\mathbf{\theta}}_{d(0)}|$, $|\overline{\mathbf{\theta}}_{k(0)}|$ record the number of $0$ entries in the $d$-th row vector and $k$-th column vector of matrix $\overline{ \mathbf{\theta}}$ respectively, and $\eta_0$, $\alpha$ are specified hyper-parameters.

\textbf{\textit{Sampling $\mathbf{\theta}$}}: as we obtain the sparse document-topic affinity matrix $\overline{\mathbf{\theta}}$, the topic proportion $\mathbf{\theta}_d$ for document $d$ is sampled from its conditional posterior distribution as,
\begin{equation}
\label{eq:theta}
\mathbf{\theta}_d \sim Dir_K(\mathbf{\overline{\theta}}_d \odot (\alpha + x_{d\cdot k_t})),
\end{equation}
where $x_{d\cdot k_t}$ records the number of words in the document $d$ assigned to the topic $k_t$.

\textbf{\textit{Sampling $x_{dwk_t}$}}: the observed word $w$'s occurrence in document $d$ is denoted as $x_{dw}$, and we augment it as $x_{dw} = x_{dw \cdot} = \sum_{k_t} x_{dwk_t}$, indicating the number of word $w$ in the document $d$ assigned to topic $k_t$, which is sampled as,
\begin{equation}
\label{eq:x_dwk}
x_{dwk_t} \sim Mult \left(x_{dw \cdot}, (\frac{\theta_{dk_t}\phi_{k_tw}} {\sum_{k_t}^{K_t}\theta_{dk_t}\phi_{k_t w}})_{k_t =1}^{K_t} \right).
\end{equation}

\begin{figure}[!htb]
	\centering
	\subfigure[A motivating example to decode the coupling relationship between consecutive topics.]{
		\includegraphics[width=0.42\textwidth]{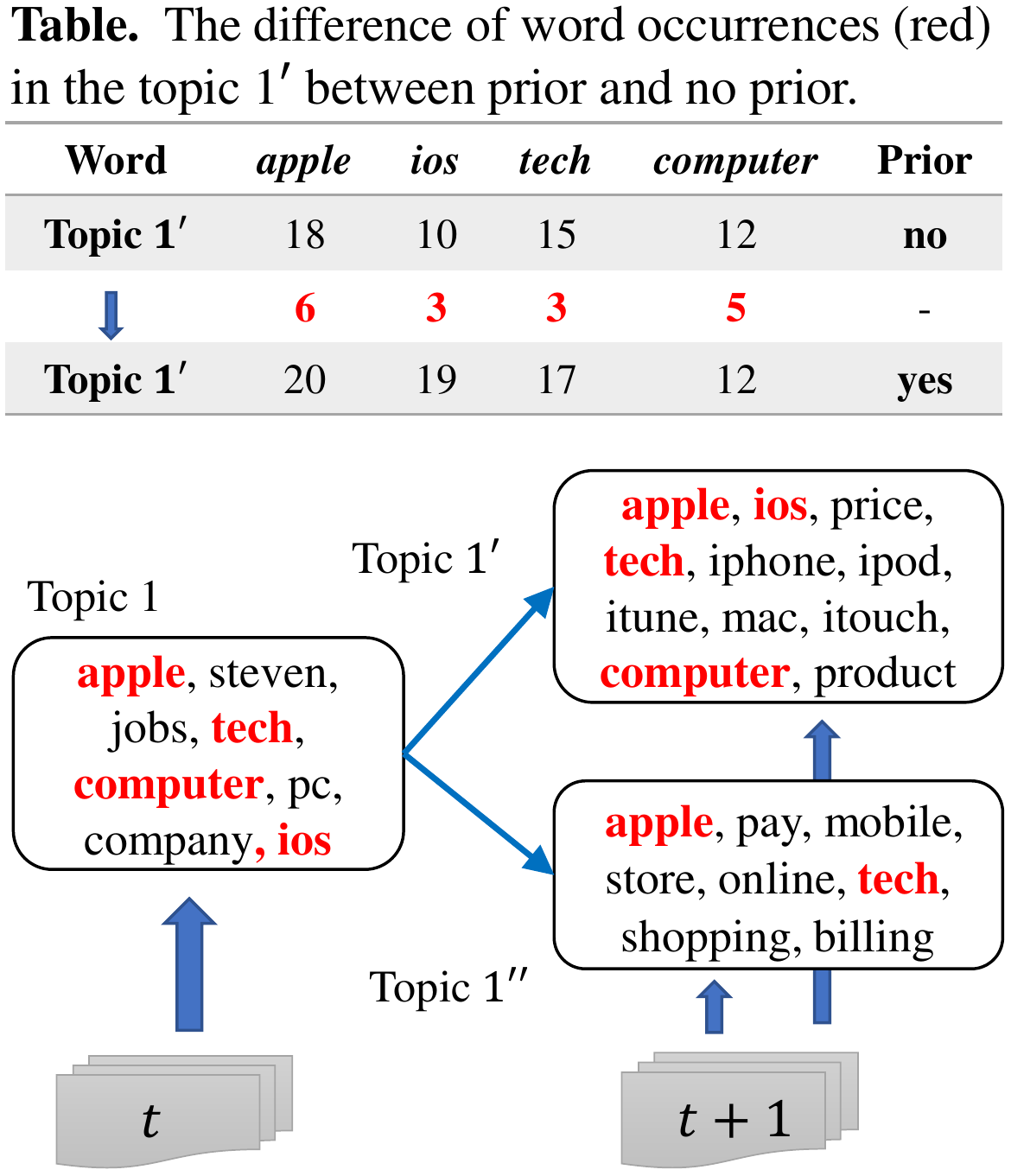}}
	\hspace{0.1in}	
	\subfigure[The inference process with backward propagation.]{
		\includegraphics[width=0.53\textwidth]{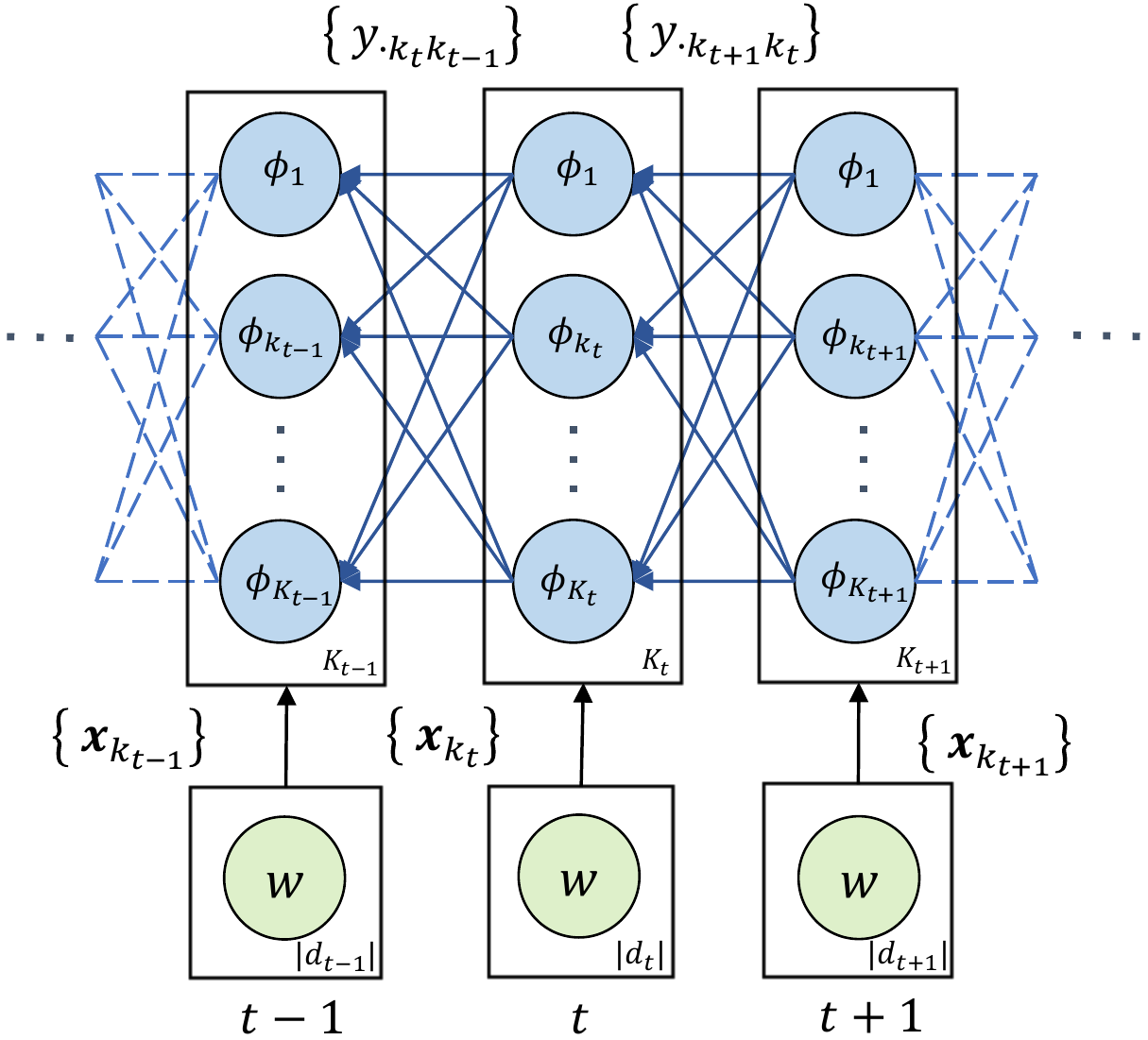}}
	\caption{In Fig. (a), each topic is represented by a set of description words, their occurrences with and without prior are respectively listed in black font, and the shared common word occurrences are denoted in red  in the table. In Fig. (b), arrows between consecutive topics denote their shared latent word counts in the backward filter, which are annotated by $\{y_{\cdot k_t k_{t-1}}\}$ in blue, and $\mathbf{x}_{k_t}$ summarizes the inferred vector of word counts assigned to topic $k_t$ in black.}
	\label{fig:backward}
\end{figure}

The vector $\mathbf{x}_{k_t}$ is defined as $\mathbf{x}_{k_t} = [x_{\cdot 1k_t}, x_{\cdot2k_t}, \cdots, x_{\cdot Vk_t}]$, indicating the vector of all word occurrences from document chunk $\mathbf{d}_t$ assigned to the topic $k_t$, which is illustrated in Fig. ~\ref{fig:backward} (b).

\stitle{Challenges of Inference.} We proceed to infer the latent parameters in the component of coupled topic evolution, which is the core part of the solution. There remain two demanding challenges to be solved for a tractable inference, which significantly challenge the existing inference approaches.  
\begin{itemize}
	\item To obtain an independent inference to coupling weight $\beta_{k_{t-1} k_t}$, it is vital to decompose the joint multiple dependencies into individual relationships associated with each prior topic.
	\item The essence of non-negative weight $\beta_{k_{t-1} k_t}$ connecting topic $k_{t-1}$ to $k_t$ remains unknown.
\end{itemize}

To solve these challenges, we induce a motivating example to illustrate it. As shown in Fig. ~\ref{fig:backward} (a), Topic 1 naturally evolves into two different threads from slice $t$ to $t+1$ via the proposed generative process, and their contents are represented by a set of frequent words. Though their word representations differ, it is found that the shared common words, such as `apple', `tech' and `computer', naturally chain the prior Topic 1 and its thread Topic $1'$ together. Indicated by the Table in Fig. ~\ref{fig:backward} (a), the occurrences of these common words in Topic $1'$ with the prior influence of Topic 1 are distinguished from Topic $1'$ without such prior dependency. An insightful fact is found that occurrences of common words in Topic $1'$ with the prior could be decoded into two parts. One part of these occurrences is directly from the documents at time $t+1$, and the other is implicitly contributed from the prior Topic 1, denoted by the numbers in red from the Table. Without such implicit word-level sharing, the dependency relationship between Topic 1 and its thread would not exist. Hence, it is concluded that coupling weight $\beta_{k_{t-1} k_t}$ connecting topic $k_{t-1}$ and its subsequent thread $k_t$ is essentially summarized by their shared latent word occurrences.

Based on the above insightful observations, it's of significance to derive the shared latent word counts between consecutive topics. Since topics at slice $t ( 1 < t <T)$ are recursively chained and inter-dependent on prior topics at $t-1$, the conventional inference techniques \cite{griffiths2004finding, liang2016dynamic}, which are implemented as independent at each slice, ignore such \textit{recursive dependency} between slices and they are inapplicable for such an inference task. Therefore, we design a novel back-forward filter to fully solve it and achieve the tractable inference. In the backward filter, the smart data augmentation techniques unfreeze the limitation of \textit{recursive dependency}, and derive the shared latent word counts between consecutive slices. Then the time-evolving parameters are naturally inferred in a closed-form in the forward filter.

\stitle{Backward propagating the latent counts.} 
We start from time slice $T$ since no more latent variables depend on it. By integrating out $\mathbf{\phi}_{k_T}$, we obtain the likelihood of latent word counts $(x_{\cdot wk_T})_{w=1}^V$ according to the conjugacy between the Dirichlet and multinomial distributions. 
\begin{equation}
\label{eq:psi}
\mathcal{L}(x_{\cdot wk_T})_{w=1}^V \sim DirMult(\mathbf{\psi}_{k_T}), \qquad \mathbf{\psi}_{k_T} =\sum_{k_{T-1}=1}^{K_{T-1}}\beta_{k_{T-1} k_T} \mathbf{\phi}_{k_{T-1}},
\end{equation}
where the multi-dependency $\mathbf{\psi}_{k_T}$ associated with all prior topics always appears in the sum form as the parameter of Dirichlet. Since we could not directly obtain the individual dependency $\beta_{ k_{t-1} k_t}$ associated with each prior topic, we introduce an auxiliary variable $\xi_{k_T}\sim B(x_{\cdot \cdot k_T}, \psi_{\cdot k_T})$, and further augment $DirMult$. The joint likelihood of $(x_{\cdot wk_T}, \xi_{k_T})$ takes the following form \cite{acharya2015nonparametric},
\begin{equation}
\begin{split}
\mathcal{L}((x_{\cdot wk_T})_{w=1}^V, \xi_{k_T}) 
& = \mathcal{L}(x_{\cdot wk_T})_{w=1}^V \times B(\xi_{k_T}| x_{\cdot \cdot k_T}, \psi_{\cdot k_T}) \\
& \propto \prod_{w=1}^{V} NB(x_{\cdot wk_T} |\psi_{vk_T}, \xi_{k_T}),
\end{split}
\end{equation}
where $NB(-,-)$ is the negative binomial distribution, and $B(-,-)$ is the beta distribution. With the auxiliary variable introduced, the variable $x_{\cdot wk_T}$ now follows the negative binomial distribution, which plays a critical role in bridging the Dirichlet and Poisson distributions. Hence, Lemma \ref{lm:2} is defined in the following, which presents the transformation relationship from a negative binomial to a Poisson distribution. The property of the Poisson distribution is thus able to be enjoyed when disentangling the joint dependency relationship after the transformation.

\begin{lemma}\label{lm:2}
	If $m\sim NB(r, p)$ represents that $m$ follows a negative binomial distribution, then the conditional posterior of $l$ given $m$ and $r$ is denoted as $(l|m, r) \sim CRT(m, r)$, a Chinese restaurant table ($CRT$) count random variable, which can be generated via $l=\sum_{n=1}^{m}z_n$, $z_n \sim Bern(r/(n-1+r))$. It can also be augmented under a compound Poisson representation as $m=\sum_{t=1}^{l}u_t$, $u_t \sim Log(p) $, $ l \sim Pois(-rlog(1-p))$ \cite{schein2016poisson,acharya2018dual}.
\end{lemma}

According to Lemma \ref{lm:2}, a new variable $y_{wk_T} \sim CRT(x_{\cdot wk_T}, \psi_{wk_T})$ is obtained based on $x_{\cdot wk_T}$ and $\psi_{wk_T}$. With further data augmentation applied, an equivalent representation of $y_{wk_T}$ under a compound Poisson form is expressed as,
\begin{equation}
\label{eq:y_wkt}
y_{wk_T} \sim Pois(-\psi_{wk_T} ln(1-\xi_{k_T}).
\end{equation}
Since $\psi_{wk_T} = \sum_{k_{T-1}=1}^{K_{T-1}}\beta_{k_{T-1} k_T}\phi_{vk_{T-1}}$ as defined by Eq.~(\ref{eq:psi}), we feed it into the above equation, which is extended as 
\begin{equation}
\label{eq:ywk}
y_{wk_T} \sim Pois(-\sum_{k_{T-1}=1}^{K_{T-1}}\beta_{k_{T-1} k_T} \phi_{vk_{T-1}} ln(1-\xi_{k_T})). 
\end{equation}
We now introduce another auxiliary variable $y_{wk_T k_{T-1}}$ which is augmented from $y_{wk_T} = y_{wk_T \cdot} =\sum_{k_{T-1}=1}^{K_{T-1}} y_{wk_T k_{T-1}}$, and the above Eq.~\ref{eq:ywk} is thus represented as follows according to the property of Poisson distribution, 
\begin{equation}
\label{eq:y_wkk}
y_{wk_T k_{T-1}} \sim Pois(-\beta_{k_{T-1} k_T}\phi_{wk_{T-1}} ln(1-\xi_{k_T})),
\end{equation}
where the joint coupling dependency is successfully decomposed into separated relationships thanks to the merit of data augmentation technique and the property of Poisson distribution. Since the auxiliary variable $y_{wk_T k_{T-1}}$ is augmented from the variable $y_{wk_T}$, now we define Lemma \ref{lm:1} in the following to present the relationship between Poisson and multinomial distributions. 
\begin{lemma}\label{lm:1}
	If $y. = \sum_{n=1}^{N} y_n$, where $y_n \sim Pois(\theta)$ are independent Poisson-distributed random variables, then $(y_1,..., y_N)\sim Mult(y.,(\frac{\theta_1}{\sum_{n=1}^{N}\theta_n},..., \frac{\theta_N}{\sum_{n=1}^{N}\theta_n}))$ \cite{zhou2013negative}.
\end{lemma} 

Thus, $y_{wk_T k_{T-1}}$ is distributed as $Mult$ via Lemma \ref{lm:1}, which is expressed as,
\begin{equation}
\label{eq:ywkk}
y_{wk_T k_{T-1}} \sim Mult \left( y_{wk_T},(\frac{\beta_{k_{T-1} k_T} \phi_{wk_{T-1}}}
{\sum_{k_{T-1}=1}^{K_{T-1}}\beta_{k_{T-1} k_T}\phi_{wk_{T-1}}})_{k_{T-1}=1}^{K_{T-1}} \right),
\end{equation}
where $y_{wk_T k_{T-1}}$ is successfully obtained to denote the shared latent word $w$' occurrence between topic $k_T$ and $K_{T-1}$ , which indicates the numbers to be inferred denoted in red from the Table of Fig. ~\ref{fig:backward} (a).

We now induce the auxiliary variable $z_{wk_{T-1}}$, which is defined as $z_{wk_{T-1}}=\sum_{k_T=1}^{K_T} y_{wk_T k_{T-1}}$. It is viewed as latent word counts propagated from topic set at slice $T$. Thus the vector $\mathbf{z}_{k_{T-1}}$ is defined as $\mathbf{z}_{k_{T-1}} = [z_{1k_{T-1}}, z_{2k_{T-1}}, \cdots, z_{Vk_{T-1}} ]$. Another highlighted variable is $y_{\cdot k_T k_{T-1}}$ obtained via $y_{\cdot k_T k_{T-1}}= \sum_{w=1}^V y_{wk_T k_{T-1}}$ to summarize the sum of shared latent word counts between topic $k_T$ and $k_{T-1}$, which is computed in advance and cached to be used in the forward filter. 

As we continue propagating backward from $t= T-1,\cdots,2$, the latent word count vectors $\mathbf{z}_{k_{T-2}},\cdots, \mathbf{z}_{1}$ are sequentially obtained. It's worth noting that since no more document chunks after slice $T$, there is no propagated word count at slice $T$ such that $\mathbf{z}_{k_{T}} = \mathbf{0}$.

In conclusion, the propagating process between consecutive evolving topics from slice $t+1$ to $t$ is summarized as: (1) the latent word count $y_{wk_{t+1}}$ is firstly derived from $x_{wk_{t+1}}$ via the $CRT$ distribution, (2) then we distribute $y_{wk_{t+1}}$ according to the $Mult$ distribution to obtain the latent count $y_{wk_t k_{t+1}}$, and (3) finally  $y_{wk_t k_{t+1}}$ is aggregated to form the latent counts $z_{wk_{t}}$ at slice $t$. This process is illustrated in Fig. \ref{fig:backward} (b).



\stitle{Forward sampling the latent variables.}
Conditioned on the propagated auxiliary values $ \{\mathbf{z}_{k_t}\}_{t=1}^{T-1}$, $ \{y_{\cdot k_2 k_1,},\cdots, y_{\cdot k_T k_{T-1}} \}$ and $\{\xi_{k_2},\cdots, \xi_{k_T}\}$ obtained via the backward propagating filter. We start sampling the latent variables by performing a forward sampling pass from $t=1,\cdots, T$.

\textbf{\textit{Sampling $ \mathbf{\Phi}$}}: based on the conjugacy between the Dirichlet and multinomial distributions, the topics $\mathbf{\phi}_{k_1}$ ($ k_1 \in \{1, \cdots, K_1\}$) at slice $t=1$ is marginalized from its conditional posterior,
\begin{equation}
\label{eq:phi_1}
\mathbf{\phi}_{k_1} \sim Dir_V(\eta + \mathbf{x}_{k_1} + \mathbf{z}_{k_1}),
\end{equation}
where $\mathbf{x}_{k_1} =[x_{\cdot1k_1}, x_{\cdot2k_1},\cdots, x_{\cdot Vk_1} ]$ is the word occurrence vector inferred from document chunk $\mathbf{d}_1$ at slice $1$ via Eq. \ref{eq:x_dwk}, and $\mathbf{z}_{k_1} = [z_{1k_1}, z_{2k_1}, \cdots, z_{Vk_1}] $ denotes the propagated word count vector from slice $2$ to slice $1$.

At time $ 1 < t \leq T $, $\mathbf{\phi}_{k_t}$ ($ k_t \in \{1, \cdots, K_t\}$) is sampled as,
\begin{equation}
\label{eq:phi_k}
\mathbf{\phi}_{k_t} \sim Dir_V(\sum_{k_{t-1}=1}^{K_{t-1}}\beta_{k_{t-1} k_t} \mathbf{\phi}_{k_{t-1}} + \mathbf{x}_{k_t} + \mathbf{z}_{k_t}),
\end{equation}
where $\mathbf{x}_{k_t}$ is the word count vector inferred from $\mathbf{d}_t$ at slice $t$, and $\mathbf{z}_{k_t}$ denotes the propagated word count vector from slice $t+1$ to $t$. It is noted $\mathbf{z}_{k_T} = \mathbf{0}$ at time slice $T$.

\textbf{\textit{Sampling $\mathbf{B}$}}: indicated by Eq.~(\ref{eq:y_wkk}), $\beta_{k_{T-1}k_T}$ still entwines with $\phi_{wk_{T-1}}$ in the parameter of Poisson distribution. However, via $\sum_{w=1}^{V}\phi_{wk_t} =1$, it is marginalized as,
\begin{equation}
\label{eq:y_kk}
y_{\cdot k_T k_{T-1}} \sim Pois(-\beta_{k_{T-1}k_T} ln(1-\xi_{k_T})).
\end{equation}
 
Recall the prior distribution of $\beta_{k_{t-1}k_t}$, which is defined as $\beta_{k_{t-1} k_t} \sim Gam(r_{k_{t-1}}, 1/c_t)$ in Eq.~(\ref{eq:topic_sequence}), thus it is marginalized via the conjugacy between the Poisson and Gamma distributions,
\begin{equation}
\label{eq:beta}
\beta_{k_{t-1} k_t} \sim Gam(y_{\cdot k_t k_{t-1}} + r_{k_{t-1}}, 1/(c_t -ln(1-\xi_{k_t}))),
\end{equation} 
where $y_{\cdot k_t k_{t-1}}$ is the sum of propagated word counts and cached in the backward filter via  $y_{\cdot k_t k_{t-1}}= \sum_{w=1}^V y_{wk_t k_{t-1}}$, and the auxiliary variable $\xi_{k_t}$ is also induced in the backward filter.

Through a series of novel data augmentation techniques, the inference of recurrent topic $\mathbf{ \phi}_{k_t} $ and its coupling strength $\{\beta_{ k_{t-1} k_t}\}_{k_{t-1}}^{K_{t-1}}$ are finally tractable at each slice. According to the expectation of Gamma distribution, it is derived that $\beta_{k_{t-1} k_t}\approx (y_{ \cdot k_t k_{t-1}} + r_{k_{t-1}})/(c_t -ln(1-\xi_{k_t}))$, implying that the sum of propagated common word counts between consecutive topics $k_t$ and $k_{t-1}$ is an important indicator to the coupling weight $\beta_{k_{t-1} k_t}$.

\textbf{\textit{Sampling $r_{k_{t-1}}$}}: recall $r_{k_{t-1}}$ is drawn from its prior distribution via $r_{k_{t-1}} \sim Gam(r_0/K_{t-1}, 1/c_0) $ (cf. the subsection \ref{sub}), and its likelihood distribution is defined as $\beta_{k_{t-1} k_t} \sim Gam(r_{k_{t-1}}, 1/c_t)$ in Eq.~(\ref{eq:topic_sequence}), the inference of $r_{k_{t-1}}$ incurs the non-conjugate problem between the Gamma and Gamma distributions, hence, we induce Lemma \ref{lm:4} in the following to help solve it.
\begin{lemma}\label{lm:4}
	if $x_i \sim Pois(m_i r_2)$, $r_2 \sim Gam(r_1, 1/c_0)$, $r_1 \sim Gam(a_0, 1/b_0)$, then 
	$(r_1|-) \sim Gam(a_0 + l, 1/(b_0 -log(1-p)))$, 
	where $(l|x, r_1) \sim CRT(\sum_i x_i, r_1)$ and $p = \frac{\sum_i m_i}{c_0 + \sum_i m_i}$ \cite{acharya2015nonparametric,acharya2018dual}.
\end{lemma}

Based on Eq.~(\ref{eq:y_kk}) and the above prior and its likelihood distribution,  $r_{k_{t-1}}$ is sampled via  Lemma \ref{lm:4},
\begin{equation}
\label{eq:r_k}
\begin{split}
& r_{k_{t-1}} \sim Gam(r_0/K_{t-1} + l_{k_{t-1}}, 1/(c_0 - log(1-p)) ), \\
& l_{k_{t-1}} \sim CRT(\sum_{k_t=1}^{K_t}y_{ \cdot {k_t  k_{t-1}}}, r_{k_{t-1}}), \qquad  p = \sum_{k_t =1}^{K_t} m_{k_t} / (\sum_{k_t =1}^{K_t} m_{k_t} + c_0),
\end{split}
\end{equation}
where $m_{k_t} = -ln(1-\xi_{k_t})$.

\textbf{\textit{Sampling $c_t, c_0$}}: given the conjugacy between the Poisson and Gamma distributions, $c_t$ and $c_0$ are sampled respectively as,
\begin{equation}
\label{eq:c}
\begin{split}
&c_t \sim Gam(e_0 + \sum_{k_{t-1}}^{K_{t-1}}r_{k_{t-1}}, 1/(\sum_{k_{t-1}}^{K_{T-1}}\beta_{k_{t-1} k_t} + d_0 )), \\
&c_0 \sim Gam(e_0 + r_0/ K_{t-1}, 1/(r_{k_{t-1}} + d_0 )),
\end{split}
\end{equation}
where $K_{t-1}$ is inferred topic number from a topic proportion process via the latent IBP compound distribution, $r_{k_{t-1}}$ and $\beta_{k_{t-1}k_t}$ are sampled from the prior steps, and the rest variables $e_0, d_0, r_0$ are specified hyper-parameters.

The whole Gibbs sampling with a backward-forward filter is presented in Algorithm \ref{al:infer}. At each iteration, $\mathbf{x}_{k_t}$ is firstly sampled via the $Mult$ distribution from the document chunk $\mathbf{d}_t$ at each slice. During the backward filter, the auxiliary variable $\xi_{k_t}$ and $y_{w k_t}$ are induced, and the sequence of propagated word counts $\mathbf{z}_{k_t}$ is obtained sequentially from slice $T$ to slice $2$ via repeated data augmentation and marginalization techniques. Conditioned on latent counts from the above filter procedure, the recurrent topics $\mathbf{\Phi}_{t}$ and their coupling matrix $\mathbf{B}_{t-1,t}$ are updated in a closed-form at each slice. These steps are repeated $MaxIteration$ times until the joint posterior distribution converges. The latent parameters are thus estimated based on the stable samples.

By far, we have introduced our novel recurrent multi-topic modeling and the corresponding novel and effective inference method. In the backward filter, the adoption of novel $NB$ augmentation into the dynamic Dirichlet chain is non-trivial, which bridges the gap between the Dirichlet and Poisson distributions. Such an infusion plays an important role in unfreezing the limitation of \textit{recursive dependency} and deriving the shared latent word counts between consecutive topics, leading to an efficient and tractable inference for recurrent topics and their multi-dependencies. Note that none of the existing work on the temporal topic modeling proposes such assumptions that naturally fit the complex sequential data, facilitates the interpretability of latent states, and yields a closed-form and straight-forward update in the inference.

\begin{algorithm}
	\caption{Gibbs Sampling with the backward-forward filter}
	\label{al:infer}
	Initializing topic assignments randomly for all documents in $ \{\mathbf{d}_t\}_{t=1}^{T} $ \;
	Initializing variables $ \mathbf{B}_{t-1,t} $, $ (r_{k_t})_{k_t=1}^{K_t}, c_t$ and $ c_0 $ at each slice\;
	\For {$ iter \in {1,2, \cdots, MaxIteration} $ }
	{
		\For {$ d \in \{\mathbf{d}_t\}_{t=1}^{T} $}
		{
			Sampling the document-specific ${\overline{\theta}_d}, \mathbf{\theta}_d$ by Eq.~(\ref{eq:theta_}) and Eq.~(\ref{eq:theta}) \;
			\For {$ w \in \{w_1,..., w_{|x_d|}\}$}
			{
				Sampling the latent word count $ {x_{dwk_t}} $ by Eq.~(\ref{eq:x_dwk})  \;			
			}	
		}	
		Backward propagating: initialize $t=T$ \;
		\While{$t>0$}
		{
			Sampling the auxiliary variable ${\xi_{k_t}}$ from beta distribution \;
			Sampling the latent count $y_{wk_T}$ via $CRT$ \;
			Sampling the latent count $y_{wk_T k_{T-1}}$ by Eq.~(\ref{eq:ywkk}) \;
		    Caching the auxiliary variable $y_{\cdot k_t k_{t-1}}$ to use in the forward pass \;
			$t = t-1$ \;
		}
		Forward sampling: initialize $t=1$ \; 
		\While{$t \leq T$}		
		{
			Sampling ${\phi_{k_t}}$ by Eq.~(\ref{eq:phi_1}) or Eq.~(\ref{eq:phi_k}) \;
			Sampling ${\beta_{k_{t-1}k_t}}$ by Eq.~(\ref{eq:beta}) \;
			Sampling ${r_{k_t}}$ by Eq.~(\ref{eq:r_k}) \;
			Sampling ${c_t}, c_0$ by Eq.~(\ref{eq:c}) \;
			$t = t+1$ \;
		}
	}
\end{algorithm}

\section{Experiments}
\subsection{Experiments with Synthetic Data}
To verify whether rCTM is capable to capture the multi-thread coupling weights between recurrent topics, we manually create a synthetic dataset with predefined coupling relationships over three slices. 

\begin{figure}[!htb]
	\centering
	\subfigure[The true coupling weights between topics at $t=1$ and $t=2$]{
		\includegraphics[width=.22\columnwidth]{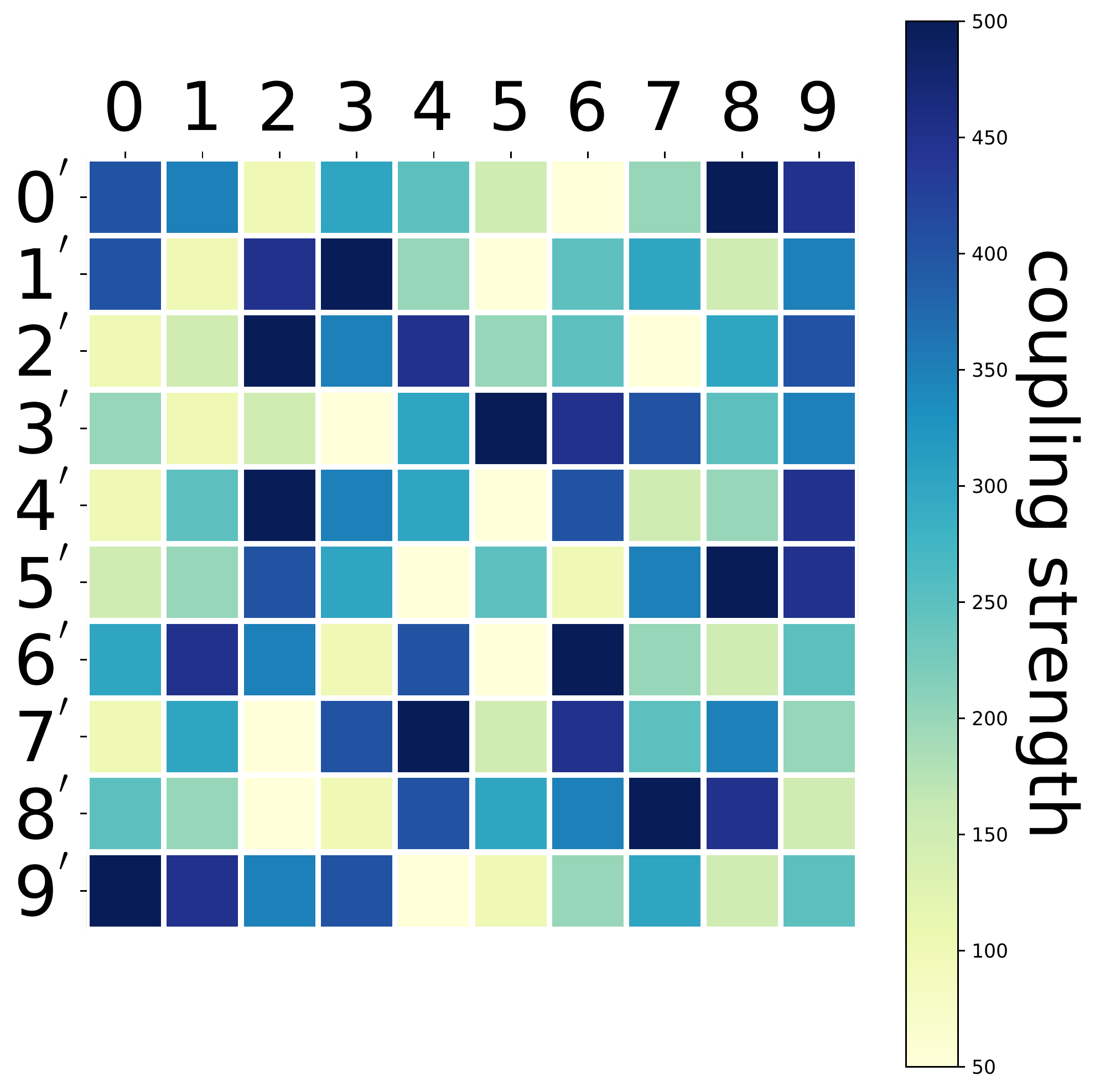}} 
	\hspace{.1in}	
	\subfigure[The estimated coupling weights between topics at $t=1$ and $t=2$]{
		\includegraphics[width=.22\columnwidth]{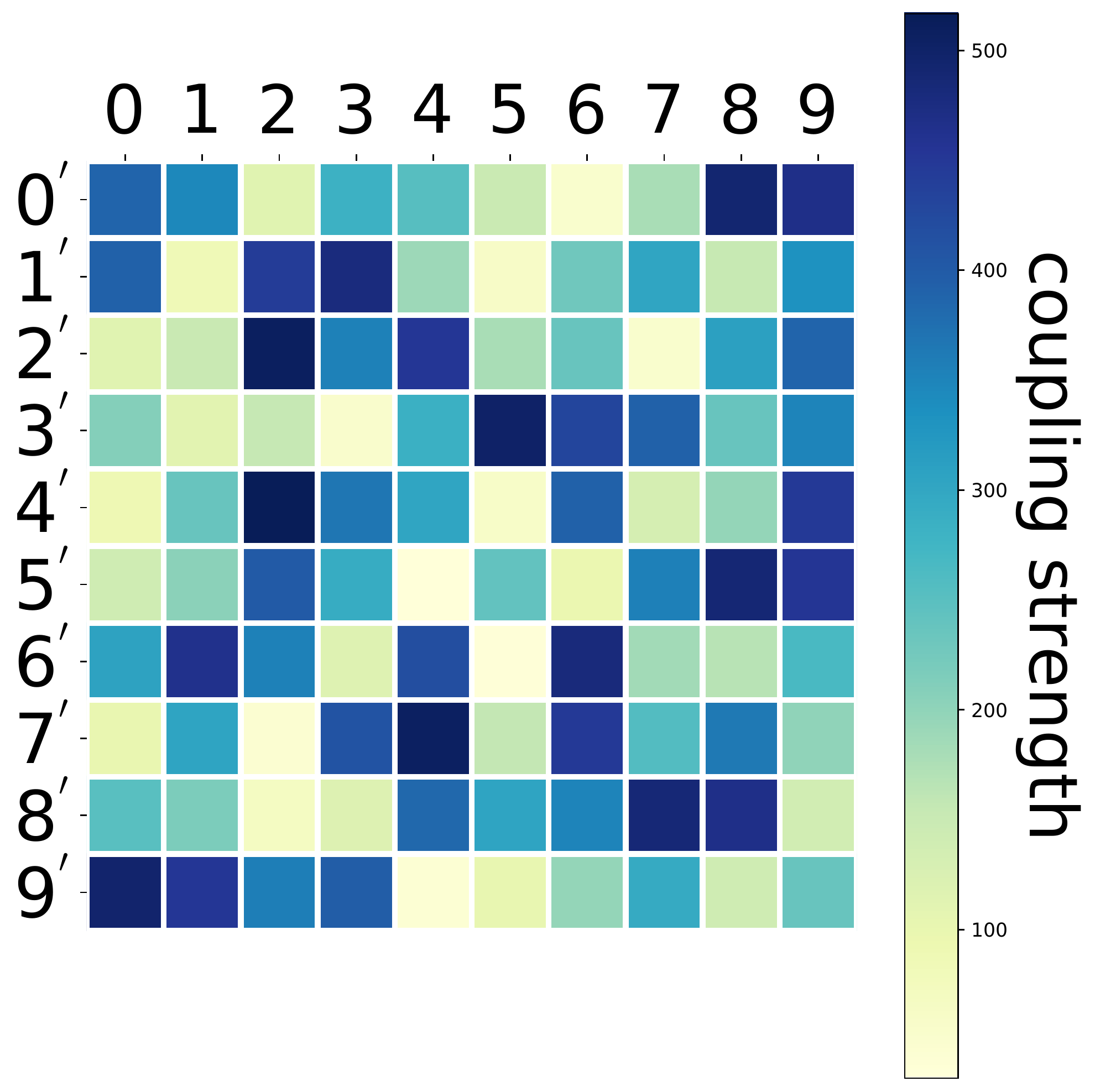}}
	\hspace{.1in}
	\subfigure[The true coupling weights between topics at $t=2$ and $t=3$]{
		\includegraphics[width=.22\columnwidth]{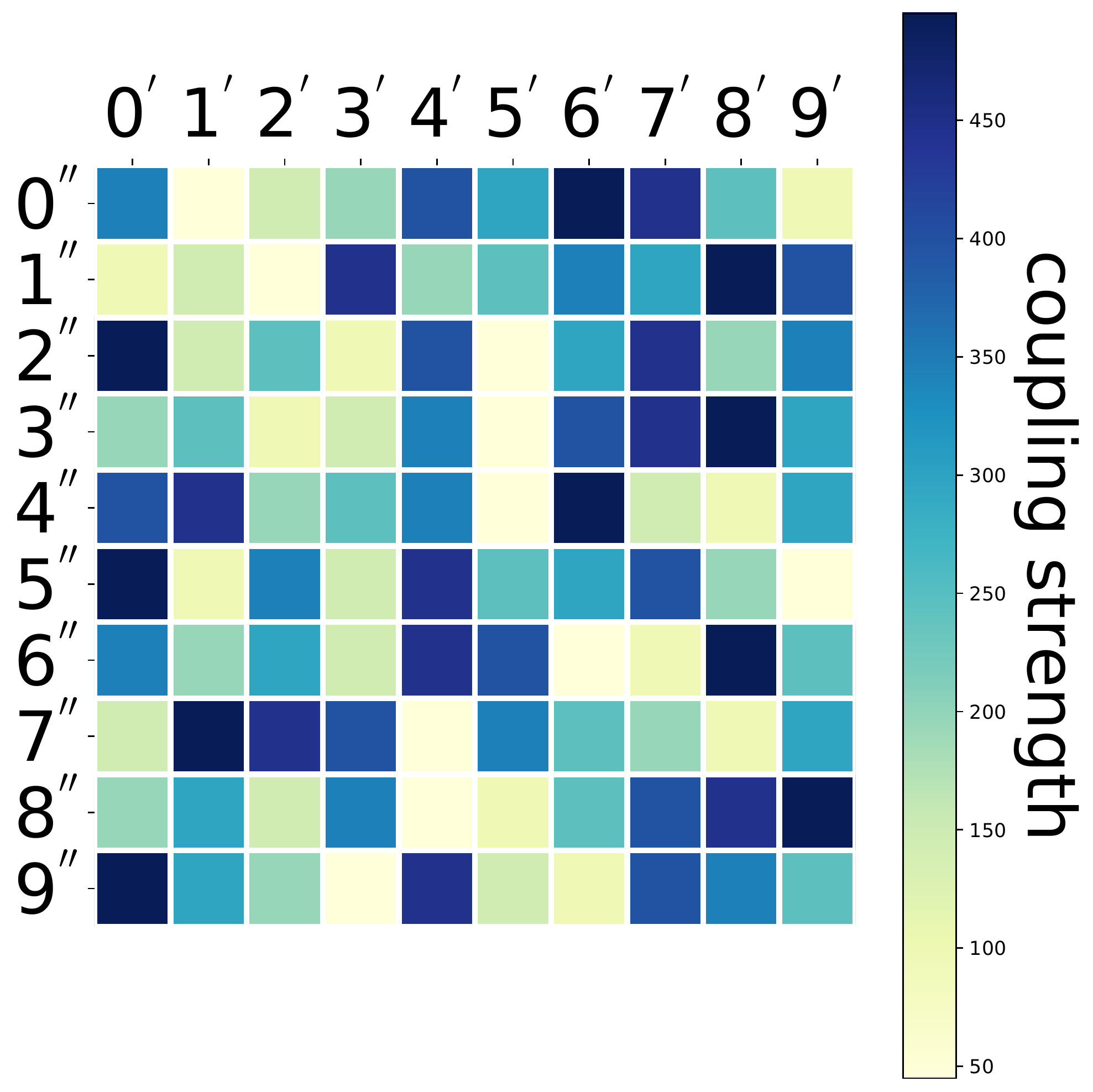}} 
	\hspace{.1in}
	\subfigure[The estimated coupling weights between topics at $t=2$ and $t=3$]{
		\includegraphics[width=.22\columnwidth]{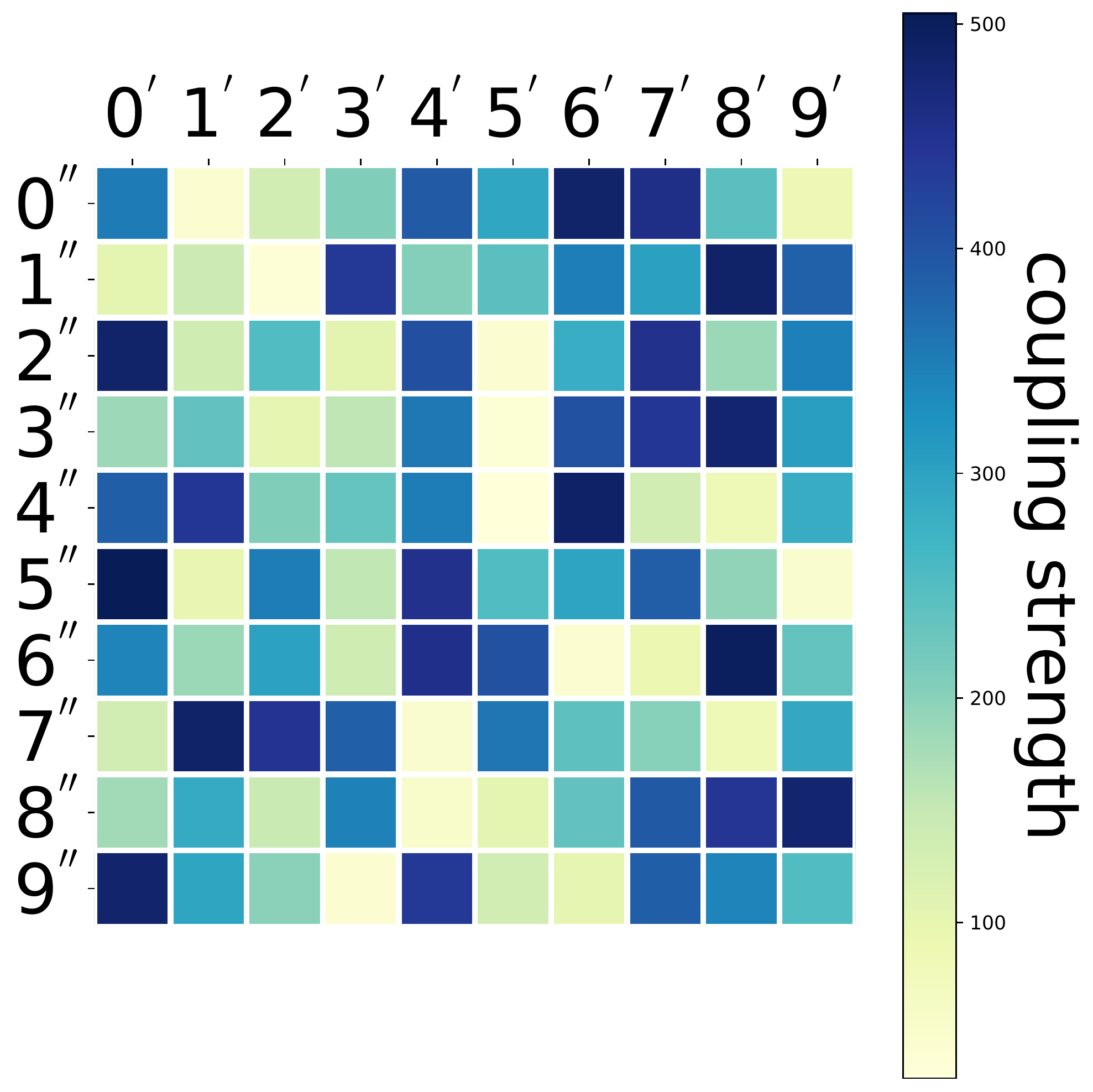}}
	\caption{The comparison of the true and estimated coupling weights between consecutive topics.}
	\label{fig:beta}
\end{figure}

Referring to the emprical study \cite{acharya2018dual}, three $1000 \times 1000$ document-word matrices with $1000$ documents over the vocabulary size of $1000$ are created sequentially at each slice according to the following steps. At slice $t=1$, we initialize ten topics $\{0, 1, \cdots, 9\}$ via the Dirichlet distributions, and randomly use them to generate the first $1000$ documents. Given the specified coupling weight matrix in Fig.\ref{fig:beta} (a), topics $\{0', 1', \cdots, 9' \}$ at slice $t=2$ are produced according to the proposed evolutionary process, and randomly to generate the second $1000$ documents. Similarly, topics $\{0'', 1'', \cdots, 9''\}$ at slice $t=3$ and the corresponding document chunk are also created.



Based on three synthetic document-word matrices, we utilize the proposed rCTM to recover the recurrent topics and their coupling weights to see whether the proposed model is able to decode the intricate dependency relationship between topics. Due to the space limit, only the comparison between the true coupling weights and the estimated weights by the rCTM is indicated by Fig. \ref{fig:beta}.
Noted from Fig. \ref{fig:beta} (a) and (b), the estimated $10\times10$ coupling weights between topics $\{0, 1, \cdots, 9\}$ at slice $t=1$ and topics $\{0', 1', \cdots, 9'\}$ at slice $t=2$ precisely match with the true matrix. Similarly, the estimated coupling weights between topics $\{0', 1', \cdots, 9'\}$ at slice $t=2$ and topics $\{0'', 1'',\cdots, 9''\}$ at slice $t=3$ also highly resemble the true weights denoted by Fig. \ref{fig:beta} (c) and (d). Furthermore, not only strong coupling but also weak dependency relationships between consecutive topics are successfully discriminated by rCTM. As the coupling weights between consecutive topics are precisely captured, the discovery of precise topics are naturally followed.
This comparison highlights rCTM is capable to decode the intrinsic multi-thread dependency between evolving topics.

\subsection{Experimental Setup with Real-world Data}
We use five real-world datasets from different domains to evaluate all algorithms. The statistics of datasets are summarized in the Table \ref{tab:dataset}. 
\begin{table}[tb]
	\centering
	\caption{The statistics of five real-wold datasets.}
	\resizebox{0.9\textwidth}{!}{%
		\begin{tabular}{@{} c c c c c @{}}
			\toprule
			Dataset & \#document & \#vocabulary  & time span & \#average document length \\ \midrule
			NIPS & 6,753 & 4,434  & 1987-2017 & 50\\ \midrule
			Flickr & 2,1435 & 1,887 &  Jun 1,2010 -- Aug 22, 2010 & 10 \\ \midrule
			News & 29,247 & 12,204 & April 17, 2014 - May 25, 2018 & 20 \\ \midrule
			ACL & 974 & 9,494 & 2016-2018 & 87 \\ \midrule
			SOTU & 227 & 6,570 &  1790 - 2016 & 1,152 \\ \bottomrule
		\end{tabular}%
	}
\label{tab:dataset}
\end{table}

\begin{figure}[!htb]
	\centering
	\subfigure[NIPS dataset]{
		\includegraphics[width=.28\columnwidth]{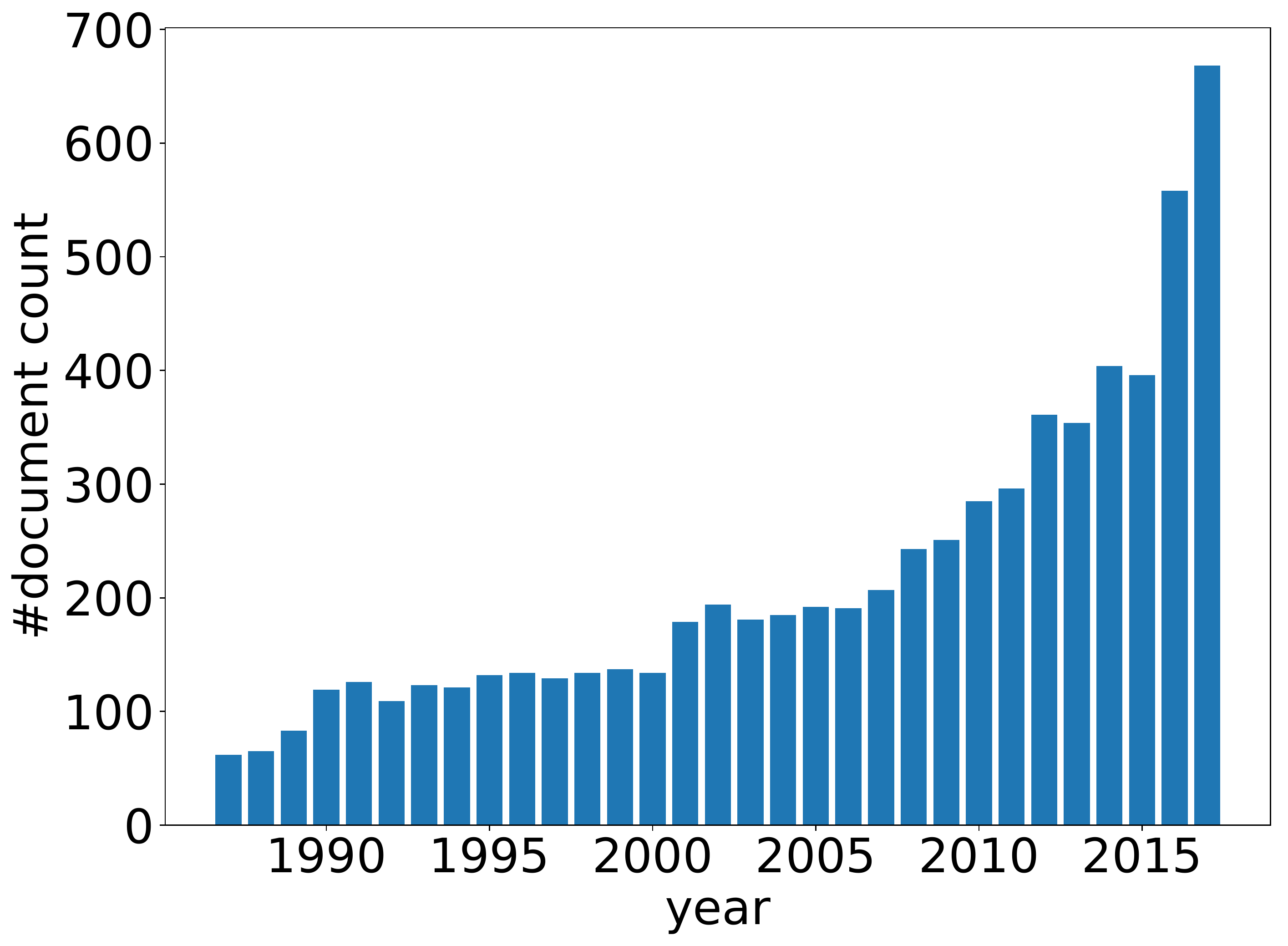}}
	\hspace{0.1in}	
	\subfigure[Flickr dataset]{
		\includegraphics[width=.28\columnwidth]{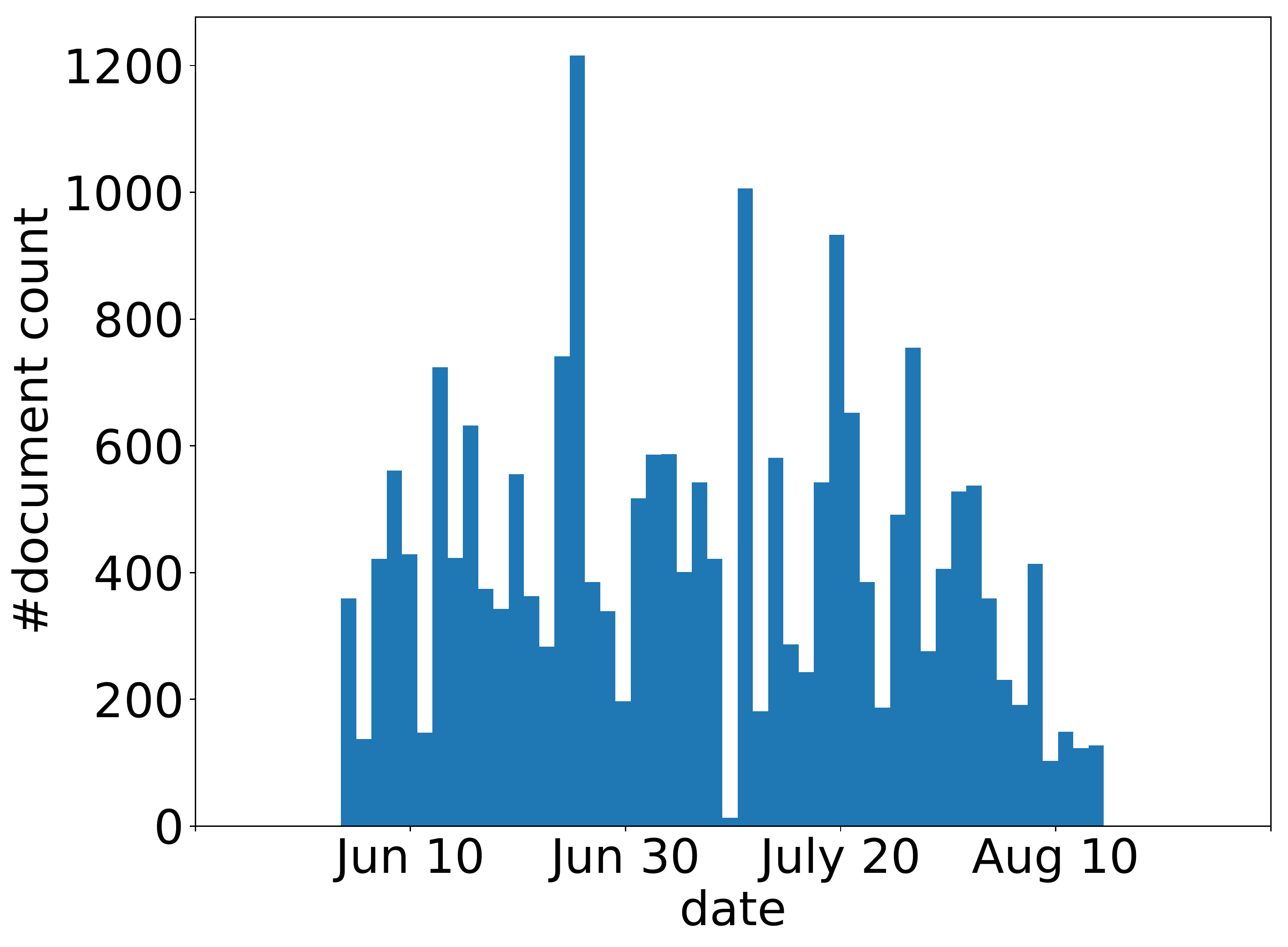}} 	
	\hspace{0.1in}
	\subfigure[News dataset]{
		\includegraphics[width=.28\columnwidth]{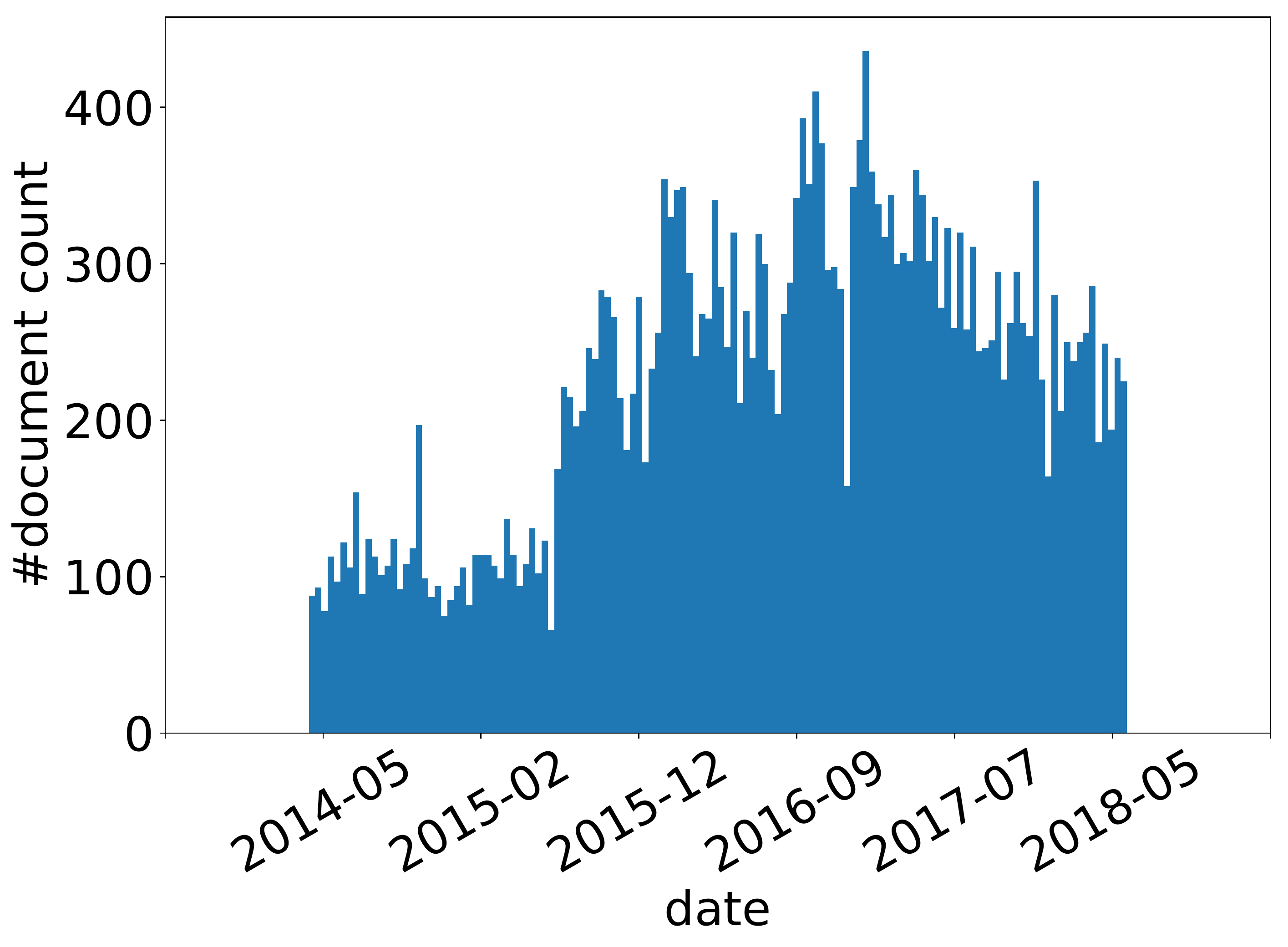}} 	
	\caption{The temporal densities of the real-world datasets.}
	\label{fig:dataset_density}
\end{figure}

\begin{itemize}
	\item NIPS corpus \cite{ben_hamner}. This benchmark dataset consists of the abstracts of papers appearing in the NIPS conference from the year 1987 to 2017. After the standard pre-processing and the removal of the most frequent and the least words, the size of corpus is reduced to 6,753 documents and 4,434 unique vocabularies, and the average document length is about 50.
	\item Flickr dataset. We collect the Flickr images within the city of Paris from June 1, 2010 to August 22, 2010. Since each image is annotated by a set of tags denoted by users, we assume those tags associated with an image make up a document. After the pre-processing, we obtain 21,435 documents with 1,887 vocabularies. Each document averagely contains 10 words.
	\item News \cite{dataset} dataset. In this dataset, only the news labeled politics is used. There are 29,247 documents with 12,204 unique vocabularies from April 17, 2014 to May 25, 2018 after the standard pre-processing. Each document averagely contains 20 words. 
	\item ACL dataset. It consists of the accepted papers from ACL Anthology in the three consecutive years. After the preprocessing, there are 974 documents with the dictionary size of 9,494. Each document averagely contains 87 words.
	\item SOTU dataset. It contains the text of annual speech transcripts delivered by the President of United States from 1790 to 2016. After the preprocessing, we obtain 227 documents with 6,570 vocabularies. Each document is averagely composed of 1,152 words.
\end{itemize}
In addition, the temporal densities of documents from NIPS, Flickr and News datasets are presented in Fig. \ref{fig:dataset_density}. On the ACL Anthology dataset, there are 265 papers in 2016, 324 papers in 2017 and 550 papers in 2018 respectively, and SOTU dataset is composed of the annual transcripts. The various densities of document arrivals in these datasets as well as the time spans thus form a good testing environment for the proposed rCTM and the other baselines. 

We compare the proposed model with the following state-of-the-art algorithms.
\begin{itemize}
	\item[--] \textbf{DTM}, short for the dynamial topic modeling \cite{blei2006dynamic}, is the seminal dynamic model for topic evolution discovery, where the dynamics of both topic proportions and word distributions are captured via the state space models. 
	\item[--] \textbf{DCT}, short for the Dynamic Clustering Topic model \cite{liang2016dynamic}, is one of the existing models which dynamically learns the topic evolution along the time slices, where both topic popularity and word evolution is captured by Dirichlet chains.
	\item[--] \textbf{rCRP}, a recurrent Chinese Restaurant Process \cite{ahmed2008dynamic}, is regarded as one of the benchmark algorithms in modeling dynamics topics where evolving topics are chained by using a recurrent Chinese Restaurant Process.
	\item[--] \textbf{ST-LDA}, short for Streaming LDA \cite{amoualian2016streaming}, originally learns the dynamic topic evolution between consecutive individual documents via a Dirichlet distribution. We extend it to capture the topic dependencies between the consecutive document chunks. In this approach, the topic evolution is chained by the Dirichlet distribution with a balanced scale parameter.
	\item[--] \textbf{DP-density} \cite{guo2017density}, explores the density of document arrivals to detect the dynamic topics based on the social media data streams, where a Dirichlet Process is used to infer the topic number and density estimation technique is exploited to learn the dynamics of topics.
	\item[--] \textbf{MStream}, a model-based text stream clustering algorithm \cite{yin2018model}, deals with the concept drift problem for the short text streams. To accommodate the topic drift in the long text, we revise its assumption of one-topic proportion to multi-topic proportions for each document. 
	\item[--] \textbf{DM-DTM} is short for the Dual Markov Dynamic Topic Model \cite{acharya2018dual} that exploits two Markov chains to capture both topic popularity and topic evolution. In this approach, the topic popularity is captured by the Gamma Markov chain, and topic evolution is modeled by the Dirichlet chain. 
	\item[--] \textbf{RNN-RSM} is an abbreviation for Recurrent Neural Network-Replicated Softmax Model \cite{gupta2017deep}, where the topic discovery and sequential documents are jointly modeled in the undirected raplicated softmax (RSM)\cite{hinton2009replicated} and the recurrent neural network (RNN) conveys the temporal information for the bias parameters of RSM.
	
\end{itemize}
Besides these competitive baselines, the following are our proposed model and its variations.
\begin{itemize}
	\item[--] \textbf{rCTM} is the proposed recurrent Coupling Topic Model, where a new proposal of the multi-topic-thread is induced to describe the topic evolution, and the IBP compound distribution is exploited to infer the topic number as well as sparse topic proportions for each document.	
	\item[--] \textbf{rCTM-D} refers to the variant model with the dropout technique. In this approach, the dropout is facilitated over the coupling connections of topics to randomly drop out connections with the given probability, which is used to validate the significance of multi-dependency between evolving topics.
	\item[--] \textbf{rCTM-F} is another variant of rCTM, where the topic number is specified by a fixed number without resorting to the latent IBP compound distribution, and a customized topic proportion of each document is replaced by the fixed common topics at each slice. 	
\end{itemize}

In the experiment, we divide a dataset into a sequence of equidistant time slices chronologically, and each document chunk corresponds to a slice. NIPS, Flickr, News and ACL dataset are divided per three years, per fortnight, per month and per year respectively, and SOTU dataset is divided into 5 slices and each slice spans 45 years. At slice $t=1$, topics are directly learned from the document chunk $\mathbf{d}_1$ without prior dependency. When time $t>1$, the evolution of topics depends on their prior states and their coupling relationships. The experimental settings for all models are as follows. (1) In each dataset, the time division is the same for those document chunk-based models including DTM, DCT, rCRP, ST-LDA, DM-DTM, RNN-RSM, the proposed rCTM and its two variants, and document stream-based models DP-density and MStream do not require this setting. 
(2) Regarding the topic number setting, the nonparametric models including rCTM, rCTM-D, DM-DTM, DP-density and rCRP are able to automatically learn topic number without such a setting, while DTM, DCT, ST-LDA, MStream, RNN-RSM and rCTM-F are specified with the same topic number as those nonparametric models in each dataset. (3) In terms of the document length, the one-topic assumption from DCT, MStream and DP-density is retained on the NIPS, Flickr and News datasets, and the assumption is extended to a multi-topic assignment to adapt to the long text of ACL and SOTU dataset. (4) In rCTM, the hyper-parameter is given as $\eta_0 = 0.1$, $\alpha = 0.1$ in the component of topic proportion construction, while the rest are tuned as $\eta = 0.1$, $a_0 = b_0 = 1$, $e_0=1$, $d_0 = 10$, $r_0=1$ by grid search based on the metric of perplexity during the topic evolution process. We run $1,000$ Gibbs samplings to implement the inference process. The parameter settings in other baselines firstly refer to their original papers if available, otherwise we set them at their optimal performance.

\subsection{Quantitative Results}
Traditionally, \textit{perplexity} \cite{blei2003latent} is defined to measure the goodness-of-fit of topic modeling by randomly splitting the dataset into training set and testing set, and it is popularly used in the recent topic modeling work \cite{acharya2018dual,zhao2018dirichlet,guo2020deep,hua2020probabilistic}. In addition, several new metrics of topic coherence evaluation have been proposed for a comparative review. Among all the competing metrics, the topic coherence \cite{lau2014machine,roder2015exploring} matches human judgement most closely, so we adopt it in this work. We also report perplexity, primarily as a way of evaluating the generativeness of different approaches.

\subsubsection{\textit{Perplexity} over Held-out Set}
Following the setting in \cite{gupta2017deep}, we randomly hold out $p$ fraction of the dataset ($p \in \{0.6, 0.7, 0.8, 0.9\}$) at each time slice, and train a model with the rest and predict on the sum of held-out sets. A lower perplexity indicates a better generation of the model. For comparing multiple modelings with different assumptions as well as different inference mechanisms, the per-word perplexity on the sum of held-out sets is formally defined as
\begin{equation*}
	perplexity = exp(\frac{- \sum_{t=1}^{T}\sum_{d=1}^{|\mathbf{d}_{t}|} \sum_{w=1}^{V} log ~P(\sum_{k_t=1}^{K_t} \theta_{dk_t}\mathbf{\phi}_{k_tw})} {\sum_{t=1}^{T}\sum_{d=1}^{|\mathbf{d}_{t}|} \sum_{w=1}^{V} x_{tdw}}),
\end{equation*}
where $|\mathbf{d}_{t}|$ records the number of documents at slice $t$,  $x_{tdw}$ indicates the observed occurrence number of word $w$ in the document $d$ at slice $t$, while $\mathbf{\theta}_d$ and $\mathbf{\phi}_{k_t}$ are estimated in the inference procedure.

\begin{table}[tb]
	\centering
	\caption{Perplexity results of the increasing training data with varying ratios $p \in \{0.6, 0.7, 0.8, 0.9 \}$ on the NIPS, Flickr and News datasets (The best performance is highlighted in boldface, the second best is emphasized with $^\ast$ and the third best is denoted in underlined).}
	\resizebox{\columnwidth}{!}{%
		\begin{tabular}{@{} l c c c c c c c c c c c c @{}}
			\toprule
			\multirow{2}{*}{\textbf{Models}} & \multicolumn{4}{c}{\textbf{NIPS}} & \multicolumn{4}{c }{\textbf{Flickr}} & \multicolumn{4}{c}{\textbf{News} } \\\cmidrule(lr){2-5}  \cmidrule(lr){6-9} \cmidrule(lr){10-13}
			& $p= 0.6$ &  $p=0.7$ &  $p=0.8$ &  $p=0.9$ &  $p=0.6$ &  $p=0.7$ &  $p=0.8$ &  $p=0.9$ &  $p=0.6$ &  $p=0.7$ &  $p=0.8$ &  $p=0.9$  \\  \midrule
			DTM & 1658 & 1632 & 1619 & 1597 & 522 & 496 & 487 & 469 & 2880 & 2854 & 2821 & 2785 \\ \midrule
			DCT & 1718 & 1669 & 1592 & 1534 & 430 & 412 & 398 & 390 & 2682 & 2609 & 2571 & 2555  \\ \midrule
			rCRP & 1516 & 1457 & 1395 & 1360 & 413 & 398 & 388 & 381 & 2692 &  2588 & 2559 & 2499 \\ \midrule
			MStream & 1328 & 1324 & 1303 & 1297 & 419 & 395 & 380 & 375 & 2749 & 2677 & 2637 & 2578 \\ \midrule
			DP-density & 1488 & 1457 & 1450 & 1446 & 347 & 344 & 342 & 340 & 2622 & 2570 & 2522 & 2490\\ \midrule
			ST-LDA & \underline{1207} & \underline{1203} & \underline{1198} & \underline{1195} & \underline{249} & \underline{245} & \underline{240} & \underline{237} & 2864 & 2724 & 2625 & 2549\\ \midrule
			DM-DTM & 1400 & 1364 & 1333 & 1309 & 427 & 382 & 354 & 330 & \underline{2401} & \underline{2365} & \underline{2323} & \underline{2300} \\ \midrule
			rCTM & \textbf{1057} & \textbf{1045} & \textbf{1015} & \textbf{1013} & \textbf{179} & \textbf{163} & \textbf{152} & \textbf{140} & \textbf{1824} & \textbf{1773} & \textbf{1749} & \textbf{1630} \\ \midrule
			rCTM-F & 1090* & 1075* & 1040* & 1036* & 192* & 178* & 166* & 160*	& 2099* & 2002* & 1931* & 1859* \\ \bottomrule
		\end{tabular}%
	}
	\label{tab:perp1}
\end{table}
\begin{figure}[t]
	\centering
	\subfigure[NIPS dataset]{
		\includegraphics[width=.3\columnwidth]{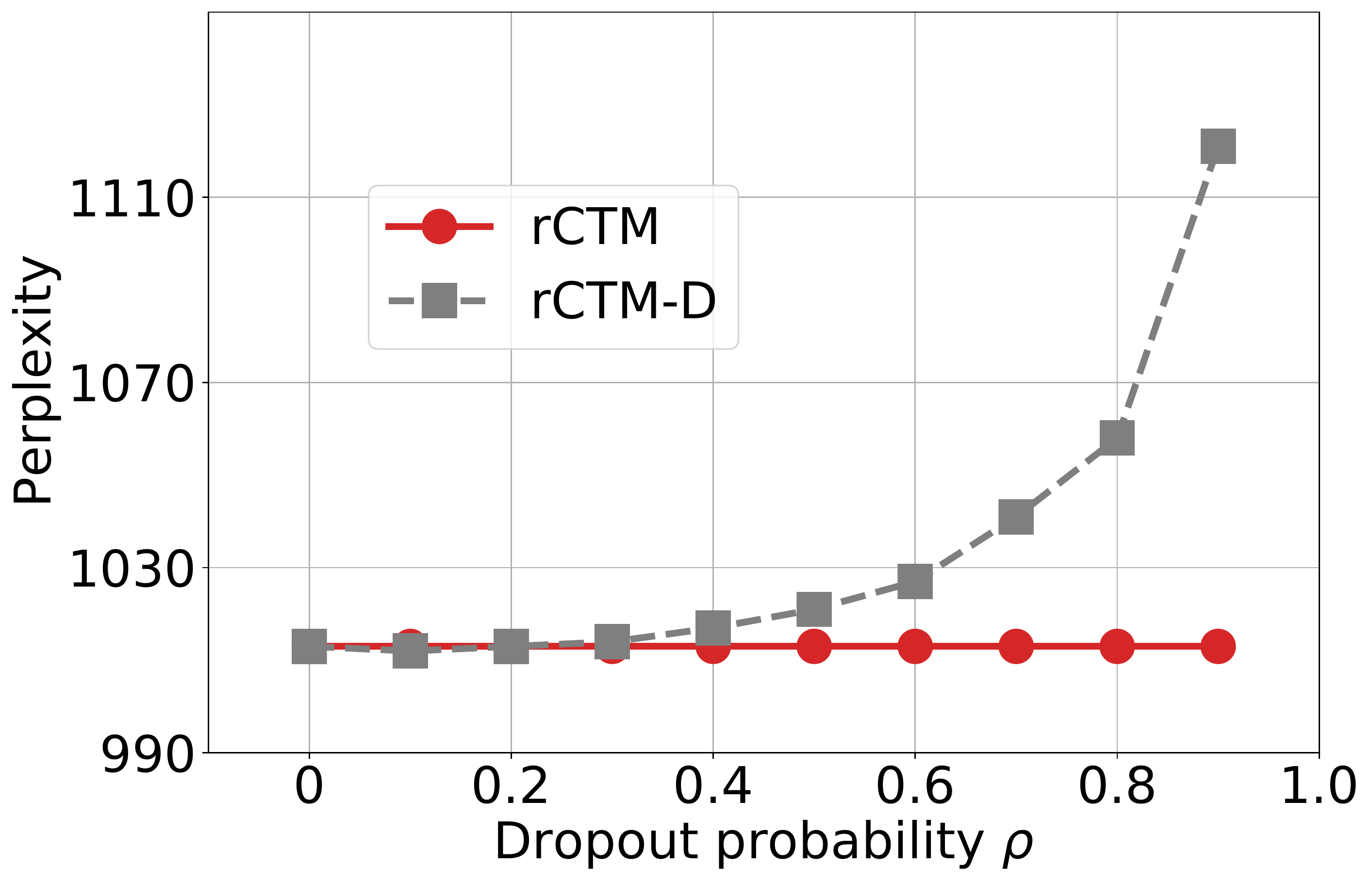}}
	\hspace{0.1in}	
	\subfigure[Flickr dataset]{
		\includegraphics[width=.3\columnwidth]{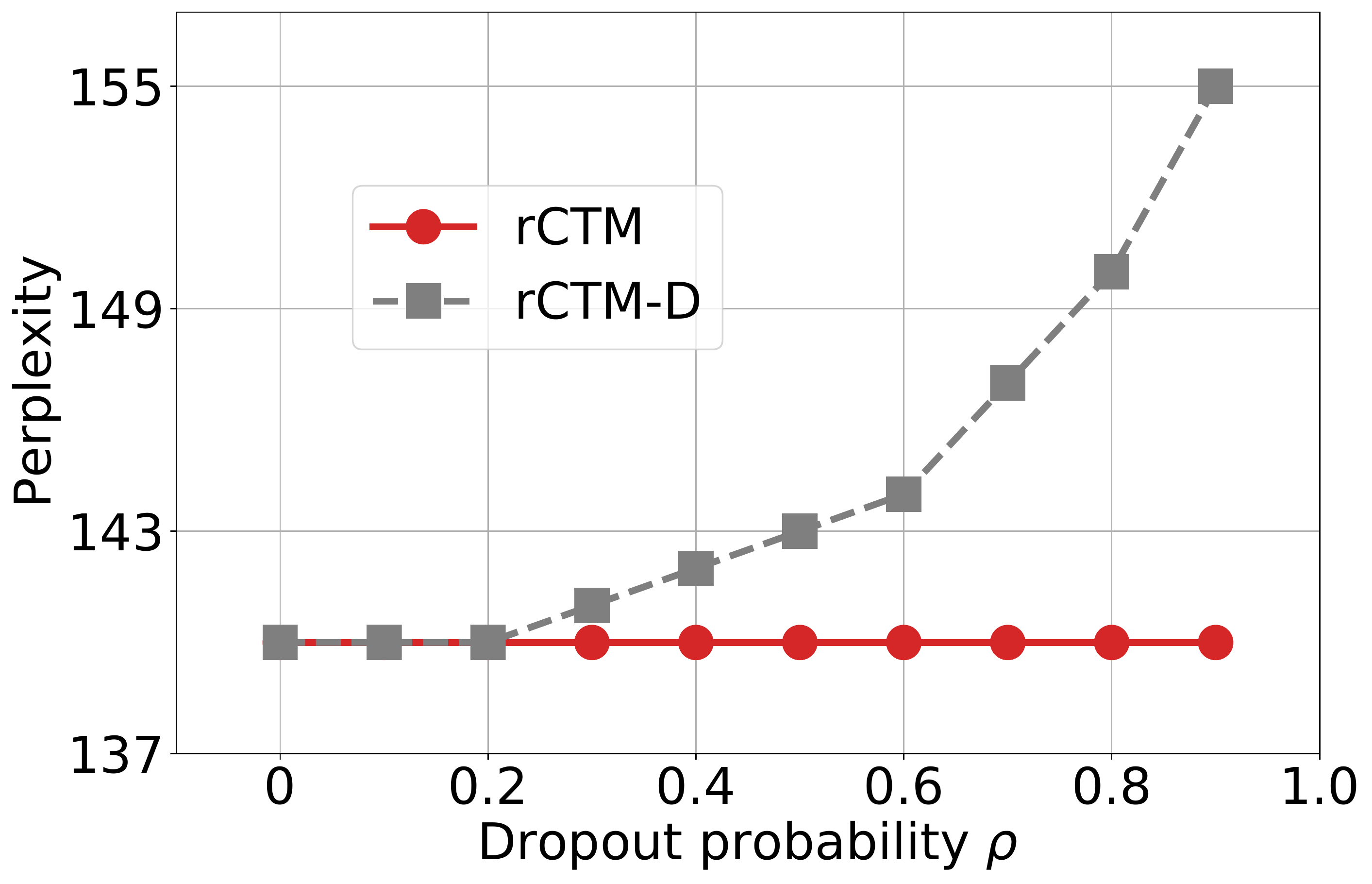}} 
	\hspace{0.1in}
	\subfigure[News dataset]{
		\includegraphics[width=.3\columnwidth]{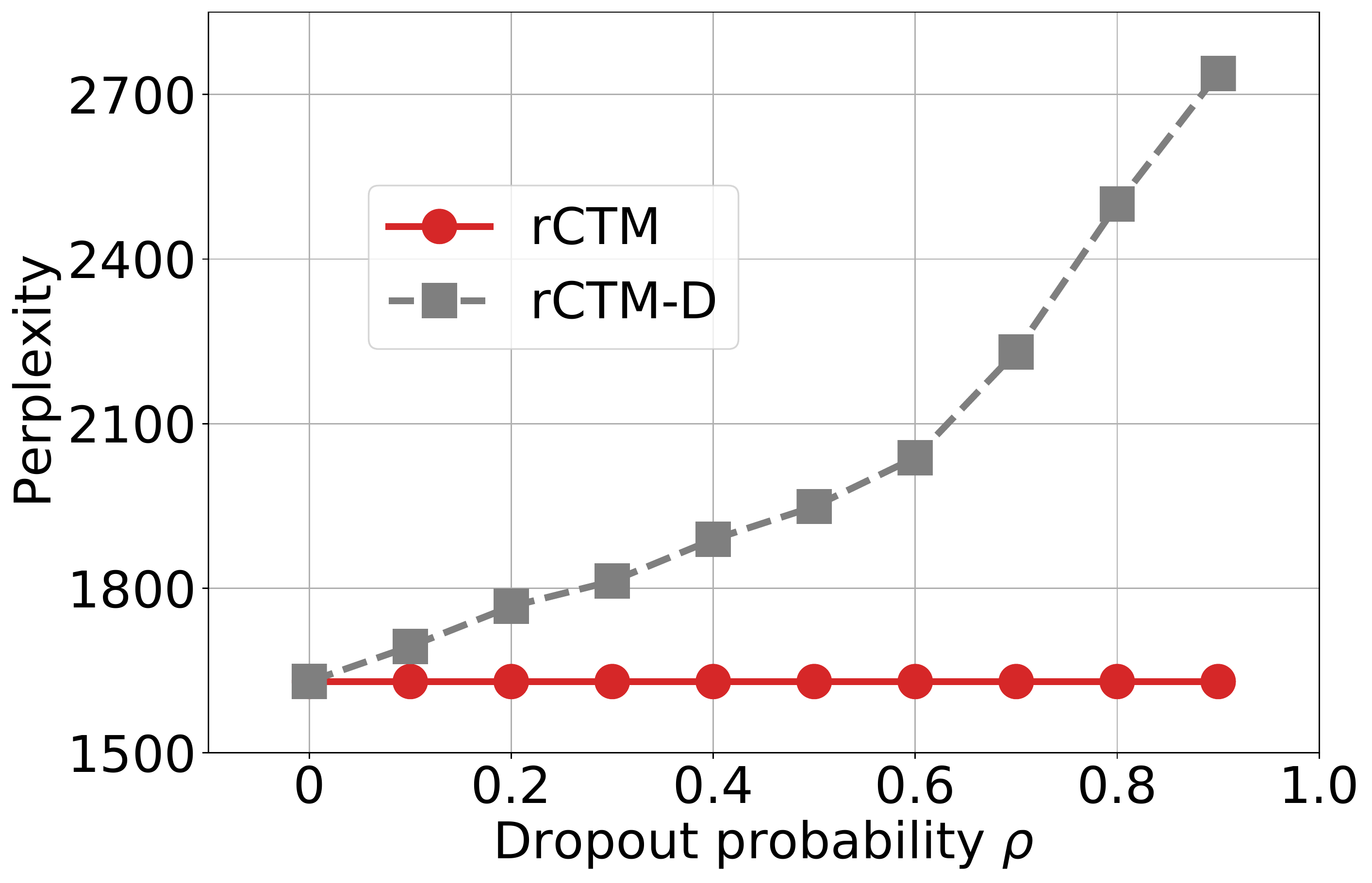}}	
	\caption{Perplexity comparison between rCTM and its variant rCTM-D with different dropout probabilities $\rho \in [0, 1)$ on the NIPS, Flickr and News datasets. Both models are measured over the training data with the split ratio $p = 0.9$.}
	\label{fig:perp_d1}
\end{figure}

Table \ref{tab:perp1} reports the \textit{perplexity} performance over varying ratios of held-out sets on the short-text datasets including NIPS, Flickr and News dataset. By examining the performance result on each dataset, we have the following remarks.

\stitle{Document chunk-based models.} (1) Though the density of document arrivals as well as time span on three datasets are very different, the proposed rCTM and its variant rCTM-F consistently outperform the other baselines with a significant decrease in the perplexity at varying ratios of held-out sets, which confirms the superiority of the proposal of multiple dependencies between evolving topic sequences. Moreover, though rCTM-F is specified with the same topic number learned from rCTM, the perplexity difference between them validates the advantage of the latent IBP compound process in the task of inferring topic number and sparse topic proportion construction. (2) Except for the rCTM and its variant, ST-LDA achieves the best performance on the varying ratios on the NIPS and Flickr datasets, while DM-DTM wins at high ratios on the News dataset, which may be explained that the Gamma Markov Chain in DM-DTM is more fit for the topic weight evolution than others on the News dataset. Among the rest document chunk-based models, the nonparametric model rCRP consistently performs better than DCT and DTM at varying ratios on the three datasets.

\stitle{Document stream-based models.} It is noted that the performance of MStream and DP-density is different on the three datasets, though both models target at the document streams. DP-density achieves a better result than MStream on the Flickr and News dataset while its performance decreases on the NIPS dataset. That's because DP-density incorporates the arriving density of document streams to determine the dynamics of topics while MStream does not, thus DP-density is more suitable for the social media data with dense arrivals of documents. Such a comparison also implies the proposed generic rCTM is robust to the datasets with different temporal densities. 

\stitle{Dropout-based model.} The comparison between rCTM and the variant rCTM-D with varying dropout probabilities on the three datasets is shown in Fig. \ref{fig:perp_d1}. Both proposed models are measured with training data ratio $p = 0.9$. In rCTM-D, the dropout indicator $m$ is drawn from the Bernoulli distribution with $\rho$. If $m = 0$, the coupling connection is preserved, otherwise it is pruned. On the NIPS dataset, it is observed that perplexity results dramatically increase when the dropout probability $\rho \ge 0.3$. The large dropout probability would drop out the most coupling connections, and topics thus evolve with little dependency from their prior states, leading to the corrupted evolving topic sequences. The comparison results imply the significance of multiple couplings between topic chains. On the NIPS dataset, when the dropout probability $\rho \le 0.3$, it implies most of the multi-coupling connections are maintained, and the perplexity results are stable and nearly approach the optimal performance. On the Flickr dataset, rCTM-D achieves the best performance when the dropout probability $\rho \le 0.2$, and it is more evident that rCTM-D on the News dataset obtains its best performance only when the dropout probability $\rho = 0$, which means all coupling relationships are preserved. The results from rCTM-D on the three datasets further confirms the proposal of multi-coupling relationships between evolving topics. 

\begin{table}[tb]
	\centering
	\caption{Perplexity performance of the increasing training data with varying ratios $p \in \{0.6, 0.7, 0.8, 0.9 \}$ on the ACL and SOTU datasets (The best performance is highlighted in boldface, the second best is emphasized with $^\ast$ and the third best is denoted in underlined).}
	\resizebox{0.83\columnwidth}{!}{%
		\begin{tabular}{@{} l @{\hspace{0.5cm}} c c c c  c c c c  @{}}
			\toprule
			\multirow{2}{*}{\textbf{Models}} & \multicolumn{4}{c}{\textbf{ACL}} & \multicolumn{4}{c }{\textbf{SOTU}} \\\cmidrule(lr){2-5} \cmidrule(lr){6-9}
			& $p= 0.6$ &  $p=0.7$ &  $p=0.8$ &  $p=0.9$ &  $p=0.6$ &  $p=0.7$ &  $p=0.8$ &  $p=0.9$ \\ \midrule
			DTM & \underline{2599} & 2581 & 2567 & 2555 & 4060 & 4181 & 4312 & 4479 \\ \midrule
			DCT & 3340 & 3295 & 3263 & 3175 & 4211 & 3910 & 3774 & 3477 \\ \midrule
			rCRP & 3314 & 3271 & 3233 & 3140 & 4089 & 3826 & 3701 & 3416 \\ \midrule
			MStream & 3109 & 3065 & 3025 & 2800 & \underline{3461} & 3400 & 3350 & 3317 \\ \midrule
			DP-density  & 3105 & 3049 & 2980 & 2803 & 3480 & 3445 & 3415 & 3389  \\ \midrule
			ST-LDA & 2612 & \underline{2549} & \underline{2512} & \underline{2488} & \textbf{3256} & \textbf{3214} & 3181* & 3160* \\ \midrule
			DM-DTM & 2766 & 2704 & 2648 & 2591 & 3502 & 3467 & 3430 & 3399 \\ \midrule 
			rCTM & \textbf{2426} & \textbf{2388} & \textbf{2371} & \textbf{2355} &  3336* & 3239* & \textbf{3165} & \textbf{3144} \\ \midrule
			rCTM-F & 2479* & 2445* & 2416* & 2401* & 3472 & \underline{3304} & \underline{3206} & \underline{3166}  \\ \bottomrule
		\end{tabular}%
	}
	\label{tab:perp2}
\end{table}
\begin{figure}[tb]
	\centering
	\subfigure[ACL dataset]{
		\includegraphics[width=.35\columnwidth]{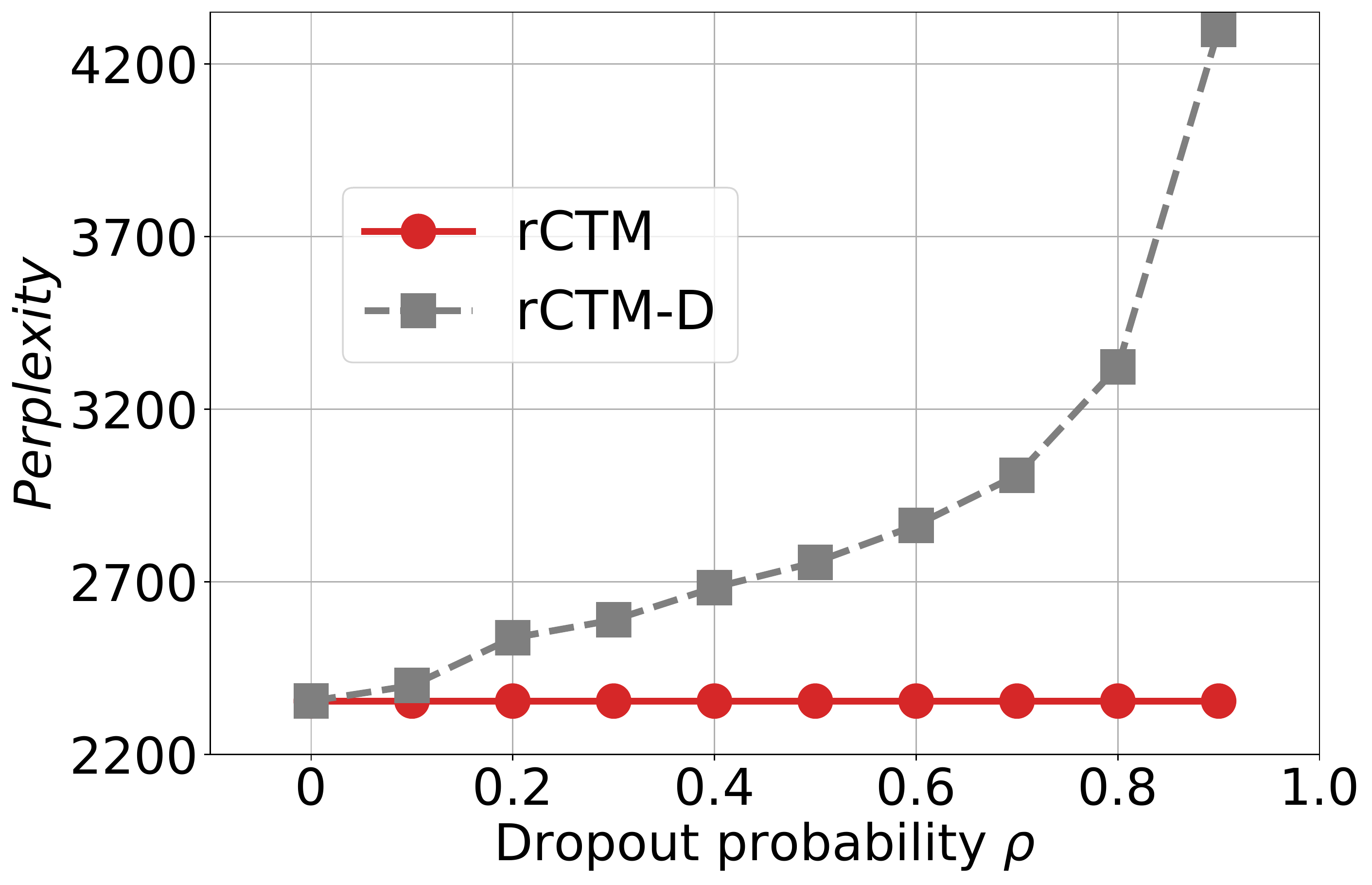}}
	\hspace{0.25in}	
	\subfigure[SOTU dataset]{
		\includegraphics[width=.35\columnwidth]{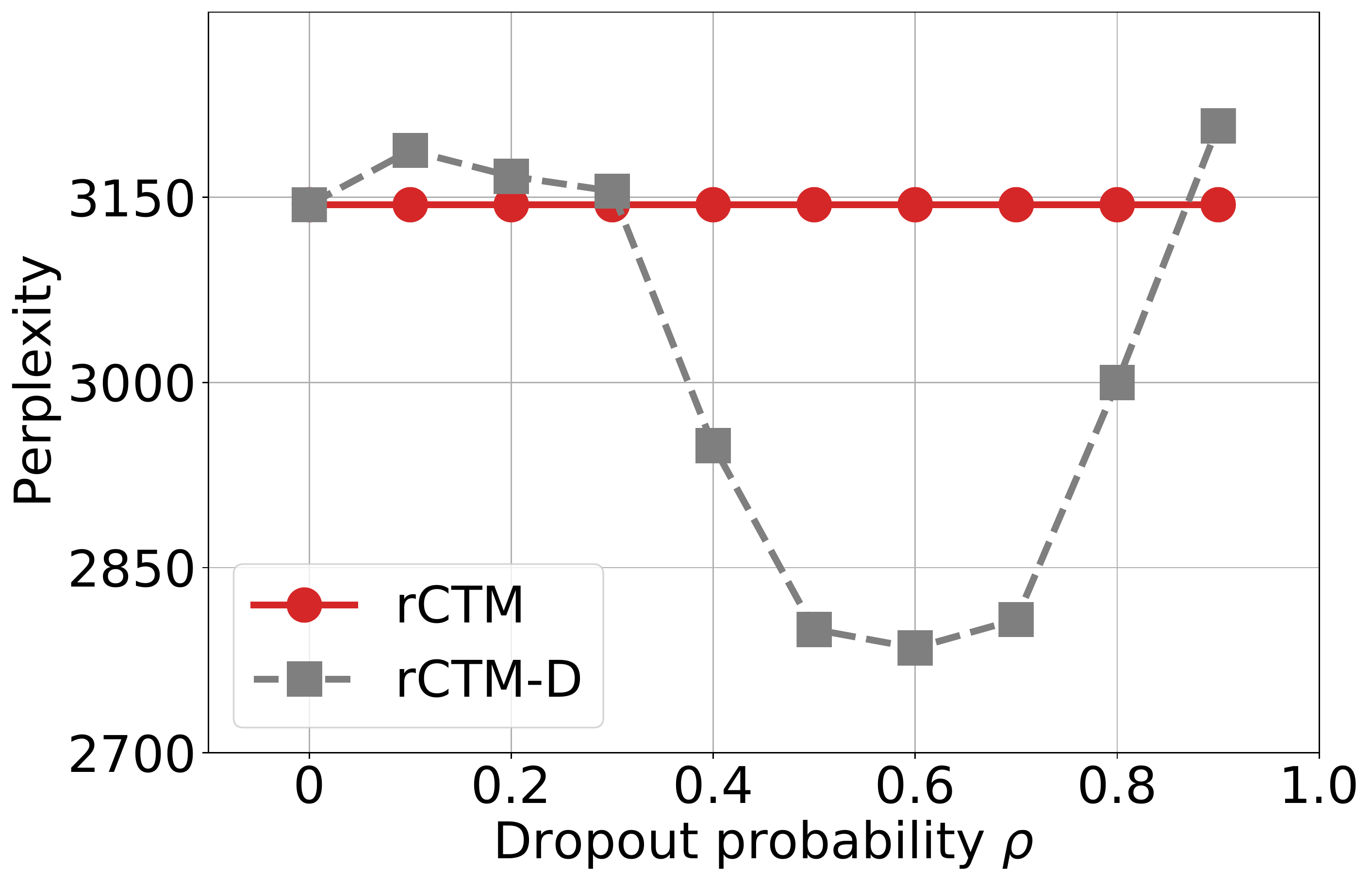}} 	
	\caption{Perplexity comparison between rCTM and the variant rCTM-D with different dropout probabilities $\rho \in [0, 1)$ on the ACL and SOTU datasets. Both models are measured over the training data with the split ratio $p = 0.9$.}
	\label{fig:perp_d2}
\end{figure}

Indicated in the Table \ref{tab:perp2} and Fig. \ref{fig:perp_d2}, the \textit{perplexity} analysis of all competitors on the two long-text datasets including ACL and SOTU dataset, is in the following.

\stitle{Document chunk-based model.} 
On the ACL dataset, (1) both rCTM and rCTM-F achieve the best performance with an evident decrease in the perplexity at different ratios, followed by the ST-LDA and DTM. Such a comparison once again validates the proposal of multi-topic-thread evolution. With the same topic number setting, the distinct difference between rCTM and rCTM-F in the perplexity is credited to the latent IBP compound process in the construction of sparsely customized topic proportions for documents. (2) Among the rest document chunk-based models, DM-DTM performs better than rCRP, followed by the performance of DCT.

On the SOTU dataset, (1) the competitor of ST-LDA and the proposed rCTM achieve comparable results at varying ratios, while the former performs better at ratio $p=0.6$ and $p=0.7$ and the latter stands out at $p=0.8$ and $p=0.9$. However, both strong methods are defeated by the variant of rCTM-D, which earns a much lower perplexity result when the dropout probability $\rho \in \{0.4, 0.5, 0.6, 0.7, 0.8\}$, indicated by Fig.\ref{fig:perp_d2} (b). Specifically, rCTM-D reaches its optimal performance at the dropout probability $\rho = 0.6$. Such results imply that the performance of rCTM improves when a large portion of topic couplings between evolving topics are dropped out. After carefully checking the word distributions of topics as well as their coupling weights, we find this phenomenon is caused by the characteristics of the long-term dataset. In this case, the SOTU dataset is divided into $5$ slices, and each slice is allocated with $45$ documents spanning $45$ years. A portion of topics crossing two slices are actually weakly coupled during the 90-year time, even though some topics seem similar by sharing common frequent words (e.g., power, president, and right). Hence, some of their dependency connections could be dropped out. This phenomenon remains at different time divisions. And not coincidentally, it also occurs in other baselines, whose performance degrades on this dataset. However, rCTM-D survives by dropping out some topic coupling connections between evolving topics on the dataset. (2) Among the rest document chunk-based models, their performance is ranked as DM-DTM > rCRP > DCT > DTM.

\stitle{Document stream-based model.} On the ACL dataset, the difference between DP-density and MStream is slight at varying ratios in terms of a 3-year timespan, while MStream performs better than DP-density on the SOTU dataset. It's because the annual transcripts from the SOTU dataset may not be a good clue to the density estimation in DP-density and its performance is thus compromised.

\stitle{Dropout-based model.} Indicated by Fig. \ref{fig:perp_d2} (a), only when the dropout probability $\rho = 0$, rCTM-D obtains its best performance on the ACL dataset when all coupling relationships are preserved, which confirms the significance of multi-coupling relationships between topic chains. Distinct from the aforementioned datasets, SOTU dataset contains the long-range annual transcripts from 1790 to 2016, which results in the weak connection between topics in consecutive slices. Therefore, rCTM-D reaches the lowest perplexity by dropping out some topics coupling connections between topics.

\subsubsection{Topic Coherence}
\begin{figure}[!htb]
\centering
\subfigure[NIPS dataset]{
	\includegraphics[width=.32\columnwidth]{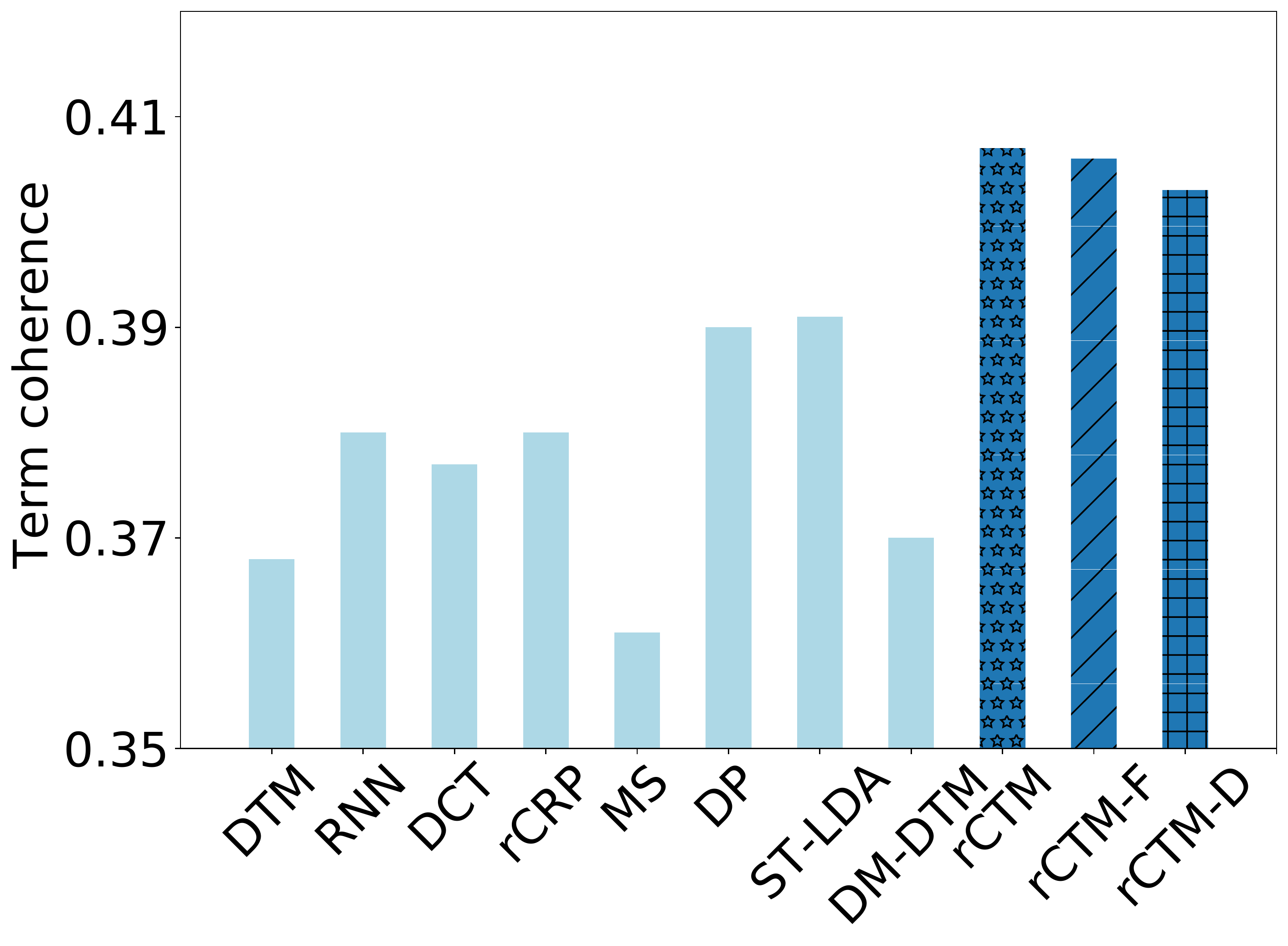}}
\subfigure[Flickr dataset]{
	\includegraphics[width=.32\columnwidth]{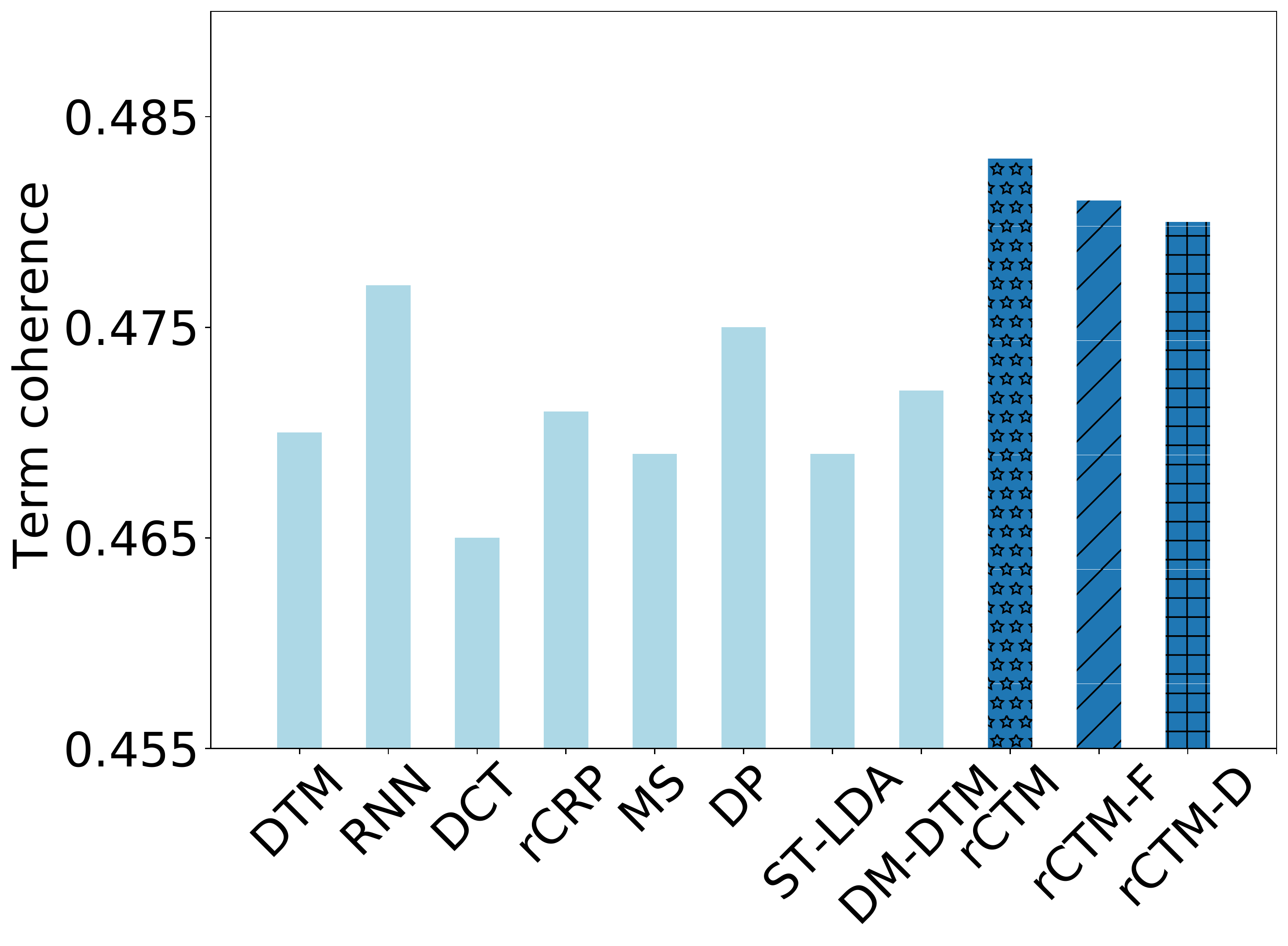}} 	
\subfigure[News dataset]{
	\includegraphics[width=.32\columnwidth]{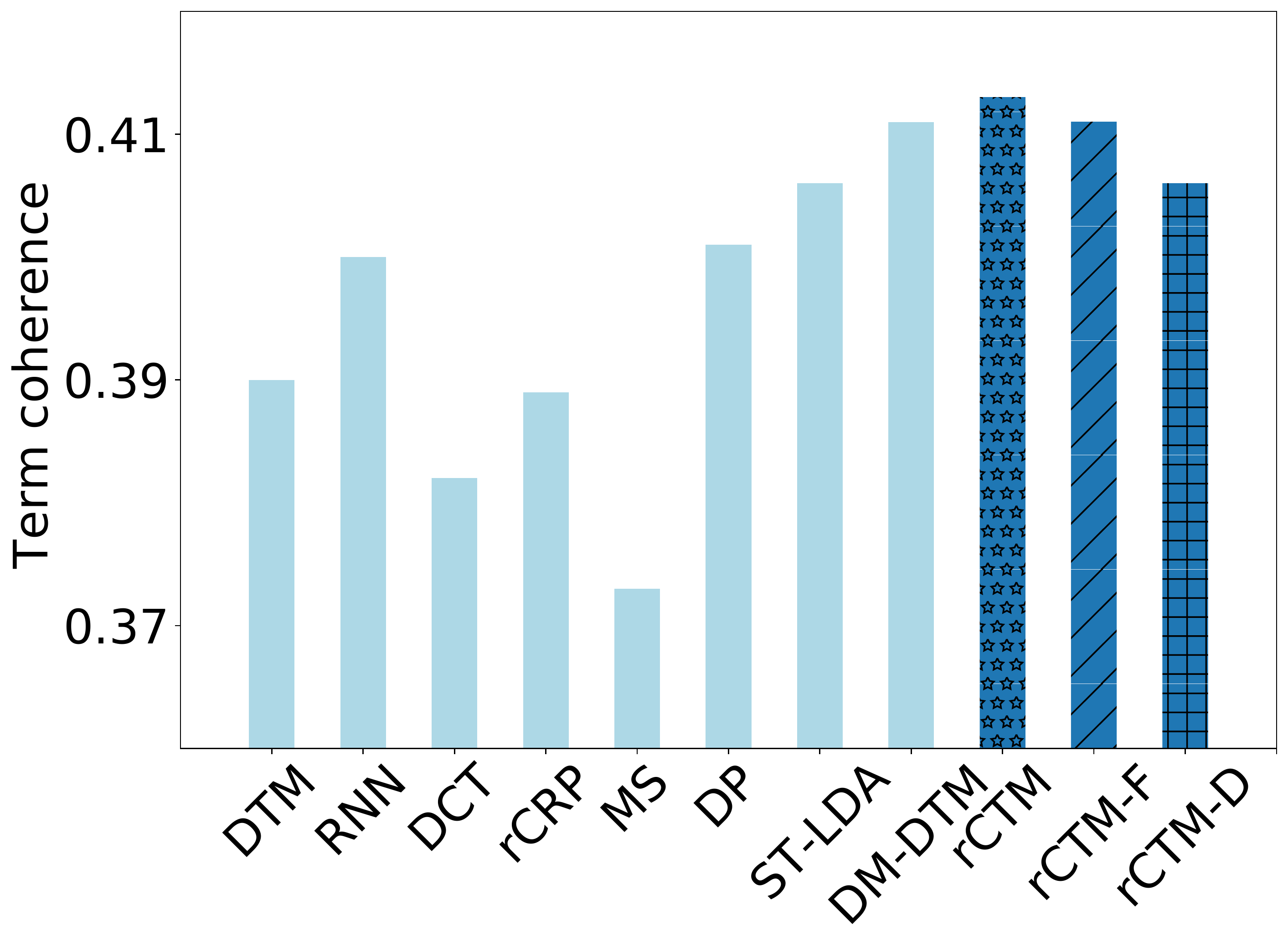}} 	
\subfigure[ACL dataset]{
	\includegraphics[width=.32\columnwidth]{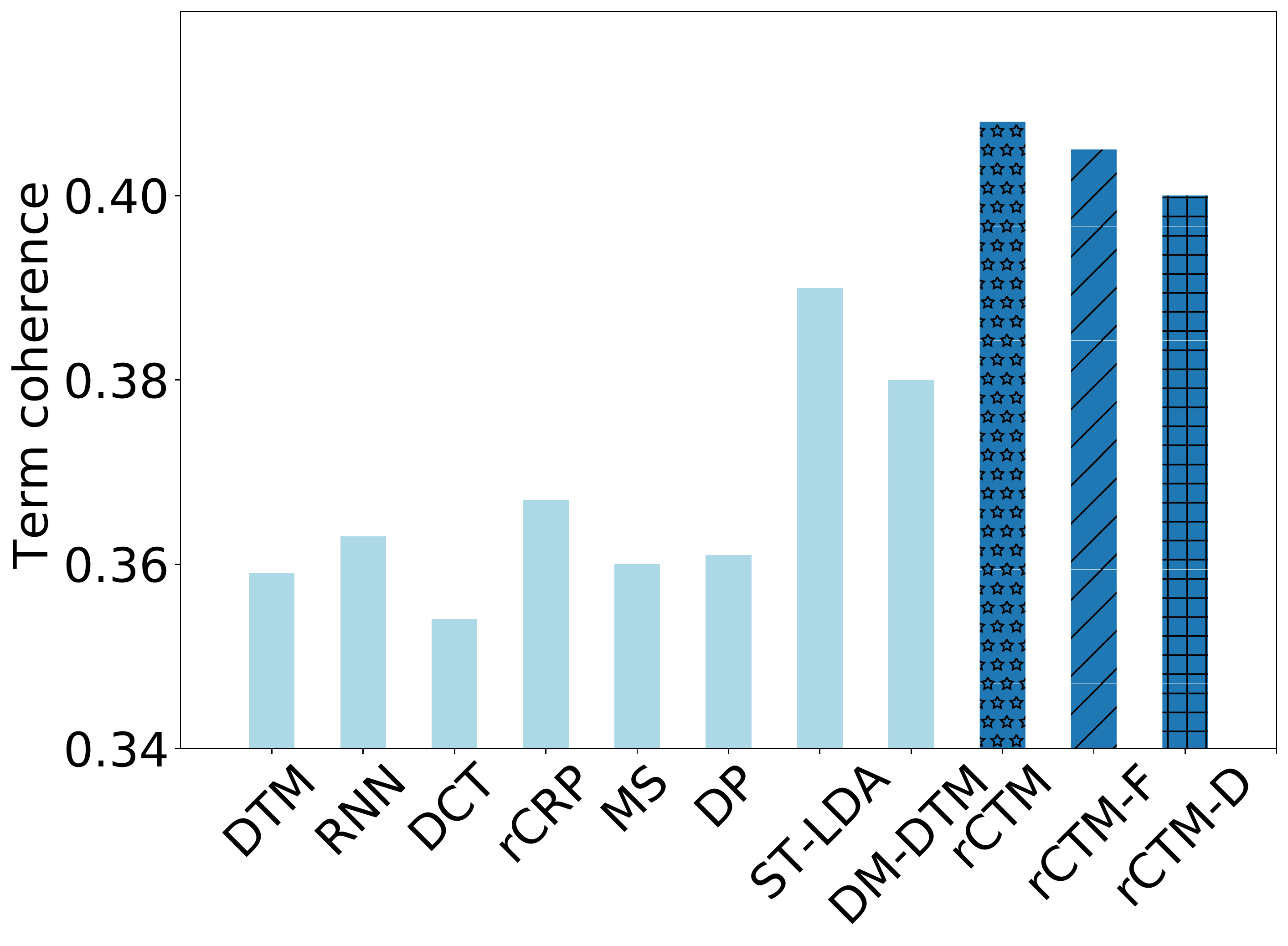}} 
\hspace{0.3in}
\subfigure[SOTU dataset]{
	\includegraphics[width=.32\columnwidth]{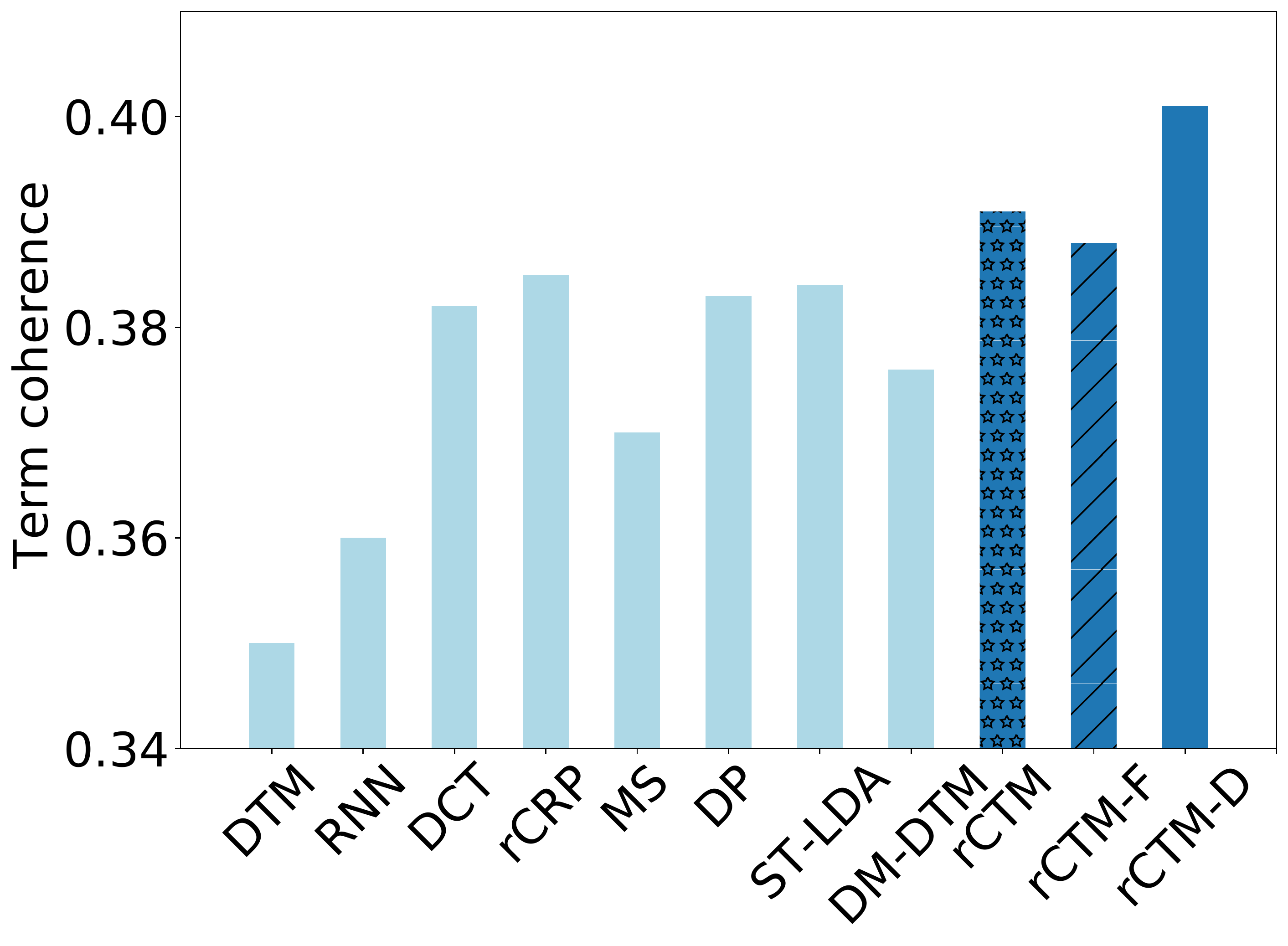}} 	
\caption{Average term coherence results from all competitor models on the five datasets, where the proposed rCTM, rCTM-F and rCTM-D are emphisized by the rightmost three bars. (For the purpose to align with their coherence bars, RNN refers to RNN-RSM, MS corresponds to MStream, DP is short for DP-density here). }
\label{fig:coherence}
\end{figure}

We proceed to evaluate the interpretability of detected topics based on the important measure of topic coherence normalized $PMI$ (Pointwise Mutual Information) \cite{lau2014machine}, which is formally defined as based on the top-$k$ terms within a topic,
\begin{equation*}
	C = \frac{2}{k(k-1)\sum_{i=1}^{k-1}\sum_{j=i+1}^{k} nPMI(w_i, w_j)}, \qquad nPMI(w_i, w_j) = \frac{log\frac{P(w_i, w_j) +\epsilon}{P(w_i)P(w_j)}}{-log(P(w_i, w_j) + \epsilon)},
\end{equation*}
where $P(w_i, w_j)$ denotes the probability of co-occurrence of $w_i$ and $w_j$ in one document and $P(w_i)$ is the probability of $w_i$ appearing in the document. A higher $PMI$ value indicates the terms within the topics are more consistent and interpretable. To obtain an unbiased result, we resort to the large-scale external Wikipedia data \cite{roder2015exploring} to measure the top-$10$ coherence values for all competitor models. The average topic coherence results based on all topics from each slice are presented in Fig. \ref{fig:coherence}, and the analysis of all competitor models is in the following.

On the NIPS and Flickr dataset, the topic coherence results are presented in Fig. \ref{fig:coherence} (a) and (b). The dropout probability of rCTM-D is set as $\rho = 0.2$ on both datasets, at which rCTM-D obtains the lowest perplexity. It is observed rCTM and its variants achieve the highest coherence scores among all competitors, and rCTM is superior to its variants with a higher coherence score, indicating more interpretable and coherent topical terms therein. The superiority of rCTM-based models over the other baselines with the single-topic-thread evolution assumption vouches for the significance of multi-thread couplings between evolving topics. In addition, DP-density is noted to retain its advantage over other baselines with a higher coherence score, while ST-LDA degrades and RNN-RSM, MStream, and DM-DTM improve their performance by a large margin on the Flickr dataset.
                                                                                                                 
On the News dataset, the dropout probability in rCTM-D is set as $\rho = 0.1$. The coherence results from all competitors  are presented in Fig. \ref{fig:coherence} (c). It's noted that rCTM still outperforms others with the highest coherence score, and baselines including RNN-RSM, DP-density, ST-LDA, DM-DTM and rCTM-F offer the competitive coherence scores. Among them, DM-DTM is tied with rCTM-F for second place and performs better than ST-LDA and rCTM-D. Besides, RNN-RSM consistently outperforms DTM by a large margin, and rCRP also performs better than DCT with a higher coherence score. In contrast, the performance of MStream degrades, implying its disadvantage on a dataset with dense arrivals. 
 
On the ACL and SOTU dataset, the dropout probability of rCTM-D in these two long-text datasets is given as $\rho = 0.1$ and $\rho = 0.6$ respectively. The coherence results are presented in Fig. \ref{fig:coherence} (d) and (e). On the ACL dataset, the proposed rCTM is superior to its variants, and ST-LDA is followed among the rest competitors. In constrast, on the SOTU dataset, rCTM-D with $\rho = 0.6$ is the winner with the highest coherence value and rCTM is the runner-up compared with other baselines. Besides the proposed model, the performance of rCRP, DCT, DP-density and ST-LDA is close in the coherence measure, while the performance of DTM, RNN-RSM decreases in these two long-text documents.

In a nutshell, the proposed rCTM and its variants exhibit superiority to other baselines in terms of the coherence metric on different datasets. Such performance is roughly consistent with the perplexity results, which once again confirms the significance of modeling multiple couplings between evolving topics as well as the sparse customization of topic proportions in rCTM. Besides, DP-density, ST-LDA, DCT and rCRP are robust on different datasets without a drastic change in the coherence values. However, RNN-RSM and DTM are advantageous on the short-text datasets, while DM-DTM more fits the datasets with densely irregular document arrivals. The performance of MStream is not satisfactory on the NIPS, News, ACL and SOTU datasets.

\subsubsection{Document Time Stamp Prediction.}
To further evaluate these recurrent modelings, referring to the empirical study \cite{gupta2017deep} we split the sequential documents at each time slice when the ratio $p=0.9$, and predict the time stamp of a document on the held-out dataset by finding the most likely location based on the topics with maximum likelihood over the timeline. The document stream-based methods are excluded due to the different settings and the results of document time stamp prediction accuracy from the rest competitors are presented in the Table. \ref{tab:time_pred}.

\begin{table}[t]
	\centering
	\caption{Document time stamp prediction accuracy results over the five datasets (The best performance is highlighted in boldface).}
	\resizebox{0.9\columnwidth}{!}{%
	\begin{tabular}{@{} l @{\hspace{0.7cm}} c c c c c c c c c @{}}	\toprule[1.2pt]
	           & DTM    & RNN-RSM & DCT    & rCRP  & ST-LDA & DM-DTM & rCTM  & rCTM-F & rCTM-D \\  \midrule[1pt]
		NIPS   &  0.50  & 0.52   &  0.45   & 0.50  &  0.45  &  0.53  & \textbf{0.55}  &  0.53  &0.54\\ \midrule
		Flickr &  0.46  & 0.48   &  0.45   &  0.46 &  0.53  &  0.50  & \textbf{0.54}  &  0.52  &0.54   \\ \midrule
		News   &  0.53  & 0.53   &  0.49   &  0.50 &  0.54  &  0.52  & \textbf{0.55}  &  0.54  & 0.54  \\ \midrule
		ACL    &  0.60  & 0.62   &  0.63   & 0.64  &  0.67  &  0.62  & \textbf{0.68}  &  0.67  & 0.66  \\ \midrule
		SOTU   &  0.40  & 0.43   &  0.45   &  0.45 &  0.47  &  0.44  & 0.48  &  0.48  &  \textbf{0.51}  \\ \bottomrule[1pt]
	\end{tabular}
}
\label{tab:time_pred}
\end{table}

It's noted that the proposed rCTM as well as its variants rCTM-F and rCTM-D outperform the other baselines with the higher prediction accuracy over five datasets, implying the higher semantic match between the held-out documents and recognized topics along the timeframe. Among the rest of competitors, ST-LDA outperforms the other baselines on the Flickr, News, ACL and SOTU dataset. DM-DTM and RNN-RSM gain comparable results over the five datasets while enjoying the advantage in the short-text datasets, which is true of DTM. In addition, rCRP obtains a better prediction accuracy than DCT over the five different datasets.

\subsubsection{Effects of Varying $\mathbf{\eta}$}
\begin{figure}[!htp]
	\centering
	\subfigure[NIPS dataset]{
		\includegraphics[width=.42\columnwidth]{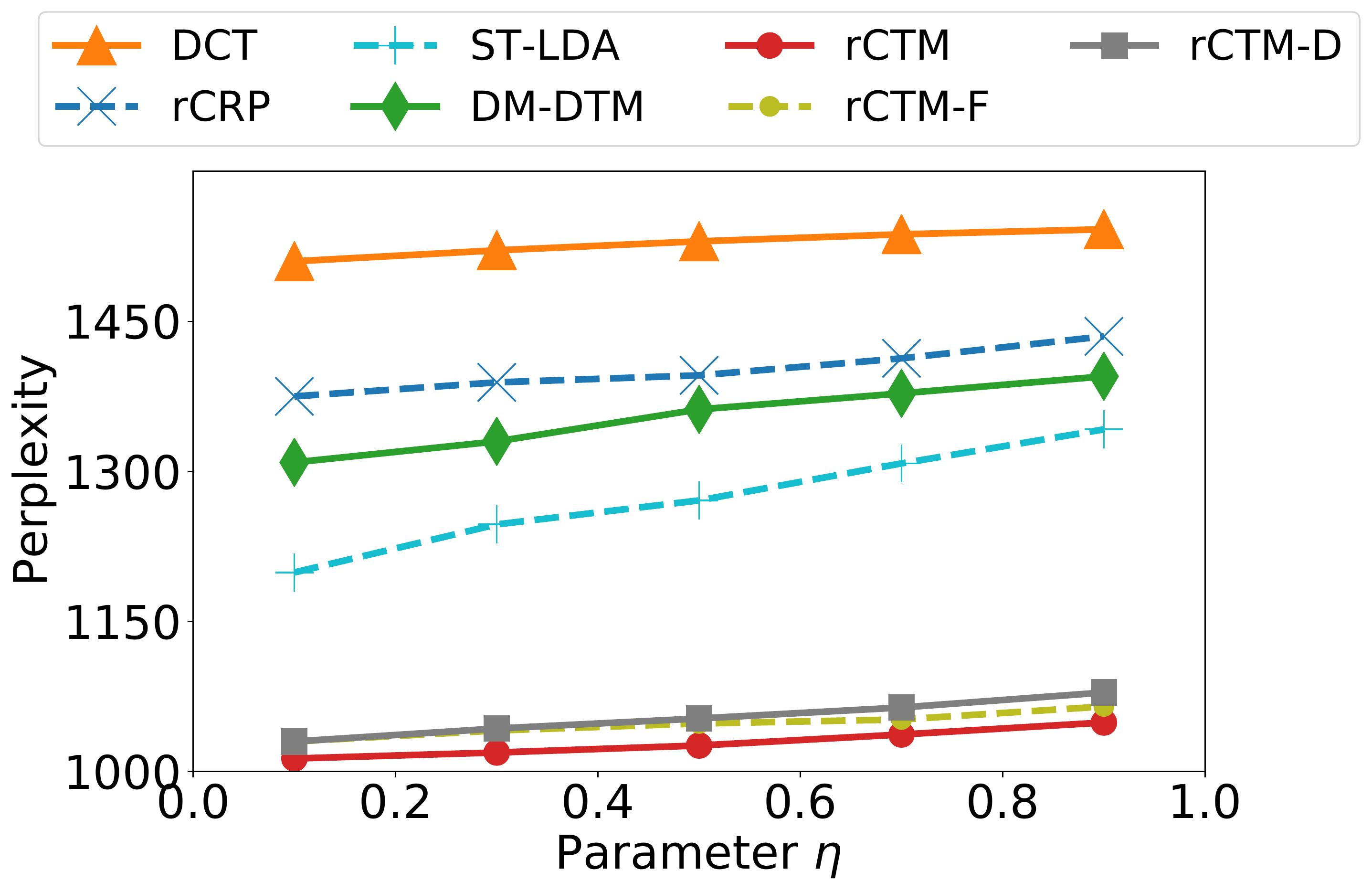}}
	\hspace{0.15in}
	\subfigure[Flickr dataset]{
		\includegraphics[width=.42\columnwidth]{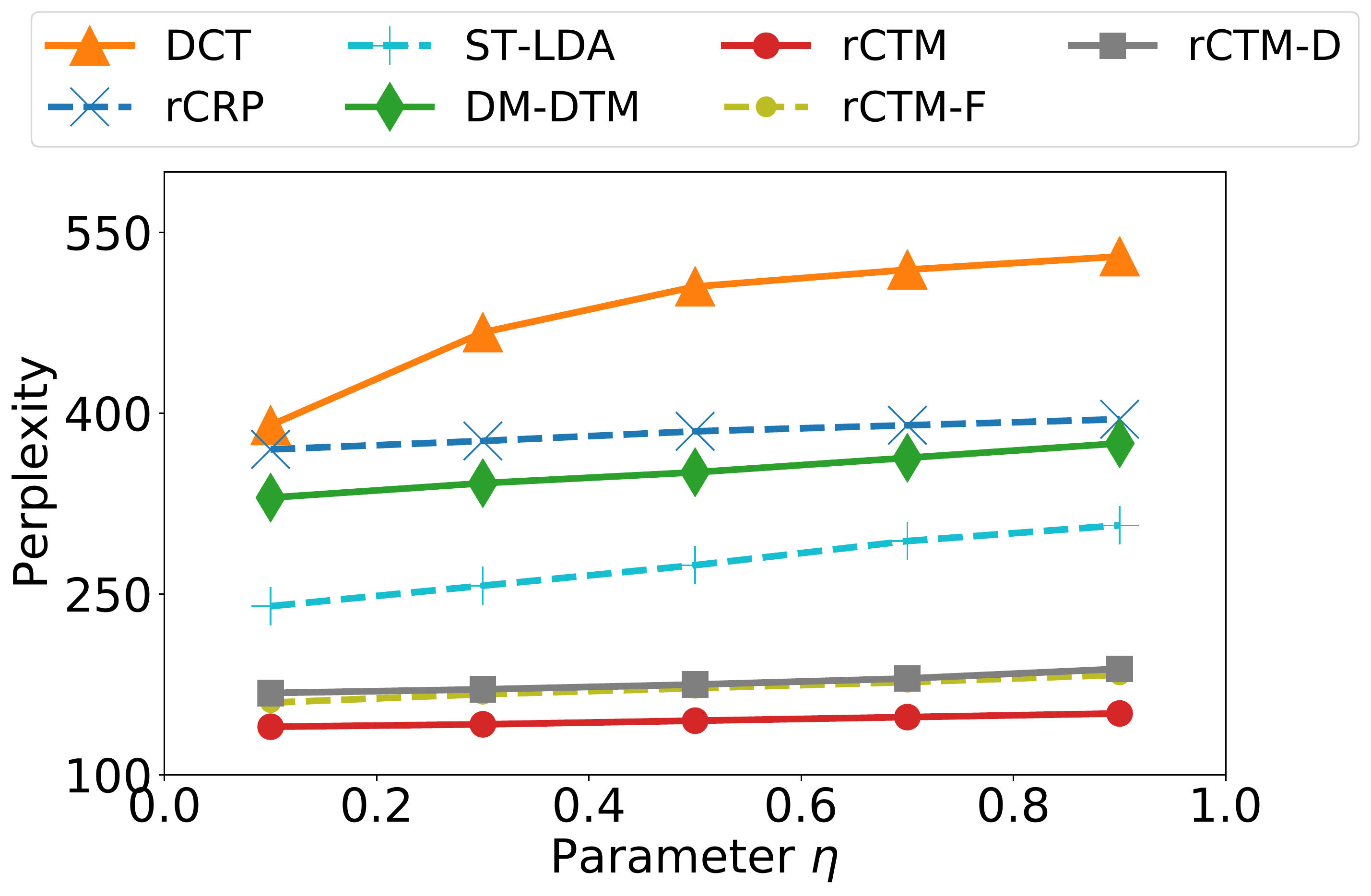}}
		\hspace{0.15in}
	\subfigure[News dataset]{
		\includegraphics[width=.42\columnwidth]{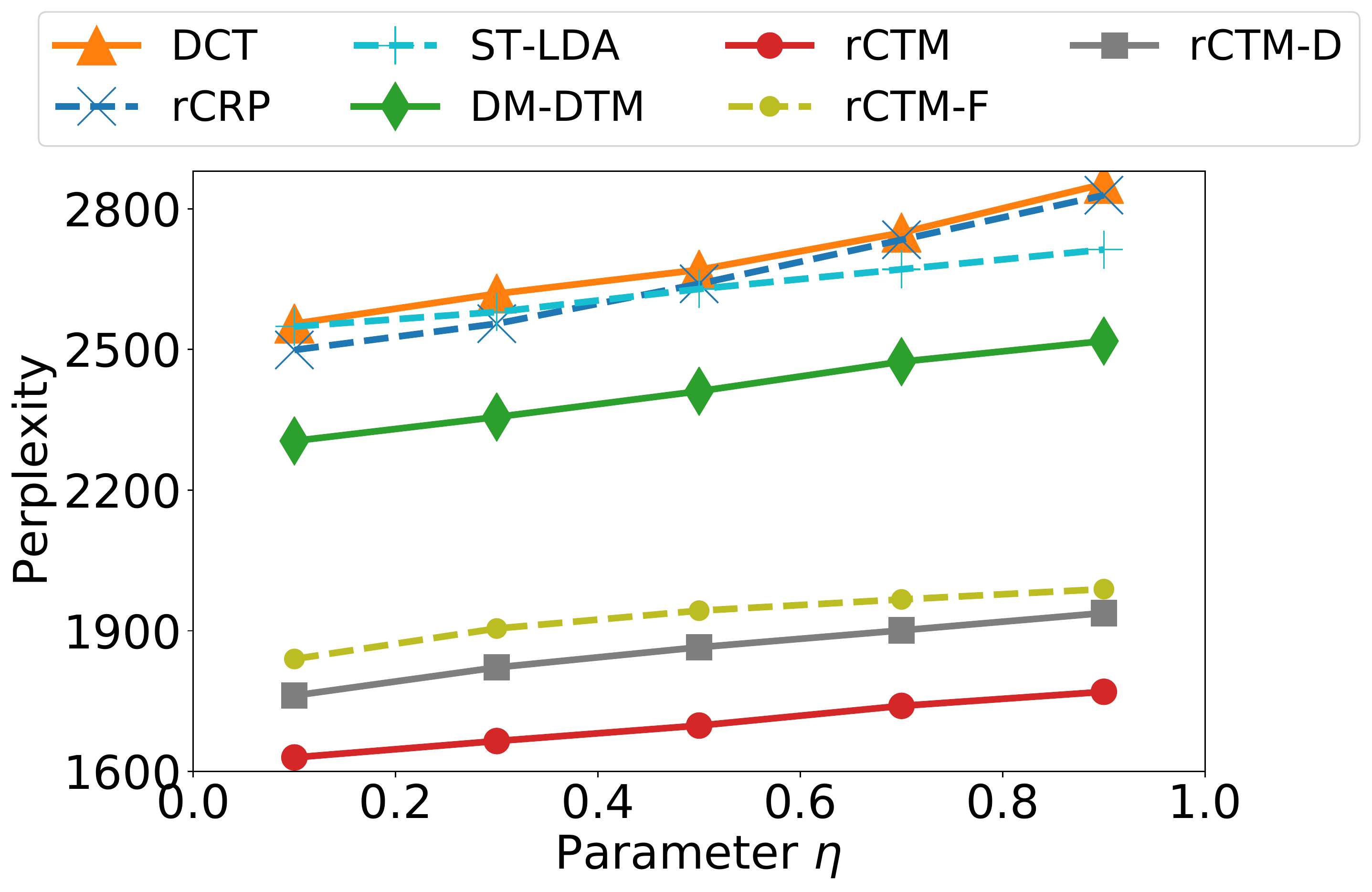}}
		\hspace{0.15in}
	\subfigure[ACL dataset]{
		\includegraphics[width=.42\columnwidth]{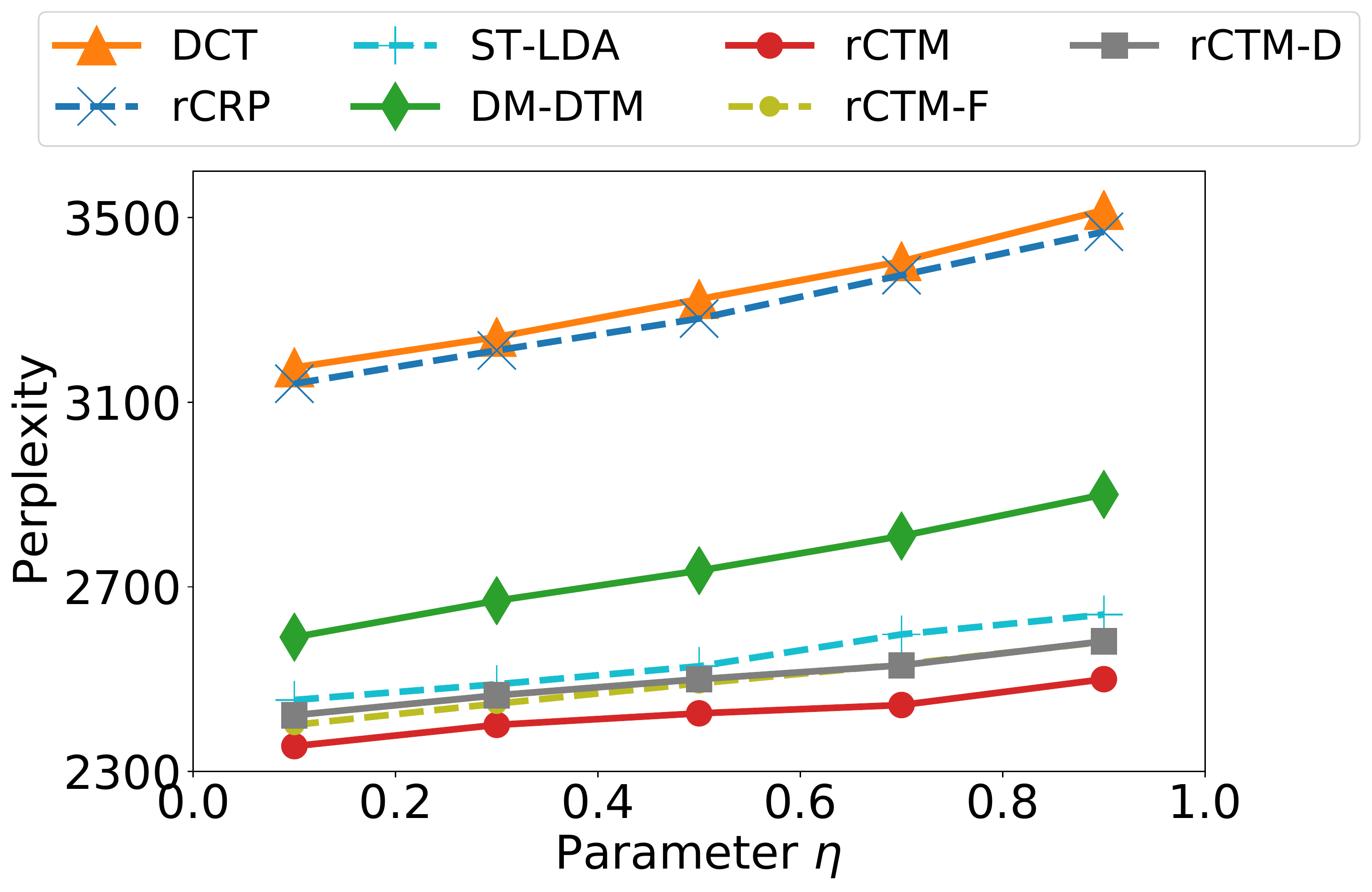}}
		\hspace{0.15in}
	\subfigure[SOTU dataset]{
		\includegraphics[width=.42\columnwidth]{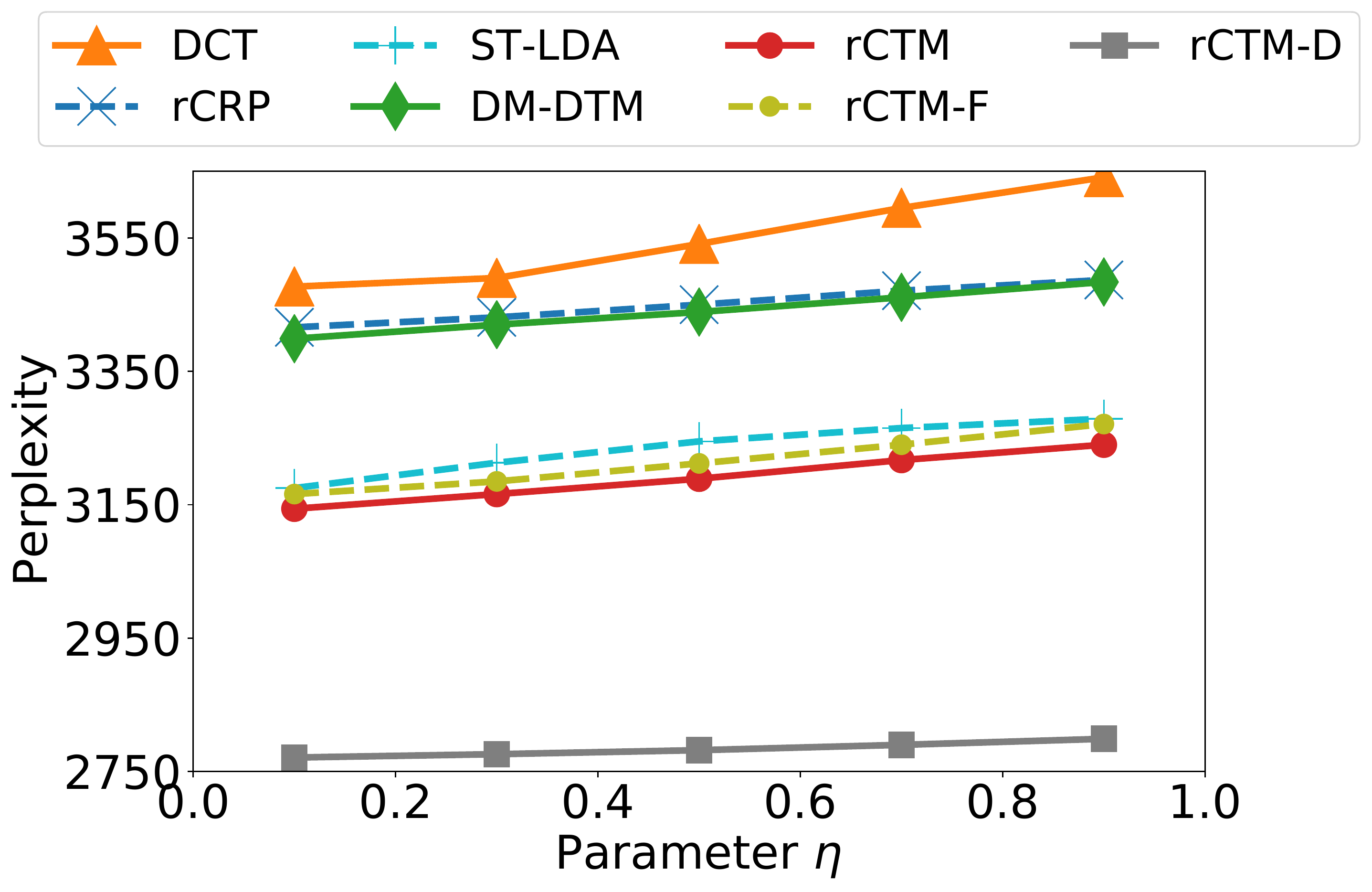}}
	\caption{Performance comparison with varying the parameter $\mathbf{\eta} \in (0, 1)$ on the five datasets. }
	\label{fig:para}
\end{figure}

Since topics at slice $t=1$ are directly learned via $\mathbf{\phi}_{k_1} \sim Dir_V(\eta)$ ($ k_1 \in \{1, 2, \cdots, K_1\}$) without prior dependency. Then they serve as the input to the recurrent coupled topic sequences, and the posterior topics as well as their coupling relationships at slice $t > 1$ are sequentially learned. Hence, the results of topics at slice $t=1$ are important for the whole topic evolutionary process, which is also true for other dynamic topic models. To see the effects of varying $\mathbf{\eta}$ on the overall performance, topics at slice $t=1$ are initialized with varying $\mathbf{\eta}$ in these competitors following the prior work \cite{liang2016dynamic,acharya2018dual}, and the overall performance on the five datasets is presented in Fig. \ref{fig:para}.

The results of the document stream-based approaches as well as DTM and RNN-RSM are excluded due to the different settings. The perplexity performance is measured on the held-out set when $p = 0.9$ on the five datasets, and the dropout probability in rCTM-D is set $\rho =0.2$ on the NIPS, $\rho = 0.2$ on the Flickr, $\rho = 0.1$ on the News, $\rho = 0.1$ on the ACL and $\rho = 0.6$ on the SOTU datasets.
We observe that, in addition to the lowest perplexity results, the proposed rCTM and its two variants acquire a slower increase than other baselines with $\mathbf{\eta}$ growing, which demonstrates the merit of rCTM and its variants that they are robust and less sensitive to the growing $\mathbf{\eta}$ with the multi-topic-thread evolution assumption.
On the other hand, the increasing perplexity from all competitors indicates that 
varying $\mathbf{\eta}$ affects their performance in the task of evolving topic sequences, and 
a small value $\mathbf{\eta}$ to initialize topics at slice $t=1$ is preferred by these document-chunk based topic modelings.

\subsection{Qualitative Results}
\subsubsection{Topic Evolving Sequence with Coupled Dependencies}
To have an intuitive understanding of evolving topics as well as their multi-dependency relationships, we present two representative examples to exhibit the evolutionary process.

\begin{figure}[!htp]
	\centering
	\subfigure[Recurrent topic sequences.]{
		\includegraphics[width=.90\columnwidth]{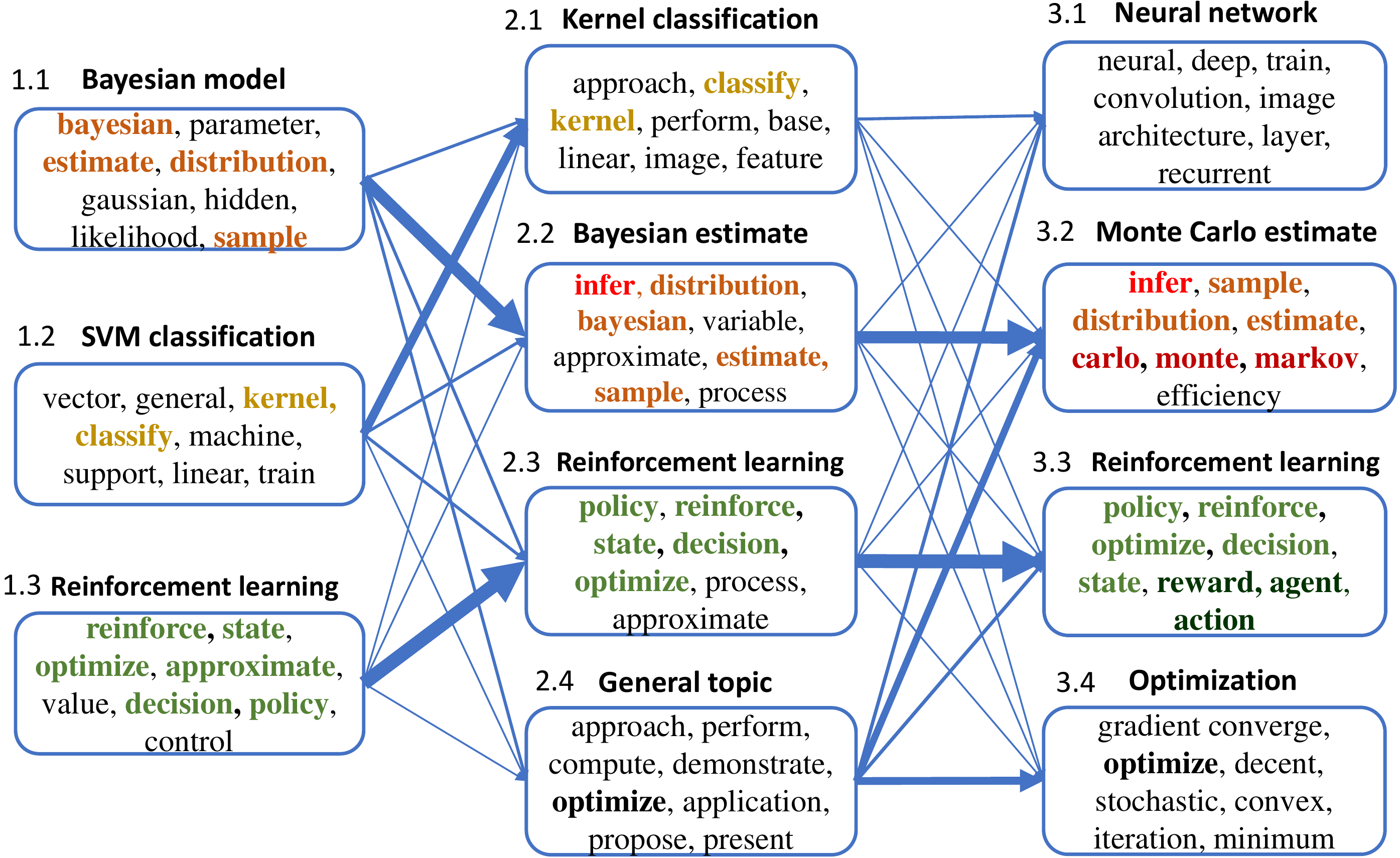}} 
    \subfigure[Corresponding coupling weights between topics in two consecutive slices.]{
    	\includegraphics[width=.90\columnwidth]{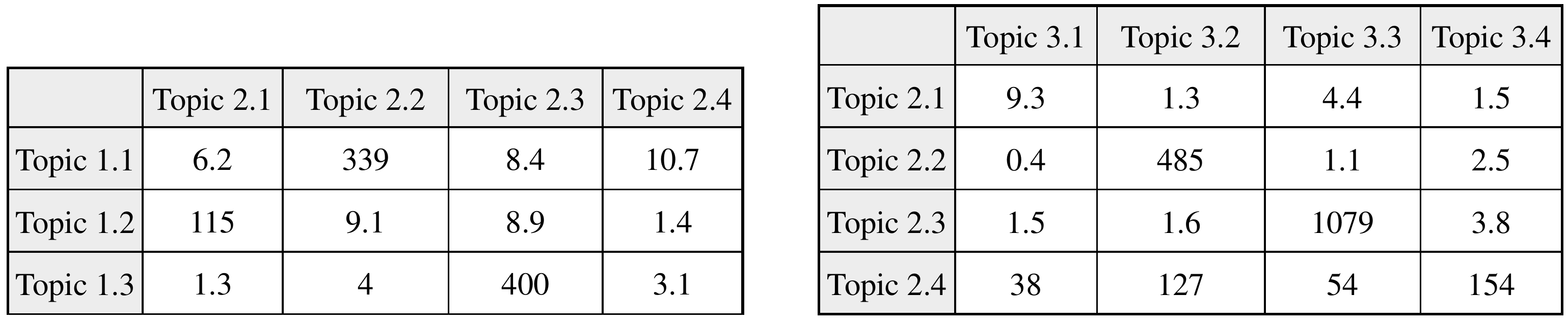}} 
    \caption{The example of topic evolving sequence discovered on the NIPS dataset. In Fig. (a), topics are annotated by their time-slice and index, each of them is represented by the top-$8$ words within the rectangular, common words between consecutive topics are highlighted by different colors and the thickness of arrows between consecutive topics indicates their coupling strengths. Fig. (b) summarizes the exact coupling weights between consecutive topics. }
    \label{fig:example_nips}
\end{figure}

Fig. \ref{fig:example_nips} (a) presents the recurrent topics on the NIPS dataset, which is divided into four equidistant time-slices, and topics in the last three slices are exhibited considering the space limit. Fig. \ref{fig:example_nips} (b) provides the corresponding weights between consecutive topics, which summarizes the sharing of latent  word counts between them. Our observations are in the following. (1) Topics in each column are semantically meaningful by the most probable words, and similar topics are closely coupled with the highlighted common words across the slices. For example, the topic sequence about Bayesian Method evolves from topic 1.1 -> topic 2.2 -> topic 3.2 across three slices with strong coupling weights indicated in Fig. \ref{fig:example_nips} (b), and the highlighted common words in their contents, such as `bayesian', `distribution' and `estimate', are shared crossing the slices.  In addition, the topic sequence about Reinforcement Learning develops from topic 1.3 -> topic 2.3 -> topic 3.3 with strong coupling dependencies across the slices. (2) Besides long-term topic sequences, topic 1.2 about SVM Classification in the first column evolves to the subsequent topic 2.1, which weakly connects with the posterior topics in the last slice. In comparison, the general topic of topic 2.4 contributes to all posterior topics with different coupling weights. (3) Coupling weights indicate that new topic 3.1 about Neural Network weakly connects with the priors and no common words are shared, which fits the fact that the topic of the neural network gets its popularity in recent years.

\begin{figure}[!htp]
	\centering
	\subfigure[Recurrent topic sequences.]{
		\includegraphics[width=.90\columnwidth]{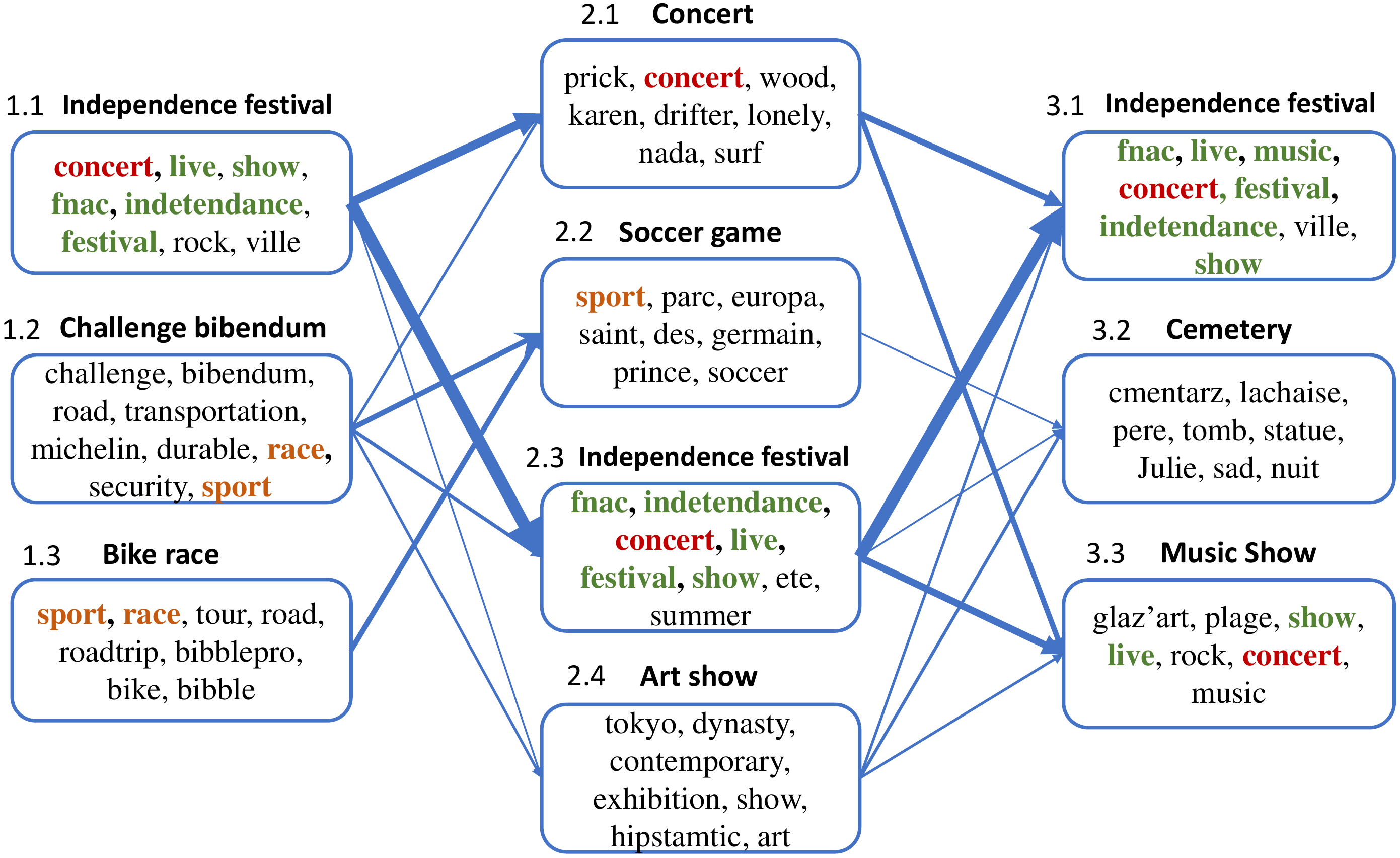}} 
	\subfigure[Corresponding coupling weights between topics in two consecutive slices.]{
		\includegraphics[width=.90\columnwidth]{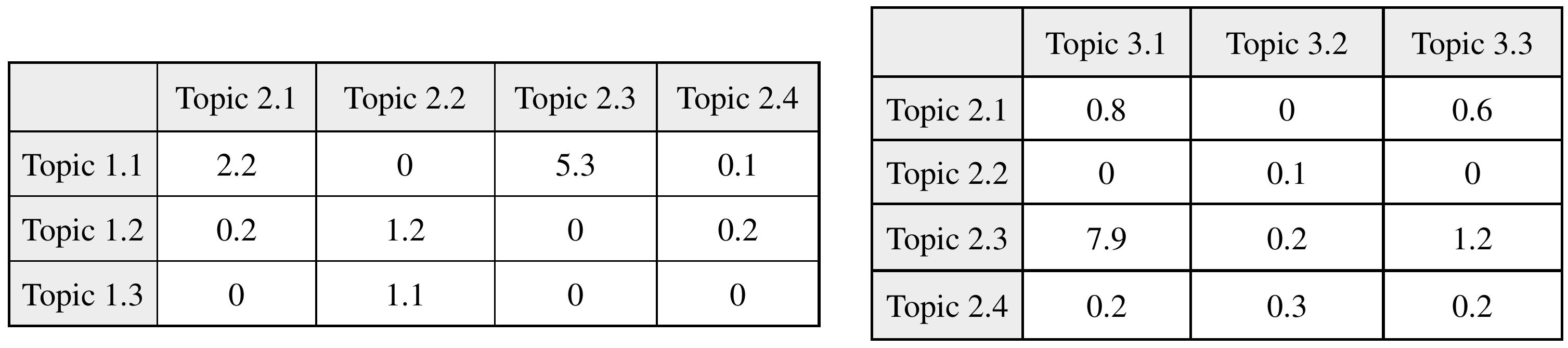}} 
	\caption{The example of topic evolving sequence discovered on the Flickr dataset. In Fig. (a), topics are annotated by their time-slice and index, each of them is represented by the top-$8$ words within the rectangular, common words between consecutive topics are highlighted by different colors and the thickness of arrows between topics indicates their coupling strengths. The arrows with coupling weight $0$ are not plotted. Fig. (b) summarizes the exact coupling weights between consecutive topics.}
	\label{fig:example_flickr}
\end{figure}

Distinct from scientific topic sequences on the NIPS dataset, the Flickr dataset records real social activities in the world and its topic sequences together with their coupling weights are presented in  Fig. \ref{fig:example_flickr}. We report the recurrent topics in the last month and each slice lasts for ten days. We observe that (1) the coupling weights between consecutive social activities on the Flickr dataset are small compared with scientific examples on the NIPS dataset, and some coupling weights are $0$. That is because each Flickr document contains fewer words and the discovered topics from Flickr are real and different activities. The coupling weights between them are thus small. (2) Relevant topics are coupled while their coupling weights and shared common words are distinguishable, for example, even though the topic sequence about Concert evolves as (topic 1.1) -> (topic 2.1, topic 2.3) -> (topic 3.1, topic 3.3) in multiple threads across three slices. Indicated by the coupling weights in Fig. \ref{fig:example_flickr} (b), topic 2.1 couples the posterior topics with the small weights, and topic 3.3 also weakly couples its prior topics. Though both of them talk about Music, they are distinguished from the other strong coupled topic sequences on Independence Festival, which shares more common frequent words, e.g., `fnac' and `indetendance', highlighted in green color. In addition, the topics about Sport (topic 1.2, topic 1.3) -> (topic 2.2) are naturally chained, and the small weights between them indicate each topic records a different sports event, which is reinforced that no more common words are shared between them except for `sport' and `race'. (3) Unrelated topics are naturally identified by the small coupling weights. For example, topic 2.2 is about Soccer Game, whose connections with the posterior topics are denoted by the small weights, and it is also true for topic 2.4, which is unrelated to the posterior topics.

Two intuitive examples from the NIPS and Flickr datasets further prove the effectiveness of multi-dependencies associated with prior topics. And the flexible weights learned via the hierarchical Gamma distribution successfully identify the evolutionary closeness between consecutive topics. 

\section{Related Work}
Most of the dynamic topic models are built under the single-topic-thread assumption that the current state of one topic solely depends on its own historical states without referring to other topics. We summarize the related dynamic models in three aspects. The first two modelings are inherited from temporal topic modeling, and the third one is founded on Poisson factor analysis. Last but not the least, we briefly compare language models with recurrent neural networks.

\stitle{State space modeling}. One of the benchmark models learning the evolution of topics is the state space model, in which the \textit{V}-dimensional topic $\mathbf{\phi}_t$ at step $t$ evolves via $\mathbf{\phi}_{k_t} \sim \mathcal{N}(\mathbf{\phi}_{k_{t-1}}, \Sigma)$ \cite{blei2006dynamic} or the linear form $\mathbf{\phi}_{k_t} \sim \mathcal{N}(\beta \mathbf{\phi}_{k_{t-1}}, \Sigma)$ \cite{schein2016poisson}. The seminal work dynamic topic model (DTM) \cite{blei2006dynamic} captures the evolution of topics by the state space models over the sequence of discrete time-slices, where Kalman filter \cite{kalman1960new} infers temporal update of the state space parameters. Comparing with the classic DTM, our model significantly differs in three aspects. Firstly, instead of fixed topic number setting in DTM, our model is capable to automatically learn the topic number at each slice as well as the sparse topic proportions of each document and thus accommodate new topics along the time. Second, the topic evolves in a single thread in DTM and it fails to identify the influences from other related topics. However, our model breaks such a limitation and supposes topic evolves in multiple threads with corresponding dependencies on the previous. Furthermore, our model induces a tractable and efficient inference method with data augmentation techniques, and such an inference problem cannot be solved by DTM.
The later continuous time dynamic topic model (cDTM) \cite{wang2008continuous} replaces the discrete state space model and detects the evolution of topics over continuous documents using Brownian motion, where variational Kalman filtering is exploited to infer the parameter in the continuous time setting. To relieve the manual setting of the topic number, the work \cite{ahmed2008dynamic, ahmed2011online, ahmed2010timeline} facilitates the nonparametric prior of a Dirichlet process to automatically derive the topic number for the sequential documents. Among them, topics (storylines) in \cite{ahmed2008dynamic, ahmed2011online} are chained via a recurrent Chinese Restaurant Process (rCRP), which allows topics to evolve with genesis and death. While the base measure of topics in \cite{ahmed2010timeline} is tied via the rCRP, documents are generated from an epoch-specific hierarchical Dirichlet process. In these scenarios, the number of topics or themes are flexibly learned rather than predefined and their topic transitions are chained by Gaussian state space models. Successful as state space modelings are, one of the main deficiencies is that these models suffer from a heavy computational cost due to the non-conjugate problem, and their scalability would be prohibitive in the high dimensional data. To this end, a line of research work is thus developed to mitigate the problem. The work \cite{linderman2015dependent} employs the P\'{o}lya-Gamma augmentation trick to provide a conditionally conjugate scheme for Gaussian priors. To mitigate the scalability limit from the state space modeling, the work \cite{jahnichen2018scalable} presents a generalized class of tractable priors and scalable approximate inference to explore both long-term and short-term evolving topics, while the work \cite{bhadury2016scaling} proposes a parallelizable inference using Gibbs sampling with Stochastic Gradient Langevin Dynamics to scale up the dynamic topic modeling in both single and distributed environments. Besides the scalability, the work \cite{nallapati2007multiscale, iwata2010online, chen2018ims} focuses on the evolving topics with various time-scales of the resolution, which allows topics to evolve in the different scales. 

Even though significant progress has been made in state space-based models in the task of topic evolution, such models restrict the evolving topics under the single-topic-thread assumption and fail to capture the potential multiple dependencies between evolving topics. Without thoroughly encoding the complex temporal relationships between time-evolving topics, the learned evolution of topics might be defective.

\stitle{Dynamic modeling with the Dirichlet chain.} Extensive studies exploit the Dirichlet distribution to chain the dynamics of topics over the sequence of discrete slices, in which the tractable inference of sampling topics becomes the advantage over the state space models due to benefit of the Dirichlet distribution.
The topic tracking model (TTM) \cite{iwata2009topic} and dynamic clustering topic model (DCT) \cite{liang2016dynamic} harness a Dirichlet distribution to chain the consecutive evolving topics over the text stream, where the evolution of topic popularity and word distributions depend on their prior states via two Dirichlet Markov chains. 
In comparison, the dual Markov dynamic topic model (DM-DTM) \cite{acharya2018dual} employs two different Markov chains to detect the topic evolution in the count data, where the topic popularity is modeled by the Gamma Markov chain, and the evolution of topics is also captured by the Dirichlet distribution. The work \cite{liang2019collaborative} turns the problem of user interest drifting to be the evolution of topics, where the dynamics of topics over time is also captured via a Dirichlet chain. Despite the wide application of the Dirichlet chain, little attention is paid to the inference of evolutionary weight between consecutive topics, which is still intractable and incurs heavy computation. Furthermore, approaches with a Dirichlet chain ignore multi-dependency relationships between time-evolving topics. In contrast, though the proposed rCTM also exploits a Dirichlet distribution to chain evolving topics, it breaks the limitation of the single-topic-thread evolution and proposes a new framework where the current topic evolves from all prior topics with the corresponding coupling weights. To avoid the confusion with correlated topic modeling (CTM) \cite{blei2006correlated,he2017efficient}, we clarify the major difference in two aspects. First, the correlation between topics in CTM indicates the existence of correlation in their proportions via the logistic normal distribution. In comparison, the couplings between evolving topics are defined as the coupling closeness between their word distributions via 
the hierarchical Gamma distributions. Second, our proposed model aims at encoding the complex temporal correlations between evolving topics in the dynamic context while CTM is limited to a static text dataset.

Besides the above dynamic modelings in the context of discrete slices, a line of studies captures the dynamics of topics from the continuous document streams. The work \cite{du2015dirichlet} combines Dirichlet and Hawkes processes to capture the dynamics of topics from sequential documents, in which the Hawkes process learns temporal density of topics with multiple predefined Gaussian kernels. The work in \cite{guo2017density} further mitigates the restriction from the predefined kernels and exploits the density estimation technique to incrementally learn the dynamics of topics with the sliding window. In addition, the Temporal LDA \cite{wang2012tm} aims at predicting the transition of topic weight in the future documents while ignoring the transition of its word distribution. In comparison, the work in \cite{amoualian2016streaming} puts forward a Bayesian model named streamLDA to learn the transition of topic weight as well as its word transition between consecutive documents. In addition, a large number of researches focus on continuous streaming short texts from social media to reveal the topic drift. The work in \cite{yin2016text, yin2018model} makes an effort to incrementally cluster the short text streams from social media and uncover the dynamic clusters (topics) by assigning one topic to each short text. A joint model in \cite{xu2018topic} handles the Chinese streaming short text by integrating the prior of rCRP and biterm topic model \cite{yan2013biterm} to detect the dynamic topics. Given the meta features of social media data, the work in  \cite{zhang2016geoburst,zhang2017triovecevent} incrementally groups the continuous tweets into different varying topic sets according to the combination of textual contents, spatial and temporal features.

\stitle{Dynamic Poisson factor analysis.} Targeting at the count data, it is a matrix factorization method for the discrete sequential count data under the Poisson factor analysis (PFA) \cite{acharya2015nonparametric}. Though some applications of PFA are not the focus of this paper, for the sake of completeness, we discuss some representative work to introduce how the latent variables evolve over the count data. The work \cite{schein2016poisson} proposes a Poisson-Gamma dynamic system (PGDS) for sequentially count data, where the latent states of topic proportions are chained via the Gamma shape parameter. Its later deep variant, \cite{guo2018deep}, extends the Poisson-Gamma dynamic system by constructing a hierarchical latent structure for the topic proportions, which allows both first-order and long-range temporal dependencies. We credit the data augmentation technique in the proposed model to these approaches. The recent work in  \cite{schein2019poisson} closely relates with PGDS and presents the Poisson-randomized Gamma dynamic system for the sequential biased data with sparsity or burstiness. In addition, the work in \cite{do2018gamma} models the evolution of latent factors in terms of user preferences and item features in the context of recommender system via the Gamma scale parameters. Some studies based on the dynamic relational data \cite{yang2018dependent,li2020recurrent} learn the evolution of node membership by leveraging data augmentation technique under the framework of Poisson factor analysis.

\stitle{Comparison with the RNN-based language models.} In addition to the Bayesian approaches, a line of studies \cite{Dieng2016TopicRNN,Lau2017Topically,wang2018topic, guo2020deep} integrates topic models and language models and inherits merits from both sides. Among them, the work in \cite{Dieng2016TopicRNN} develops the model TopicRNN, where the global semantics is captured by the topic modeling while the local dependency between words within a sentence is detected by a recurrent neural network (RNN). The work in \cite{Lau2017Topically} integrates two components to jointly learn topics and word sequence, where a word sequence is predicted via the RNN. The work in \cite{wang2018topic} simultaneously learns the global semantics of a document via a neural topic model and uses the learned topics to build a mixture-of-experts language modeling based on RNN. The model of RNN-RSM in \cite{gupta2017deep} also aims at recurrent topic discovery, and it leverages the Restricted Boltzmann Machines (RBMs) to define the interaction between topics and words and the RNN is used to convey the temporal information and update the bias parameters of RBMs. In this solution, consecutive topics are not directly connected, which is a marked contrast to the stochastic multi-topic-thread assumption between topics in our model. Additionally, it adopts the contrastive divergence algorithm to estimate the parameters, which also differs from the Gibbs sampling in our Bayesian network.
The most recent paper \cite{guo2020deep} uses a recurrent deep topic model to guide a stacked RNN for language modeling, and thus the words from a document are jointly predicted by the learned topics via the topic modeling and its preceding words via the RNN. It is noted that most of the RNN-based language models are typically applied at the word level and learn the local temporal dependency between the words. Such a task is quite different from ours. First, the topics (word distributions) capture the global semantics of the corpus by word occurrences across the documents. Such long-range dependency and global semantics may not be captured well by the RNN-based language models \cite{Dieng2016TopicRNN,Lau2017Topically,wang2018topic,guo2020deep}. In addition, our proposed work aims at encoding the temporal dependency between two sets of latent topics across the time steps, which is distinct from the syntactic dependency between words in the RNN-based models. Though the task of encoding dependency is quite different in dynamic topic modelings and RNN-based language models, they could work together to cooperatively capture both global semantics and local dependency for language generation.

\section{Conclusion and Future Work}
We introduce a novel nonparametric Bayesian model, a \textit{recurrent Coupled Topic Modeling} (rCTM) over sequentially observed documents. The multi-fold contributions are summarized in the following. (1) This model breaks the limitation of single-topic-thread evolution from most of the existing work and induces a new and flexible proposal of the multi-topic-thread evolution. Accordingly, the current topics evolve from all prior topics with the corresponding topic coupling weights. Such a flexible proposal naturally adapts to the sequential documents with complex relationships. (2) To tackle the unexplored and intractable inference challenge, we present a novel solution with data augmentation and marginalization techniques to decompose the joint multi-dependencies between topics into separated relationships. A novel Gibbs sampler with a backward-forward filter algorithm is exploited to efficiently infer the fully conjugate model in a closed-form. (3) Without tuning the topic number in sequential documents, we leverage the latent IBP compound distribution to automatically infer the overall topic number and customize the sparse topic proportions for each document, where both short text and long documents are flexibly adapted. To further validate the significance of topic couplings, we borrow the dropout technique from deep learning and incorporate it into the proposed rCTM as a counterpart. Evaluation on both synthetic and real-world datasets demonstrates that  rCTM infers a highly interpretable dynamic structure, and the multi-coupling relationships learned between time-evolving topics are significant to infer the topical structure in future. Further, the experimental results also indicate rCTM is superior over the competitive baselines in terms of low per-word perplexity, high topic coherence and high time prediction accuracy.

Although the analytic posterior of rCTM results in an efficient Gibbs sampling, rCTM is limited by two main disadvantages: 1) Gibbs sampling is a time-consuming batch method when inferring high-dimensional latent parameters compared with the gradient-based optimization methods in the neural networks. 2) It is not easy to plug the valuable side information into the Bayesian network with a predefined structure, e.g. document labels or promising word embeddings \cite{mikolov2013distributed,das2015gaussian,gupta2019document}, otherwise, the structure of Bayesian network has to been reformulated. Therefore, one future attempt is to marry rCTM with neural networks and incorporate the variational Autoencoder \cite{srivastavas17} into the proposed model, where the pretrained word embeddings are possibly incorporated. Another promising direction is to extend the evolving topics into the hierarchically recurrent coupled topics, where not only the coupled topic evolution but also the hierarchical topics from general to specific could be captured.

\begin{acks}
This work was supported by National Key D\&R Program of China (2019YFB1600704), FDCT (FDCT/0045/2019/A1, FDCT/0007/2018/A1), GSTIC (EF005/FST-GZG/2019/GSTIC), University of Macau (MYRG2018-00129-FST), and GDST (2019B111106001).

\end{acks}

\bibliographystyle{ACM-Reference-Format}
\bibliography{reference}

\appendix

\end{document}